\documentclass[manuscript,nonacm]{acmart}

\acmYear{2026}
\setcopyright{none}
\renewcommand\footnotetextcopyrightpermission[1]{}

\usepackage{array}
\usepackage{booktabs}
\usepackage{multirow}
\usepackage{algorithm}
\usepackage{algorithmic}
\usepackage{subcaption}
\usepackage{xcolor}
\usepackage{placeins}


\newcommand{\bpe}{\textsc{bpe}}
\newcommand{\BV}{\textsc{BV}}
\newcommand{\BG}{\textsc{BG}}
\newcommand{\CS}{\textsc{CS}}
\newcommand{\CG}{\textsc{CG}}
\newcommand{\CB}{\textsc{CB}}

\begin{document}

\title{Copy, Predict, Code the Rest: Lossless Graph Compression with Decoder-Side Neighbor Prediction}

\author{Jimmy Dubuisson}
\affiliation{%
  \institution{Vantino}
  \city{Geneva}
  \country{Switzerland}
}
\email{jimmy@vantino.com}
\orcid{0009-0003-4697-8848}

\begin{abstract}
We present a lossless compressor for directed graphs. It copies neighbors from an earlier row, predicts further targets from the part of the graph already decoded, and codes what remains. Its main coder, the candidate-bit coder (\CB{}), proposes a bounded set of candidate targets, mixes their probabilities adaptively, and codes the leftover targets with a popularity-aware model. The encoder tries a fixed set of vertex orders and storage directions and keeps the smallest file that decodes exactly.

On the benchmark graphs outside our development set, the selected files are smaller than the best recorded WebGraph and Zuckerli files and the best published HyperFast results. These savings have a cost. \CB{} decodes much more slowly than the baselines and supports only sequential access. A faster companion coder, CG, is still slower than the baselines on large graphs. An indexed variant, CS-RA, gives smaller seekable files than the measured access baselines on nearly all tested graphs, but its cold queries are slow and it fails on one development graph. We also explain what successful and failed predictions cost, and when copying a repeated row saves information.

\end{abstract}

\ccsdesc[500]{Information systems~Data compression}
\ccsdesc[500]{Theory of computation~Graph algorithms analysis}
\ccsdesc[300]{Theory of computation~Data structures design and analysis}
\ccsdesc[300]{Mathematics of computing~Graph theory}
\ccsdesc[300]{Information systems~Data encoding and canonicalization}
\ccsdesc[100]{Theory of computation~Design and analysis of algorithms}

\keywords{graph compression, vertex ordering, community detection,
  order selection, Layered Label Propagation, reference encoding, bits per edge,
  algorithm engineering, WebGraph}

\maketitle

\renewcommand{\thefootnote}{\fnsymbol{footnote}}
\footnotetext[1]{Preprint. Comments welcome at \texttt{jimmy@vantino.com}.}
\renewcommand{\thefootnote}{\arabic{footnote}}

\section{Introduction}
\label{sec:intro}
A directed graph can be stored as one sorted list of neighbors per vertex. Similar lists need not be written twice. A reference encoder copies shared neighbors from an earlier list and writes the remaining targets separately. The challenge is to use structure beyond that single reference while keeping every decoding decision reproducible.

Our row representation follows \emph{copy, predict, code the rest}. First, an earlier neighbor list supplies copied targets. Second, the graph decoded so far supplies further proposals, and a probability model codes whether each proposal is present. Third, a remainder model writes missing targets explicitly. A positive proposal can save an identifier. An unsuccessful proposal consumes bits. The decoder follows the same steps to reconstruct the exact list.

We implement this representation as the candidate-bit coder (\CB{}) in Adjacently~\cite{adjacently_experimental}. It combines reciprocal and shared-neighbor proposals with a deterministic adaptive probability mixture. Its remainder code uses target popularity observed during decoding. We try three community-based vertex orders and both storage directions. We call this fixed set of encoding choices a \emph{portfolio}. We keep the smallest file that decodes exactly and also report the smallest file from a companion coder.

That companion is Clustered Graph (CG), a reference-based coder that decodes faster than \CB{}. Command Stream (CS) is another reference coder that selects row operations with command tokens. Both have indexed variants for retrieving individual neighbor lists. We call these CG-RA and CS-RA, where RA means random access. Baseline Greedy (BG), an earlier coder with variable action headers and a restricted two-reference option, is covered in the appendices.

We refer to the formats in our library as native formats. They serve different needs. \CB{} gives the smallest sequential archives in the portfolio. CG decodes faster than \CB{}, but is slower than WebGraph and Zuckerli on the large inputs we tested on a common order. CS-RA produces smaller indexed files, with millisecond queries when its decoded cache is disabled. A later CS test ties the portfolio on one input, showing that the benefit of candidate prediction depends on the graph.

We call the benchmark study a \emph{campaign} and each stage of its development a \emph{round}. A round is \emph{frozen} when its encoding choices are fixed before its new measurements. Tests added later are \emph{post-freeze}. The \emph{development set} comprises graphs designated for choosing or refining the method. An input is \emph{count-matched} when its vertex and arc counts match the published input counts.

Seventeen campaign inputs lie outside the declared development set and are count-matched. For each, we compare the selected file with the smallest recorded WebGraph file, the smallest recorded Zuckerli file, and the best published HyperFast configuration. The selected files are 1.7--58.7\% smaller than the strongest baseline. The development rounds reused these benchmark graphs, so the seventeen are not an independent test set. We describe that history before presenting the results.

\paragraph{What must an explanation account for?}
Short gaps are only one source of compression. An order also changes which lists can be copied, which targets can be proposed from the decoded graph, and how informative those proposals are. An explanation must charge unsuccessful predictions and explicit remainder targets as well as copies. After the measurements, we give a cost balance for candidate coding and a statement about the dependence exploited by copying repeated rows.

Our design contribution is the way \CB{} combines these operations. It proposes a bounded set of targets from decoded structure and mixes probabilities online. Explicitly coded targets can seed further proposals, and target popularity guides the remainder code. We specify the coder and portfolio, compare their sizes and decode costs, and use the accounting to formulate testable explanations.

Section~\ref{sec:prelim} introduces the underlying methods. Sections~\ref{sec:encoders} and~\ref{sec:ordering} describe our design. Evaluation precedes mechanism accounting. The appendices give a bit-level tutorial, detailed settings and the earlier BG/CS studies.

\section{Background}
\label{sec:prelim}\label{sec:related}
Let $G=(V,E)$ have $n$ vertices and $m$ directed arcs. A \emph{row} is a vertex's sorted outgoing-neighbor list. Its length is the outdegree. A vertex permutation relabels both source rows and target identifiers. Transposition reverses all arcs and is a separate storage choice.

\paragraph{Reference coding and WebGraph.}
Gap coding writes the first target followed by differences between consecutive targets. Source-relative formats write the first target's offset from the source. WebGraph combines gaps with references, copy blocks and intervals~\cite{boldi2004webgraph,boldi2005codes}. A reference window makes preceding rows available for copying. The reference identifier, copy description and residual targets all contribute to the cost. The reference degree bounds how many targets can be copied. Reference-chain depth bounds successive dependencies during decoding.

\paragraph{Zuckerli and probability models.}
Zuckerli combines reference compression with context coding and hybrid integers~\cite{versari2020zuckerli}. Its probability tables are stored with the graph and remain fixed during decoding. The archive mode uses asymmetric numeral systems (ANS). List-access mode uses Huffman coding. A hybrid token contains magnitude information and selected high/low bits, with the remaining bits stored separately. Our coders learn probabilities online. \CB{} also codes a proposed target's presence directly. If $q$ is its predicted probability of presence, the ideal costs are $-\log_2q$ for presence and $-\log_2(1-q)$ for absence. A range coder realizes these probabilities in a finite byte stream.

\paragraph{Prediction from decoded structure.}
The Backlinks (BL) scheme exploits reciprocal links through membership bits in addition to reference copying~\cite{chierichetti2009compressing}. Zhang et al. use causal decoded context to predict graph structure in scene-graph compression~\cite{zhang2023scene}. Logistic probability mixing has precedent in PAQ~\cite{mahoney2013}, and Suel--Yuan encode popular destinations separately~\cite{suel2001}. These ideas motivate our combination of bounded graph-specific proposal pools, online mixing and causal popularity counts.

\paragraph{Orders and embedding-based compression.}
Layered Label Propagation (LLP) creates locality at several resolutions~\cite{boldi2011llp}. Recursive graph bisection, denoted BP here, uses recursive balanced bisection with a surrogate for logarithmic adjacency-gap cost~\cite{dhulipala2016rgb}. CLOVE arranges a weighted graph of communities as part of a hierarchical embedding~\cite{balogh2025clove}. Our spectral cluster order arranges those groups differently. HyperFast uses hyperbolic embeddings and codes graph structure relative to that geometric description~\cite{celinska2026hyperbolic}. It exploits a different representation of regularity from explicit copying and decoder-side target proposals.

\section{The coders: copy, predict, code the rest}
\label{sec:encoders}\label{sec:algorithms}
The campaign uses CG and the \CB{} coder for sequential archives. CG and CS also have indexed formats for neighbor-list queries. The \CB{} format is sequential-only. Table~\ref{tab:coder-roles} lists these choices. BG adds variable action headers and a restricted two-reference option. Its earlier measurements are in the appendix.

\begin{table}[t]\centering\small
\caption{Coder representations and access modes. Indexed sizes and query costs are evaluated separately from sequential archive sizes.}
\label{tab:coder-roles}
\begin{tabular}{lp{6.5cm}l}\toprule
Coder & Row representation & Access\\\midrule
CG & Reference flag, distance, adaptive copy bitmap, interval/rank residuals & Sequential; chunk indexed\\
CS & Command token selecting reference/residual mode, copy bitmap, rank residuals & Sequential; sampled index\\
\CB{} & Degree, self bit, reference controls, presence bits, remainder & Sequential only\\\bottomrule
\end{tabular}\end{table}

\subsection{Shared integer coding and graph modes}
CG and CS share an integer backend that uses a 32-bit range coder and hybrid tokens with split exponent $K=5$, zero high bits and one low bit in the token. Its alphabet has 150 symbols. Each token context combines position within a region (0, 1, 2, 3+) with the previous value's bit-length bucket (24 magnitude buckets plus a reset sentinel), giving 100 contexts. 

Counts start at one, increase by 32 and rescale at $2^{23}$. Remaining integer bits are coded uniformly. Resetting a region resets position/previous-value addressing. The learned tables persist until a chunk restart.

\emph{Plain} stores the directed adjacency as supplied. \emph{Symmetric} mode stores each undirected pair once and reconstructs both arcs. \emph{Reciprocal} mode retains every non-reciprocal arc, including backward ones. For each forward arc it codes whether the reverse exists and omits that reverse from ordinary row coding, following the BL principle~\cite{chierichetti2009compressing}. These transformations precede row coding and are distinct from transposed storage.

\subsection{CG: streams, reconstruction and cost}
The sequential campaign uses one CG cluster for the whole graph. For each row, overlap screening retains at most 1,024 reference candidates from the window. An analytical Fibonacci estimate compares them with direct coding. The chosen representation is then emitted through the streams in Table~\ref{tab:cg-streams}. The reference window is 256, 1024 or 2048 on the BP order and 256 on the other orders. The distance stream does not reset region addressing between rows. After its first three values, the position bucket is fixed and only the previous magnitude selects the context.

\begin{table}[t]\centering\small
\caption{CG's campaign streams. Residual counts delimit the row, so CG emits no separate degree field. Copy bits follow reference-entry order.}
\label{tab:cg-streams}
\begin{tabular}{lp{4.6cm}p{4.8cm}}\toprule
Stream & Values & Probability model\\\midrule
Structure & Use-reference flag & One uncoded bit\\
Reference distance & Current minus reference row & Hybrid integer; previous magnitude after the initial three events\\
Copy & One membership bit per reference entry & Binary model conditioned on previous copy bit; count increment 24\\
Residual & Interval count/starts/lengths, residual count, left count and distance gaps & Hybrid integers; region addressing restarts for each row\\\bottomrule
\end{tabular}\end{table}

For a row with a reference, the coder removes copied targets from the identifier universe before coding the residual. It maps target $x$ and source $v$ to their ranks among the remaining identifiers. Runs of at least four targets become intervals. 

A zigzag mapping turns signed offsets into nonnegative integers. The residual stream writes the interval count plus one, first interval start as a zigzag source offset plus one, later start gaps and adjusted lengths. It then writes residual count plus one and, for a nonempty residual, left count plus one. Left targets are ordered by increasing distance from the source. Right distances have a plus-one shift to represent zero. Within each half, the first distance is written directly and each later distance difference has a plus-one shift. 

The decoder inverts the rank mapping and merges residual and copied targets. Direct rows retain the original identifiers.

For a concrete row, let $v=5$, reference row $4$ be $\{1,3,7\}$, and the current row be $\{1,3,8\}$. The reference distance is one and the copy bitmap is \texttt{110}. Removing copied identifiers $\{1,3\}$ maps source $5$ to rank $3$ and residual $8$ to rank $6$. There are no intervals, one residual and zero left residuals, so the integer residual sequence is $(1,2,1,4)$. The last value is $6-3+1$.

To make the bit calculation explicit, consider this isolated row operation with fresh integer and binary model tables and an already supplied reference. Each of its five integer events (one distance and four residual integers) accesses a uniform 150-symbol context and costs $\log_2 150$. Copy bits \texttt{110} have probabilities $1/2$, $25/26$, $1/50$, because the binary model starts with counts $(1,1)$, previous bit one, and increments the observed count by 24. Including the structure flag, the ideal cost is
\[
1+5\log_2 150-\log_2\!\left(\tfrac12\tfrac{25}{26}\tfrac1{50}\right)
\approx43.845\ \text{bits}.
\]
This is a specified model-state calculation, before stream framing and termination. A row inside a continuous archive uses the probabilities learned from preceding rows. The calculation illustrates the actual bitmap and integer streams. It is not a fixed per-row byte charge.

Given the decoded reference, reconstruction scans its entries and the output row, taking $O(|N(r)|+d)$ work with bounded-alphabet model operations. Sequential decoding maintains earlier rows. Arbitrary queries may also need a reference chain or chunk. Section~\ref{sec:random-access} measures those costs.

\paragraph{CS for indexed access.}
CS writes a command token selecting whether a reference is used and whether residuals use gaps, intervals or repeated-gap runs. Five adaptive streams carry commands, binary flags, reference distances, copy bits and residual integers. Its sampled index restarts these streams every 2,048 rows, allowing a query to decode the containing chunk prefix up to the requested row by default. 

Cross-chunk references are enabled by default and are resolved recursively. The reported access measurements use whole-chunk decoding. In the openflights comparison CS has the smaller indexed file of our two formats. We report the full size and cache costs below.

\subsection{\CB{}: candidate targets and shared state}
For each row the \CB{} coder first writes the degree. An empty row ends there. Otherwise it writes a self-loop bit, then a bit saying whether the row has a reference. If it does, a second bit says whether the reference is the same as the last one. The distance is written only when the reference changes. Copy bits mark each tested reference target, skipping the self target and stopping when the degree is filled. 

Its reference window is 64. A bounded overlap heuristic chooses the reference. Copied targets initialize the known part of the row. Candidate generators then inspect already decoded structure (Table~\ref{tab:generators}). They visit a bounded number of list entries, exclude targets already found or tested, and give reciprocal proposals priority, followed by proposal count and deterministic tie rules.

\begin{table}[t]\centering\small
\caption{Candidate generators for current vertex $v$. Every inspected row has already been decoded. Found targets include copies, positive proposals and explicit remainder targets.}
\label{tab:generators}
\begin{tabular}{lp{9.5cm}}\toprule
Generator & Reproducible source of proposals\\\midrule
Reciprocal & An earlier vertex $u$ already points to $v$; propose $u$ as a target of $v$.\\
Co-citation & For an earlier $u$ pointing to $v$, propose targets in $u$'s decoded row.\\
Expansion & For a found target $w$, inspect $w$'s row if already decoded, and decoded rows of earlier vertices pointing to $w$. Propose their targets.\\\bottomrule
\end{tabular}\end{table}

Eight feature-conditioned probability inputs and one global base-rate input (the bias input) feed the mixer. Fourteen weight sets are selected by reciprocal status and a seven-way total-proposal-count bucket. Table~\ref{tab:candidate-contexts} lists the feature combinations. A fixed-point logistic mixer combines their predictions. The candidate integer streams use adaptive binary trees over hybrid tokens. CG uses direct 150-symbol tables.

After every presence bit, encoder and decoder update the same model state. A positive bit adds the target and can generate another round. When proposals are exhausted, the remainder code supplies the next target and can seed further proposals. Reaching the transmitted degree ends the row.

\begin{table}[t]\centering\small
\caption{Candidate probability inputs. Each feature is bucketed using decoder-known state; the ninth input has one global adaptive context rather than a graph-feature partition.}
\label{tab:candidate-contexts}
\begin{tabular}{ll}\toprule
Input & Context features\\\midrule
1 & Reciprocal status, proposal count, round\\
2 & Reciprocal status, counts from each of the three generators\\
3 & Observed target popularity, decoded target outdegree, reciprocal status\\
4 & Distance, popularity, whether count exceeds one\\
5 & Remaining degree, count, reciprocal status\\
6 & Popularity, count, distance\\
7 & Count/popularity contrast, reciprocal status, round\\
8 & Remaining degree, popularity\\
9 & Global base rate (bias input)\\\bottomrule
\end{tabular}\end{table}

Algorithms~\ref{alg:cand-decode}--\ref{alg:cand-aux} specify the shared traversal and the encoder's choices. \textsc{Find} inserts a target, decrements $need$, and queues it for expansion. \textsc{UpdateMix} uses the same fixed-point tables, clipping and capped count updates at both ends. The source mapping and update equations are in Appendix~\ref{app:candidate-pseudocode}.
\begin{algorithm}[t]
\caption{\CB{} \textsc{DecodeRow}$(v)$; $I(x)$ contains earlier decoded rows pointing to $x$. Persistent reference and model state carries between rows.}
\label{alg:cand-decode}\label{alg:row-choice}
\small\begin{algorithmic}[1]
\STATE Decode degree $d$ conditioned on whether the preceding degree was zero; update previous degree.
\IF{$d=0$}\STATE Set previous-reference flag false; return an empty row.\ENDIF
\STATE Decode self bit $s$; mark $v$ tested; insert $v$ if $s=1$; $need\leftarrow d-s$; $N\leftarrow\emptyset$.
\IF{$W>0$ and $need>0$}
\STATE Decode has-reference bit using the previous-reference flag.
\IF{a reference is present}
\STATE Decode same-reference bit; use $r_{last}$ if set, else decode $r=v-\mathit{distance}$; update $r_{last}$.
\STATE Set previous-reference flag true and previous-copy bit one.
\FOR{each $x$ in row $r$, excluding $v$, while $need>0$}
\STATE Mark $x$ tested; decode its copy bit using previous bit, observed popularity and remaining degree; update previous-copy bit; if present, \textsc{Find}$(x)$.
\ENDFOR
\ELSE\STATE Set previous-reference flag false.\ENDIF
\ELSE\STATE Set previous-reference flag false.\ENDIF
\STATE $first\leftarrow true$; $round\leftarrow1$; $k\leftarrow0$; $work\leftarrow0$.
\WHILE{$need>0$}
\WHILE{$need>0$ and ($first$ or $N\ne\emptyset$)}
\STATE $C\leftarrow\textsc{Propose}(v,N,first)$; $N\leftarrow\emptyset$; $first\leftarrow false$.
\FOR{each $x$ in $C$ while $need>0$}
\STATE Mark $x$ tested; form eight feature contexts, global input and weight-set index.
\STATE Compute clipped mixed probability $p$; decode $y$; \textsc{UpdateMix}$(y,p)$; if $y=1$, \textsc{Find}$(x)$.
\ENDFOR
\STATE Increment $round$.
\ENDWHILE
\IF{$need>0$}
\STATE Decode $x$ by \textsc{Remainder}; mark it tested; \textsc{Find}$(x)$; update previous explicit target/gap and $k$; set $N=\{x\}$.
\ENDIF
\ENDWHILE
\STATE Sort the row; update $I(x)$ and popularity buckets for every recovered target.
\end{algorithmic}\end{algorithm}

\begin{algorithm}[t]
\caption{\CB{} \textsc{EncodeRow}$(v,T)$ for true sorted target set $T$. The state traversal, contexts and updates are exactly those of Algorithm~\ref{alg:cand-decode}.}
\label{alg:cand-encode}
\small\begin{algorithmic}[1]
\STATE Encode $d=|T|$; if zero, set previous-reference flag false and return.
\STATE Encode $s=[v\in T]$ and initialize tested/found sets, $need=d-s$ and $N$ as in the decoder.
\IF{$W>0$ and $need>0$}
\STATE Search $q=v-1,\ldots,\max(1,v-W)$, skipping empty rows, for the strictly largest positive score $4|(T\cap N(q))\setminus\{v\}|-|N(q)|-4$.
\STATE Encode whether a reference exists; when it does, encode same-reference and, if needed, distance $v-r$.
\STATE Traverse the reference as in the decoder, emitting copy bits $[x\in T]$ and updating found/tested state.
\ENDIF
\STATE Run the decoder's proposal/remainder loop with these substitutions:
\STATE For each proposed $x$, emit $y=[x\in T]$ at the shared probability, then apply the same updates.
\STATE At a remainder step, choose the smallest $x\in T$ still unfound and encode it using Algorithm~\ref{alg:cand-aux}.
\STATE Maintain previous-reference, round, explicit-target and gap state identically; update incoming lists/popularity at row end.
\end{algorithmic}\end{algorithm}

\begin{algorithm}[t]
\caption{\CB{} shared \textsc{Propose} and \textsc{Remainder} procedures, for default candidate generators and gap/popgap modes.}
\label{alg:cand-aux}
\small\begin{algorithmic}[1]
\STATE \textsc{Propose}: expanding a decoded row requires length $\le512$ and $work<2048+1024d$; add its length to $work$.
\IF{$first$}
\STATE For up to 256 most recent $u\in I(v)$: propose $u$ as reciprocal and expand row $u$ for co-citation counts.
\ENDIF
\FOR{$w\in N$}
\STATE If $w<v$, expand row $w$; also expand rows of up to 32 most recent $u\in I(w)$, keeping counts by generator.
\ENDFOR
\STATE Discard tested/found targets and non-reciprocal proposals with count below one. Sort by reciprocal first, count descending, first-proposal order; return the list.
\STATE \textsc{Remainder}: use the degree bucket, explicit-target position and previous-gap magnitude as integer context.
\STATE Map the first explicit target by $g=2(x-v)$ if $x\ge v$, else $2(v-x)-1$; later $g=x-x_{prev}$.
\IF{gap mode}\STATE Encode/decode $g$; invert this map and return $x,g$.\ENDIF
\STATE Popgap: encode/decode token $t=g$ for $g<32$, else $t=32+\lfloor\log_2g\rfloor-5$.
\IF{$t<32$}\STATE Set $g=t$ and invert the offset/gap map to recover $x$.
\ELSE
\STATE Let $\delta=t-27$; form the in-range identifiers whose mapped gaps lie in $[2^\delta,2^{\delta+1})$, including both source sides for the first target.
\STATE Count identifiers $n_b$ in each popularity bucket, including previously tested identifiers; form integer frequencies proportional to $n_b\,hits_b/exposure_b$.
\STATE Encode/decode the popularity bucket and rank within its identifier set (Appendix~\ref{app:candidate-pseudocode}); recover $x$, then its mapped gap $g$.
\STATE Increment the selected bucket's hits and every bucket's exposure by $n_b$, with the prescribed rescaling.
\ENDIF
\STATE Return $x,g$; the calling loop updates found/tested sets and the next-round seed.
\end{algorithmic}\end{algorithm}

\subsection{A row reconstructed step by step}
\label{sec:candidate-toy}
Consider vertex $10$ with true neighbor list $\{3,5,9,14\}$. An earlier reference row is $\{3,5,8\}$. Its copy bitmap \texttt{110} supplies $3$ and $5$. Another decoded row, vertex $2$, contains $\{9,10\}$. It supplies two plausible proposals: reciprocal target $2$ and co-cited target $9$.

For this illustration, enable just those two generators, process $2$ before $9$, and use the probabilities in Table~\ref{tab:candidate-toy}. In Algorithm~\ref{alg:cand-decode}, the self bit is zero, the reference is selected, and the copy step leaves $need=2$. Targets $3,5,8$ have been tested, with $3,5$ found. The proposal loop rejects $2$, accepts $9$, and leaves $need=1$. The explicit remainder supplies $14$, completing the list. 

For the production remainder procedure, target $14$ has first zigzag offset $2(14-10)=8$. The table below retains the simpler absolute-first gamma illustration to expose its bit arithmetic. Its proposal probabilities are shared illustrative values. The implementation obtains them from its adaptive models.

\begin{table}[t]\centering\small
\caption{A complete copy/predict/remainder example. Presence outcomes are exact. Costs are ideal predictive bits and a toy gamma-coded remainder, before common headers and range-coder termination.}
\label{tab:candidate-toy}
\begin{tabular}{lp{3.3cm}lp{3.6cm}}\toprule
Step & Decoder action & Known targets & Charged information\\\midrule
Copy & Apply \texttt{110} to the reference & $3,5$ & Reference and copy description\\
Propose $2$ & Read absence, $q=1/4$ & $3,5$ & $-\log_2(3/4)\approx0.415$ bits\\
Propose $9$ & Read presence, $q=3/4$ & $3,5,9$ & $-\log_2(3/4)\approx0.415$ bits\\
Remainder & Read target $14$ & $3,5,9,14$ & Gamma$(14)$: $2\lfloor\log_2 14\rfloor+1=7$ bits\\\bottomrule
\end{tabular}\end{table}

Without proposals, the post-copy targets $9,14$ have absolute-first gamma gaps $(9,5)$, costing $7+5=12$ bits. The illustrated proposal path costs $0.415+0.415+7\approx7.830$ bits for the same targets, saving about $4.170$ payload bits. Both paths must also pay their copy, mode and model costs. Appendix~\ref{app:candidate-details} carries the arithmetic through the common fields and explains the distinction between ideal bits and final bytes.

\subsection{Popularity-aware remainder coding}
The remainder has two alternatives. The gap model codes the first explicit target by its zigzag offset from the source and subsequent explicit targets by ascending gaps. The \texttt{popgap} model writes short gaps as exact tokens. For a large gap it writes a distance bucket and a popularity bucket. It then writes the target's rank among all identifiers in that pair of buckets. 

For the first remainder, the distance bucket spans both sides of the source. Its counts include already tested identifiers. The rank universe is not filtered by candidate outcomes. Popularity is the incoming count observed in decoded rows, so the decoder can maintain the counts itself.

For intuition, suppose one distance bucket contains 64 identifiers. Four belong to a popularity group assigned half the probability. A target in that group costs one bit to select the group plus two bits for its rank. A flat index over all 64 would cost six bits. These costs are conditional on the distance bucket. The bucket and its probability model have their own costs. The implementation learns bucket probabilities from hits and exposure and uses Fenwick trees, which maintain cumulative counts, for rank and selection. Parameters and boundary rules are in Appendix~\ref{app:candidate-details}.

The candidate state spans the decoded graph and its online models, requiring $O(n+m)$ storage in this implementation. Each row limits list-expansion work to a threshold $2048+1024d$, with at most one eligible-row overshoot. The coder sorts proposals per round and evaluates nine probability inputs per tested target. Popgap needs up to 22 bucket queries and rank/select operations of logarithmic cost per explicit large-gap target. 

There are no independently seekable \CB{} checkpoints. The indexed-reader entry point materializes a plain \CB{} file through full decoding. It rejects files carrying symmetric, reciprocal or transposed flags.

\section{Encoding strategies and file selection}
\label{sec:ordering}
An encoding arm specifies a vertex order, storage direction, coder and parameters. The method constructs the arms in Table~\ref{tab:arms}, encodes them, checks exact adjacency reconstruction, and keeps the smallest successful file. It also retains the smallest CG file as an alternative with a different decoding cost.

\subsection{Three orders}
The \emph{community order} groups vertices with Louvain-style local moves, connected-component refinement and aggregation, then applies LLP within groups~\cite{blondel2008louvain,boldi2011llp}. The implementation's historical function name contains ``Leiden'', but the implementation does not run the full Leiden algorithm~\cite{traag2019leiden}.

The \emph{BP order} refines the community order by recursive balanced graph bisection using a gap-cost surrogate. The \emph{spectral cluster order} merges small communities, builds the weighted graph of inter-community arcs, and sorts whole groups by a two-dimensional spectral angle. Each group keeps its internal LLP order. Moving groups changes between-group distances while preserving within-group relative positions. The exact partitioning, spectral and BP settings are recorded in Appendix~\ref{app:strategy-details}.

\subsection{Storage direction}
An asymmetric graph can be stored as outgoing rows of $G$ or as rows of $G^T$, followed by transposition during reconstruction. We abbreviate these directions as G and T in the tables. The choice changes row degrees, references and the structure visible to the candidate generators. The \CB{} arms test both directions. For new CG window-256 arms, a sample-based code-length score selects a direction when its estimated margin is large enough. A small margin retains both. Symmetric inputs require one direction. The protocol gives the threshold and states which graph each file reconstructs.

\begin{table}[t]\centering\small
\caption{The complete portfolio. The introduction round is included for the cumulative portfolio expansion, rather than defining separate methods. Earlier successful files remain eligible.}
\label{tab:arms}\label{tab:pipelines}
\begin{tabular}{p{2.5cm}p{2.6cm}p{4.6cm}l}\toprule
Order & Direction & Coder/settings & Introduced\\\midrule
BP refinement & Original & CG 256/1024/2048, plain/reciprocal & 1\\
Community & CG score & CG 256, plain/reciprocal & 2\\
Community, BP & Both & \CB{}-gap & 2\\
BP refinement & CG score, additional direction & CG 256, plain/reciprocal & 2\\
Spectral & CG score & CG 256, plain/reciprocal & 3\\
Community, BP, spectral & Both & \CB{}-pop & 3\\\bottomrule
\end{tabular}\end{table}

\subsection{Selection and exact reconstruction}
Let $\mathcal A$ be the declared successful arms and $B(a)$ the counted bytes for arm $a$. The output is $\arg\min_{a\in\mathcal A}B(a)$ with equal-length files sharing the same size objective. An adjacency mismatch aborts that graph. Failed and timed-out arms supply no \CB{} file. An eight-hour timeout bounds each new arm's execution. The selection rule minimizes size. Runtime and access capabilities are separate criteria, and the CG alternative offers a different speed/size choice.

Input order and plain global LLP are outside this portfolio. The guarantee concerns the files that the portfolio actually retains. Adding successful arms cannot increase the selected size. This property also explains the non-increasing columns of the cumulative portfolio expansion.

\section{Evaluation}
\label{sec:experiments}\label{sec:results}
\subsection{Protocol and provenance}
\label{sec:campaign-protocol}
\paragraph{Development history.}
The campaign used three frozen rounds. Round 1 was specified and measured first, with enwiki already a development input. Rounds 2 and 3 were designed after examining the preceding results on these same graphs. Google's BP behavior informed round 2. Patents, an evaluation loss in round 2, became a development graph for round 3. The citation graphs also informed round 3.

We fixed each round's choices before measuring it, but used earlier benchmark results to design later rounds. The rounds are not independent holdouts. Appendix~\ref{app:campaign-protocol} gives the dated manifests and per-round records.

\paragraph{Inputs and byte accounting.}
We remove loops and take the largest weakly connected component. We represent each undirected edge by two arcs and compact the vertex identifiers. Inputs used for the published comparison match its vertex and arc counts. The local enron-nat input is the loop-free largest weakly connected component, with 67,521 vertices and 273,649 arcs. Its arc count matches the publication, but its vertex count differs from the reported 67,251. We exclude it. Five other source inputs could not be reproduced.

Bits per arc (bpe) is eight times counted bytes divided by directed arcs. External-ID maps are excluded. Matching counts establishes the scope of the published comparison. Exact checks of our files use the processed adjacency itself. Our sequential sizes count the actual files, including the Modified Graph Storage (MGS) transpose extension on transposed winners. Those files decode back to the original directed graph. The extension adds two or three bytes.

Two files are exceptions. The superseded enwiki round-2 CB file is unflagged. The post-freeze in-2004 CS file is checked against its explicitly transposed input. Neither is a current portfolio winner. Appendix~\ref{app:campaign-protocol} gives the accounting and restoration details.

\paragraph{Baselines.}
We use Java BVGraph for WebGraph, with window/depth settings $(7,3)$ and $(16,2^{31}-1)$. The shorthand w7/m3 means window 7 and maximum reference depth 3. Zuckerli uses the recorded command-line defaults in archive and access modes. We did not use the publication's tuned settings in this campaign. On the earlier five matched inputs, those settings reduce sizes by 0.5--2.3\% when we take the smaller mode in each configuration (Appendix~\ref{app:matched}).

If the campaign Zuckerli values were three percent smaller, the selected file would still win on every graph where Zuckerli is the strongest baseline. We did not measure that configuration.

Each baseline column takes the smallest recorded file across those settings and the tested community/BP orders, in the original direction G only. WebGraph counts its \texttt{.graph} payload. Zuckerli counts its single codec file. The baseline search omits the spectral order and storage-direction selection in our portfolio. Direction selection is a feature of our encoder and is not applied to the baselines.

For HyperFast, we take the smallest central estimate in the publication's Table 6. We compare its four embedding methods, labelled BFKL, LPCS, HMCS and CLOVE, at bucket counts $B_r=B_d\in\{8,16\}$. Appendix~\ref{app:hf-configurations} gives the winning configurations and the HMCS-16 values.

The same table~\cite{celinska2026hyperbolic} also reports WebGraph on embedding-derived orders. Its best such value on uk-2002 is 1.67 bpe. Our selected file is 29.7\% smaller than that. Against HyperFast itself, at 2.99 bpe, it is 60.7\% smaller. Values that we reconstructed on other graphs carry an R marker and are discussed with that comparison. Appendix~\ref{app:baseline-provenance} identifies each selected baseline configuration and its order.

\paragraph{Evidence snapshot and timing.}
Tables use the 23 November snapshot. All round-3 selections are available, including uk-2002. Our selection records give bytes, hashes, parameters and successful full adjacency checks. The transpose-extension records also verify reconstruction of G.

The comparator pass ran alone on the same AMD EPYC 9655P allocation exposing 16 cores, with one application thread for our coders. Its script hashes were frozen before results. An amendment to the start trigger delayed the pass until the transpose checks finished. Each codec receives the identical community-ordered G input.

The full-decode timing rules differ between codecs.
\begin{itemize}
\item Zuckerli uses the median of three whole-process wall times, including startup.
\item WebGraph uses the best of three scans inside the Java virtual machine (JVM), including graph opening but excluding JVM startup.
\item CG uses the median of three in-process full decodes. The first includes just-in-time (JIT) compilation.
\end{itemize}
On all eight large inputs, CG's first run is the slowest. Its median is therefore the slower of the two warm runs. WebGraph is credited with its best run, so this rule works against CG. Zuckerli's startup inclusion favors our coders. The external decoders scan successors for a checksum, whereas our decoder materializes the adjacency. The three timings are therefore not strictly comparable.

\subsection{Size against three baselines}
\begin{table}[t]\centering\small
\caption{Campaign sizes in bits per directed arc, under Section~\ref{sec:campaign-protocol}'s accounting. WG: WebGraph; ZK: Zuckerli; HF best: best published HyperFast configuration from Table 6. \CB{} selected is the portfolio minimum (all displayed winners use \CB{}), CG the minimum over CG arms. The margin uses the smallest of the three baseline columns. All selections are complete; enwiki and patents are development inputs. Baseline configurations and orders are listed in Appendix~\ref{app:baseline-provenance}.}
\label{tab:campaign-main}\label{tab:campaign-development}
\begin{tabular}{lrrrrrr}\toprule
Graph & WG & ZK & HF best & \CB{} selected & CG & Margin (\%)\\\midrule
Human1 & 3.674 & 3.828 & 3.610 & 1.709 & 1.756 & -52.7 \\
Drosophila2 & 7.257 & 7.370 & 5.000 & 3.324 & 3.347 & -33.5 \\
openflights & 4.889 & 4.861 & 3.830 & 2.133 & 2.192 & -44.3 \\
grqc & 5.476 & 5.293 & 3.720 & 2.079 & 2.220 & -44.1 \\
Mouse3 & 2.945 & 2.665 & 2.420 & 1.144 & 1.305 & -52.7 \\
facebook & 3.140 & 2.935 & 2.800 & 1.326 & 1.414 & -52.6 \\
oregon & 6.699 & 6.510 & 4.200 & 2.721 & 2.861 & -35.2 \\
hepph & 3.084 & 2.923 & 3.210 & 1.212 & 1.321 & -58.5 \\
wordassociation-2011 & 10.276 & 9.757 & 7.480 & 7.350 & 8.420 & -1.7 \\
condmat & 6.631 & 6.017 & 5.470 & 2.364 & 2.664 & -56.8 \\
astroph & 4.989 & 4.642 & 5.480 & 1.919 & 2.178 & -58.7 \\
enron & 6.941 & 6.578 & 5.610 & 2.705 & 3.046 & -51.8 \\
slashdot & 11.197 & 10.047 & 8.320 & 5.927 & 6.024 & -28.8 \\
amazon & 8.563 & 7.472 & 5.110 & 2.867 & 2.929 & -43.9 \\
google & 4.584 & 3.830 & 4.090 & 3.118 & 3.386 & -18.6 \\
uk-2014-tpd & 9.963 & 8.940 & 8.750 & 7.643 & 8.678 & -12.7 \\
patents & 13.905 & 12.231 & 10.570 & 10.228 & 10.308 & -3.2 \\
enwiki & 11.546 & 10.346 & 10.140 & 7.770 & 8.866 & -23.4 \\
uk-2002 & 1.770 & 1.303 & 2.990 & 1.175 & 1.301 & -9.8 \\
 
\bottomrule\end{tabular}\end{table}

Across all nineteen rows of Table~\ref{tab:campaign-main}, the selected file is smaller than each recorded baseline, with a median reduction of 43.9\%. This descriptive total includes the two development inputs. Seventeen inputs are count-matched and lie outside the declared development set. On these, savings against the strongest baseline range from 1.7\% to 58.7\%. The baseline with the smallest file varies with the graph.

We treat size differences within 0.5\% as ties.

Uk-2002 illustrates the effect. The selected file occupies 1.175 bpe on 292 million arcs, against Zuckerli's 1.303 and HyperFast's 2.990. Its saving against the strongest measured baseline is 9.8\%. 

\begin{figure}[t]\centering
\includegraphics[width=\linewidth]{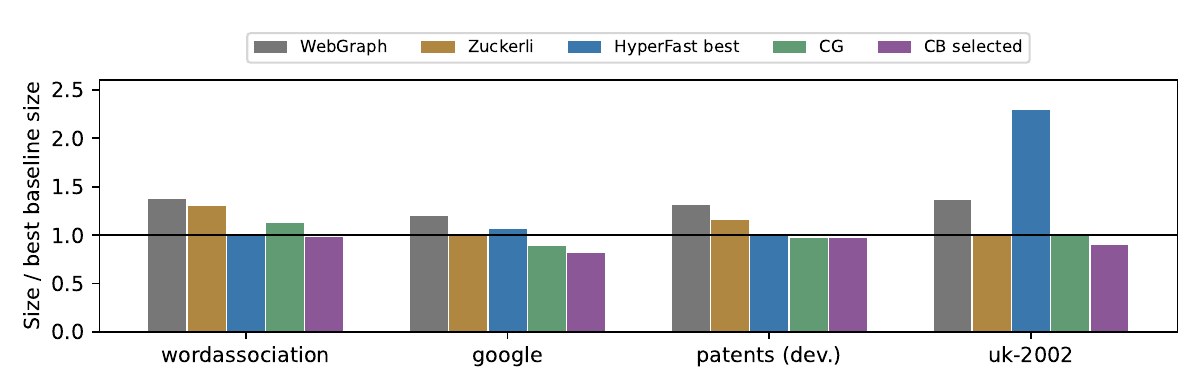}
\caption{Each file size divided by the smallest baseline size on that same input. One means equality with the strongest baseline; lower is better. Sizes follow the declared byte accounting; WebGraph excludes offsets. Patents is a development input; all native bars use the final selections.}
\Description{Grouped size ratios for WebGraph, Zuckerli, HyperFast, CG and the selected \CB{} file on four inputs. Zuckerli is the strongest baseline on google and uk-2002.}
\label{fig:campaign}
\end{figure}

The CG alternative occupies 1.301 bpe, a 0.17\% reduction relative to Zuckerli and therefore a tie under the 0.5\% band. On wordassociation the selected \CB{} file also beats the strongest baseline while CG loses. Figure~\ref{fig:campaign} places those comparisons alongside google and patents.

\subsection{Cumulative portfolio expansion}
\begin{table}[t]\centering\small
\caption{Selected examples from the cumulative portfolio expansion on completed, count-matched campaign inputs; full history is in Appendix~\ref{app:round-history}. Round 2 jointly adds \CB{}-gap coding, direction selection and additional order arms. Round 3 jointly adds spectral cluster order and popgap. These are bundled additions, not single-factor causal ablations. Each winning arm names the order (Community, BP or Spectral), direction G/T, and coder CB-gap, CB-pop, or CG followed by plain/reciprocal P/R.}
\label{tab:method-ablation}
\begin{tabular}{lrrrl}\toprule
Graph & Round 1 & Round 2 & Round 3 & Selected arm\\\midrule
Human1 & 1.801 & 1.709 & 1.709 & Community/G/CB-gap \\
facebook & 1.461 & 1.328 & 1.326 & Spectral/G/CB-pop \\
wordassociation-2011 & 8.767 & 7.418 & 7.350 & Community/G/CB-pop \\
google & 4.588 & 3.345 & 3.118 & Spectral/G/CB-pop \\
uk-2014-tpd & 8.678 & 7.810 & 7.643 & Spectral/G/CB-pop \\
patents & 12.224 & 11.153 & 10.228 & Spectral/G/CB-pop \\
 
\bottomrule\end{tabular}\end{table}

The progression separates the final method from its development history (Table~\ref{tab:method-ablation}). \CB{}-gap arms account for substantial improvements when round 2 becomes available. The last round can improve further by changing the remainder, the cluster order, or both. Some winners remain from round 2, showing the value of keeping previously successful files. Patents changes from a loss against its best baseline to a smaller file after the final bundle is added. Because several components changed together, the winning-arm labels identify a successful combination. They do not isolate a component's causal contribution.

\subsection{Decode-time trade-off}
\begin{table}[t]\centering\small
\caption{Full sequential decode on identical community-ordered G inputs above three million arcs, in seconds. ZK max/access: Zuckerli archive/access modes; WG: WebGraph window/depth settings w7/m3 and w16/unlimited; CG: plain window 256. These fixed configurations differ from the size-selected arms. Timing scopes and repetition rules are given in Section~\ref{sec:campaign-protocol}; all 23 inputs are listed in Appendix~\ref{app:completed-timing}.}
\label{tab:baseline-decode}
\begin{tabular}{lrrrrr}\toprule
Graph & ZK max & ZK access & WG w7 & WG w16 & CG\\\midrule
google & 0.0716 & 0.0615 & 0.1147 & 0.1153 & 0.2091 \\
cnr-2000 & 0.0261 & 0.0228 & 0.0394 & 0.0356 & 0.0796 \\
amazon0601 & 0.0620 & 0.0503 & 0.0770 & 0.0771 & 0.1748 \\
in-2004 & 0.0932 & 0.0810 & 0.1730 & 0.1563 & 0.4257 \\
patents & 0.3538 & 0.2792 & 0.4539 & 0.4642 & 1.2564 \\
enwiki & 1.6987 & 1.2654 & 1.9628 & 1.9201 & 6.7648 \\
uk-2014-tpd & 0.2543 & 0.1833 & 0.2859 & 0.2821 & 0.8866 \\
uk-2002 & 1.4788 & 1.3022 & 2.2267 & 2.1315 & 9.0595 \\
 
\bottomrule\end{tabular}\end{table}

On the eight inputs above three million arcs, CG takes 2.8--6.1 times Zuckerli's archive-mode decode time (median 3.5) and 1.8--4.1 times WebGraph w7/m3's time (median 2.6). On uk-2002 the times are 1.48, 2.23 and 9.06 seconds for Zuckerli, WebGraph and CG respectively (Table~\ref{tab:baseline-decode}). 

Using Zuckerli's faster access mode as the denominator instead gives a median CG slowdown of 4.7. CG is the faster native alternative to \CB{}, while both external baselines decode faster on these inputs.

\begin{table}[t]\centering\small
\caption{Dedicated round-3 timing pass for four large selected \CB{}/CG pairs, in seconds. Each is one full decode after a 300-vertex same-codec warm-up. Raw timed files match the selected payload hashes. Their selected orders and modes differ from Table~\ref{tab:baseline-decode}; transposed payload timings exclude restoration to G. These are single observations, not process-repeat medians.}
\label{tab:campaign-timing}
\begin{tabular}{lrrr}\toprule
Graph & \CB{} & CG & \CB{}/CG\\\midrule
google & 6.53 & 2.44 & 2.7 \\
patents & 92.89 & 3.20 & 29.1 \\
enwiki & 6355.16 & 13.04 & 487.2 \\
uk-2002 & 406.90 & 10.07 & 40.4 \\
 
\bottomrule\end{tabular}\end{table}

\CB{} is much slower. It must build candidate pools, evaluate contexts and run popularity rank/select queries. The dedicated pass records 407 seconds on uk-2002 and 6,355 seconds (1.77 hours) on enwiki (Table~\ref{tab:campaign-timing}). The two timing tables answer different questions. The size-selected archives show what the smallest native files cost to decode. The fixed-configuration comparison holds input order and direction equal across codecs.

\subsection{Seekable archives and random-access costs}
\label{sec:random-access}
CS-RA is a proof of concept for compact archives with occasional neighbor-list retrieval. It is not intended for latency-sensitive graph queries. It restarts command and probability streams at sampled boundaries and recursively resolves cross-chunk references. The default reader decodes the requested chunk prefix. This experiment uses whole-chunk decoding with a 256-vertex reference window and a 2,048-vertex index stride. 

CG-RA decodes and retains the containing chunk, with an unbounded decoded-chunk cache in its current reader. The \CB{} format remains sequential-only.

\paragraph{Index-inclusive size.}
On all 22 completed CS-RA inputs, files are smaller than Zuckerli's access-mode files by 1.3--29.5\% (median 7.1\%) and WebGraph w7/m3 plus offsets by 16.3--56.5\% (median 26.9\%). Every codec receives the same community-ordered original-direction graph (Table~\ref{tab:ra-completed}). 

Against the smaller of Zuckerli's archive and access modes, CS-RA wins on 20 of 22 inputs. It loses on amazon0601 (7.057 versus 6.989 bpe) and patents (12.488 versus 12.331 bpe). CS-RA still stores both arcs of an undirected edge and restarts its models, so its files on social and collaboration graphs remain substantially larger than the selected sequential archives.

\begin{table}[t]\centering\small
\caption{Completed common-input access comparison. Sizes are bpe, including native indexes and WebGraph offsets. All use community order and G orientation. ZK access is Zuckerli's access mode; WG is w7/m3. CS cold is mean query time with decoded caching disabled, in milliseconds; WG query time is in microseconds. Opening and process startup are excluded from query timings. Enwiki's CS cold-query phase failed and produced no completed result. All 22 completed CS-RA and 23 CG-RA records pass seek and full-adjacency checks using the cached readers, comparing sorted rows. The separate uncached CS-RA check covers nine graphs (Appendix~\ref{app:cold-reader-check}). Zuckerli's CLI provides no per-list query interface; HyperFast reports no index-size or query benchmark.}
\label{tab:ra-completed}
\begin{tabular}{lrrrrrr}\toprule
Graph & CS-RA & CG-RA & ZK access & WG & CS cold (ms) & WG ($\mu$s)\\\midrule
Human1 & 3.261 & 3.210 & 3.828 & 4.194 & 0.82 & 1.656 \\
Drosophila2 & 6.862 & 6.703 & 7.863 & 8.745 & 1.48 & 0.469 \\
openflights & 4.326 & 4.979 & 4.971 & 6.059 & 2.10 & 0.940 \\
grqc & 4.637 & 5.732 & 5.293 & 7.051 & 1.66 & 0.377 \\
Mouse3 & 2.651 & 2.580 & 2.833 & 3.360 & 6.22 & 7.011 \\
facebook & 2.772 & 3.051 & 2.935 & 3.574 & 5.45 & 1.778 \\
oregon & 6.157 & 10.922 & 7.282 & 10.332 & 3.19 & 0.792 \\
hepph & 2.670 & 4.081 & 2.923 & 3.763 & 7.07 & 0.994 \\
wordassociation-2011 & 9.469 & 13.489 & 9.762 & 11.744 & 4.19 & 1.134 \\
condmat & 5.583 & 9.105 & 6.017 & 8.101 & 8.87 & 1.151 \\
astroph & 4.388 & 7.008 & 4.642 & 5.807 & 14.45 & 1.443 \\
enron & 6.249 & 9.524 & 6.719 & 8.362 & 19.72 & 1.154 \\
slashdot & 10.636 & 16.504 & 10.870 & 13.179 & 49.70 & 0.780 \\
amazon & 7.346 & 10.143 & 7.754 & 10.877 & 11.71 & 0.644 \\
google & 3.783 & 6.850 & 4.471 & 6.659 & 22.31 & 1.025 \\
eat & 8.585 & 11.000 & 8.731 & 10.259 & 10.33 & 1.454 \\
cnr-2000 & 1.739 & 4.431 & 2.316 & 3.776 & 9.08 & 0.596 \\
amazon0601 & 7.057 & 9.635 & 7.192 & 10.004 & 55.50 & 0.763 \\
in-2004 & 1.251 & 3.370 & 1.774 & 2.873 & 10.19 & 0.895 \\
patents & 12.488 & 17.204 & 12.658 & 16.371 & 9.35 & 0.627 \\
enwiki & -- & 17.962 & 11.224 & 13.382 & Failed & 1.812 \\
uk-2014-tpd & 9.205 & 14.487 & 9.476 & 11.337 & 107.33 & 0.681 \\
uk-2002 & 1.265 & 3.080 & 1.779 & 2.758 & 18.73 & 0.892 \\
 
\bottomrule\end{tabular}\end{table}

\paragraph{Query latency and cache policy.}
CS-RA cold queries take 0.8--107 milliseconds per successor list, with a median of 9.2 milliseconds. On graphs with at most 2,048 vertices, such as Human1, each cold query decodes the whole graph because it occupies one chunk. 

WebGraph takes 0.38--7.01 microseconds. Mouse3 is the 7.01-microsecond outlier and every other input is at most 1.812 microseconds. The median per-graph CS-RA/WebGraph time ratio is about 8,900. CS-RA's size advantage therefore comes with much slower queries. 

Warm timings for our coders on small graphs largely measure a cache holding the graph's decoded chunks and are omitted from the performance comparison.

WebGraph runs 100,000 uniformly sampled queries after opening and a 1,000-outdegree-query warm-up. CS-RA uses 2,000 sampled queries with a zero decoded-cache budget, after a separate warm phase. Cold refers to decoded state, not the filesystem cache. The query generators use fixed seeds but different libraries and query counts. WebGraph consumes returned successors in a checksum. Our query timer retrieves the list and its length. 

Zuckerli's command-line interface exposes whole-graph decoding, so it contributes file sizes and full-decode times only. The earlier matched five-input access study, including limited-cache and breadth-first traces, remains in Appendix~\ref{sec:legacy-experiments}.

\paragraph{Reference-chain failure.}
On the development graph enwiki, CS-RA raised a \texttt{StackOverflowError} during cold queries, after the seek and warm-query phases. The run stopped before its full-adjacency check and final result record. With decoded caching disabled, the reader recursively follows references across chunks. CS bounds reference distance to 256 vertices but imposes no chain-depth bound. The compared BVGraph configuration caps reference depth at three. A chain-depth limit or an iterative resolver is needed to address this failure. The 22-input CS-RA tally excludes this failed run.

\paragraph{Why CG-RA files are larger.}
The native indexed formats refuse symmetric, reciprocal and transposed storage modes. On undirected openflights, losing symmetric storage explains most of CG-RA's size increase. On directed inputs, CG-RA confines references to fixed 2,048-vertex chunks and puts crossing arcs into an unadapted Fibonacci-coded section. Each active source/chunk-pair record repeats its full chunk-local source index, followed by a stop-terminated target-gap list, without copying or intervals.

Table~\ref{tab:cgra-layout} retains the separate layout account on the smallest indexed orders from the earlier exploratory sweep. Crossing arcs range from 14\% to 70\% of arcs and consume 55--77\% of those files. The remaining intra-chunk data, framing and index occupy 1.447 bpe on cnr-2000 and 1.150 bpe on in-2004. CG's inter-chunk representation accounts for most of these bits. This layout-specific result uses different orders from the common-community-order comparison above.

\begin{table}[t]\centering\small
\caption{Earlier CG layout account, in bpe, with window 256 and 2,048-vertex chunks. Indexed sizes minimize over community, BP and spectral orders in G; sequential CG is the frozen portfolio minimum and may use a different order/direction. Crossing-arc share and modeled inter-chunk bits use the stated indexed order. Rest is measured total minus modeled inter bits. This explains the layout cost; these are not the common-order files in Table~\ref{tab:ra-completed}.}
\label{tab:cgra-layout}
\begin{tabular}{llrrrrr}\toprule
Graph & Indexed order & CG seq. & CG-RA & Cross (\%) & Inter & Rest\\\midrule
cnr-2000 & Spectral & 1.856 & 4.289 & 38.8 & 2.842 & 1.447 \\
in-2004 & Spectral & 1.410 & 3.269 & 28.1 & 2.119 & 1.150 \\
google & Spectral & 3.386 & 6.175 & 13.8 & 3.394 & 2.781 \\
patents & Spectral & 10.308 & 15.576 & 41.9 & 11.701 & 3.875 \\
eat & BP & 7.546 & 10.614 & 70.4 & 8.140 & 2.474 \\
 
\bottomrule\end{tabular}\end{table}

\subsection{Previously studied and citation graphs}
\begin{table}[t]\centering\small
\caption{Current portfolio sizes on previously studied graphs and citation tuning inputs, in bpe. R marks reconstructed HyperFast; a dash marks an unavailable estimate. Margin uses the smallest available baseline. The upper block has raw selection and baseline-size records; the lower block comes from the rounded tuning report.}
\label{tab:campaign-historical}\label{tab:campaign-citation}
\begin{tabular}{lrrrrrr}\toprule
Graph & WG & ZK & HF$^{\mathrm R}$ & \CB{} selected & CG & Margin (\%)\\\midrule
eat & 9.276 & 8.508 & 7.477 & 7.180 & 7.546 & -4.0 \\
cnr-2000 & 2.480 & 1.894 & 2.938 & 1.541 & 1.856 & -18.6 \\
amazon0601 & 8.430 & 6.989 & 6.094 & 4.602 & 5.159 & -24.5 \\
in-2004 & 1.793 & 1.311 & -- & 1.194 & 1.410 & -8.9 \\
 
\midrule
cit-HepTh & 6.716 & 6.404 & 6.840$^{\mathrm R}$ & 5.557 & 5.856 & -13.2 \\
arxiv-hep-ph & 7.280 & 6.925 & 7.087$^{\mathrm R}$ & 5.953 & 6.294 & -14.0 \\
ogbn-arxiv & 8.830 & 8.322 & 7.754$^{\mathrm R}$ & 7.352 & 7.625 & -5.2 \\
 
\bottomrule\end{tabular}\end{table}

Table~\ref{tab:campaign-historical} extends the comparison to the previously studied and citation inputs. On in-2004, Zuckerli is smaller than CG while the selected \CB{} file is smaller than both. On EAT, the reconstructed geometric baseline lies between the \CB{} and CG sizes. The underlying orders and preprocessing differ from the earlier fixed-CS experiments, which are retained separately.

After the freeze we ran one more test, which is not in the frozen tables. Plain CS on the transposed spectral order reached 1.191 bpe on in-2004. The frozen portfolio gives 1.194 and recorded Zuckerli 1.311. This file has no transpose flag and its adjacency check is against the transposed input $G^T$. On this input a coder without candidate prediction matches the portfolio size. 

The CS file is 9.2\% below Zuckerli and 0.3\% below the portfolio, within the 0.5\% tie band. This tie is input-specific. On EAT the best measured BG/CS extension occupies 7.730 bpe, against the frozen \CB{} file's 7.180 bpe.

The rebuilt HyperFast estimates use CLOVE embeddings and coding choices calibrated against nine published graphs at two bucket-count settings. Their errors on those eighteen targets range from approximately $-4.8\%$ to $+0.8\%$. Sixteen estimates are lower than the publication. This calibration does not give a certified error interval for new inputs. 

A five-percent downward perturbation reverses ogbn-arxiv CG's apparent advantage and puts its selected \CB{} margin inside the declared half-percent tie band. The larger citation margins survive that sensitivity scenario.

Increasing CG's reference window alone leaves the three checked losses open. The follow-up tests windows 4,096, 8,192 and 16,384 in plain and reciprocal modes on wordassociation, EAT and in-2004. Every file passes full adjacency comparison. Their minima are 8.396, 7.537 and 1.476 bpe, respectively (Appendix~\ref{app:large-windows}). Uk-2002 has no retained result from this extended-window sweep. The EAT comparison still uses a reconstructed baseline.

\section{Mechanism accounting}
\label{sec:analysis}
Two statements describe the competition between copying, predicting and writing targets explicitly. The first accounts for a complete encoded stream. The second isolates the dependence that copying exploits in a repeated-row source.

\subsection{Candidate bits: savings must pay for both outcomes}
Consider a fixed encode with $N$ candidate events. Event $t$ has outcome $x_t\in\{0,1\}$ and decoder-reproducible probability $q_t\in(0,1)$. Its ideal predictive cost is
\[
\ell=-\sum_{t=1}^{N}\bigl[x_t\log_2 q_t+(1-x_t)\log_2(1-q_t)\bigr].
\]
Let $R_0$ be the baseline remainder cost and $R_1$ the \CB{} coder's remainder cost, measured consistently in bits. Let $O_0$ and $O_1$ contain all other costs, including degree/reference descriptions, model information, headers, termination and the difference between ideal costs and finalized bytes. Thus the complete lengths are $L_0=R_0+O_0$ and $L_1=\ell+R_1+O_1$.

\paragraph{Statement 1 (exact cost balance).}
With $S=R_0-R_1$ and $\Delta O=O_1-O_0$,
\begin{equation}
L_0-L_1=S-\ell-\Delta O.
\label{eq:candidate-balance}
\end{equation}
Candidate coding wins exactly when the remainder saving exceeds the candidate-bit cost plus the change in all other costs. A positive proposal replaces the conditional cost of an explicit target. Negative proposals contribute their own log loss. Measuring these terms separately connects the identity to a particular encode.

We first used a simple benefit test, called the gate, that assigns one pooled precision to each graph. Precision here is the fraction of proposals that are present. The pooled-precision calculation isolates one cost term. A controlled re-run of the encodes, which we call a replay, measures this term alongside remainder savings. Partition a fixed event sequence into decoder-known contexts $c$, with counts $N_c$ and empirical positive frequencies $p_c$. Write $\bar p=\sum_cN_cp_c/N$ and $h(p)=-p\log_2p-(1-p)\log_2(1-p)$. The ideal empirical codes satisfy
\begin{equation}
Nh(\bar p)-\sum_cN_ch(p_c)=N I(X;C)\geq0,
\label{eq:context-identity}
\end{equation}
where $I$ is mutual information under the empirical event distribution~\cite{shannon1948}. This follows by subtracting conditional entropy from entropy. For example, equally frequent contexts with positive frequencies $0.9$ and $0.1$ have pooled frequency $0.5$. The pooled code costs one bit per event. The context code costs about $0.469$ bits before model costs.

Equation~\eqref{eq:context-identity} compares ideal empirical codes for the same events. The implemented adaptive model also pays for initialization, calibration and framing. Its actual probabilities belong in $\ell$.

The replay in Appendix~\ref{app:failed-gate} applies the balance to eight tuning graphs in both storage directions. It compares the \CB{}-gap coder with the same coder's generators and remainder seeding disabled. All sixteen instrumented encodes reproduce the original bytes. Assigning range-coder information growth to each stream, including finite-precision effects and finalization, reconciles the balance with the file-size difference. In this accounting, $\ell_{\rm code}$ replaces ideal $\ell$ and the remaining overhead changes by at most $2.2\times10^{-5}$ bits per arc.

The gate predicted a benefit on two graphs. The measured benefit against its own no-candidate baseline is positive on all eight. On web graphs, the gate mainly underestimates the savings. The coder finds more targets, and they cost more to spell out than the gate assumed. 

Differences between contexts reduce the cost per successful proposal, but the coder also tests more low-probability proposals, largely cancelling that advantage in the total cost estimate. The replay therefore explains the failed sign test through measured remainder savings and both proposal outcomes. The no-\CB{} coder is a separate baseline from CG. It is substantially larger on some social graphs.

\subsection{Repeated rows and restricted product models}
Copying exploits dependence that a model of separate arcs can miss. A precise comparison must say what information the probability model already knows. Let $Z$ be side information available to both encoder and decoder. Hold $Z=z$ fixed and choose a binary row $Y\in\{0,1\}^d$. Its coordinates are independent Bernoulli variables with parameters $p_j$. The source repeats that \emph{same} row $r$ times.

\paragraph{Statement 2 (repeated-row comparison under fixed side information).}
Restrict the competing model to a product over rows and coordinates,
\[
Q(Y^{(1)},\ldots,Y^{(r)}\mid z)
=\prod_{i=1}^r\prod_{j=1}^d
q_{ij}(z)^{Y^{(i)}_j}(1-q_{ij}(z))^{1-Y^{(i)}_j},
\]
with no conditioning on earlier decoded rows beyond $z$. Then its expected ideal length on the repeated-row source is at least
\begin{equation}
r\sum_{j=1}^d h(p_j).
\label{eq:product-bound}
\end{equation}
A code that writes the first row and then references it needs only $\sum_jh(p_j)$ expected ideal bits for that row, plus the copy flags, identifiers and framing. With at most $n$ possible reference rows, a simple construction uses $O(r\log_2(n+1))$ further bits. Copying has a strict advantage whenever the saved repeated entropy exceeds this overhead.

To prove the lower bound, each factor has expected cost
$h(p_j)+D_{\mathrm{KL}}(\operatorname{Bern}(p_j)\Vert\operatorname{Bern}(q_{ij}))\geq h(p_j)$.
Sum over rows and coordinates. For the upper construction, encode the first row with its known product distribution and each later row as an exact reference. A binary row source can be realized as outgoing arcs from distinct source vertices to a disjoint target set. The repeated rows do not require parallel edges.

The conditioning restriction determines the scope. Graph-derived side information must be charged and may remove the remaining row entropy. A model that conditions on earlier rows can exploit the repetition directly. HyperFast's complete format has additional embedding and structural choices, so this restricted statement supplies no lower bound on that format.

\subsection{Accounting for cluster order}
For any order $\pi$, the arc-span quantity
$A(\pi)=\sum_{(u,v)\in E}\log_2\max\{1,|\pi(u)-\pi(v)|\}$
splits exactly into within-cluster and between-cluster sums. In the cit-HepTh check, moving whole merged clusters preserves the within-cluster term and reduces the between-cluster term from 1,120,312 to 1,051,165 bits. This is about $0.196$ bits per arc. The corresponding CG and \CB{} files improve by only $0.057$ and $0.021$ bpe. The accounting is exact for arc spans, not for consecutive-neighbor gaps or finalized files. A quotient graph records arc totals but loses which targets share a row, so it does not determine those code lengths.

\section{Limitations and conclusion}
\label{sec:discussion}
The evidence establishes size and decode trade-offs for a specified collection of files. Benchmark reuse, baseline search coverage and timing differences are documented in the protocol. Generalization calls for a fresh evaluation after the design is fixed. CS-RA is a seekable archive with slow cold queries, and its recursive reader overflows the stack on enwiki (Section~\ref{sec:random-access}). A bounded reference-chain depth or iterative resolution is needed. \CB{} files remain sequential-only.

The replay shows that the gate underestimated savings. The repeated-row comparison concerns a restricted probability model. A controlled experiment charging side information would test its relevance to complete formats. Practical prevalence and a complete experiment reproduction archive remain open.

The cited implementation revision includes the \CB{} coder and the symmetric, reciprocal and transposed format extensions. The manuscript bundle supplies its tables, figures and source/file identities. It is not a complete archive of the experimental inputs and runs.

The method combines reference copying, decoder-side neighbor prediction and explicit remainder coding. We try community-based orders and storage directions, then keep the smallest file that decodes exactly. The experiments show where this choice improves on the recorded baselines. The mechanism accounting explains which probabilities and costs determine whether a prediction saves space.

The coder family offers \CB{} for the smallest portfolio archive, CG for faster decoding than \CB{}, and CS-RA for a small, seekable archive. The post-freeze CS tie on in-2004 shows that candidate prediction's size benefit depends on the input. \CB{} supports only sequential access. Its uk-2002 file took 407 seconds to decode in a single dedicated run after warm-up. That transposed-payload timing excludes restoration to the original graph direction.

CG still decodes more slowly than both external baselines on large inputs. CS-RA trades smaller indexed files for millisecond cold-query latency. Its cold-query runs completed on 22 of 23 graphs. The remaining run, on enwiki, failed in the recursive reference resolver. A depth bound or an iterative implementation is needed. These formats suit storage-oriented workloads only when their measured decoding and access costs are acceptable.

\bibliographystyle{ACM-Reference-Format}
\bibliography{references}

\clearpage
\appendix
\section{A worked guide to graph compression}
\label{app:primer}\label{app:worked}\label{app:encoders}
This first example carries one graph through gaps, copies and ordering. The candidate/presence and popularity stages are worked through separately in Section~\ref{sec:candidate-toy} and Appendix~\ref{app:candidate-details}. It separates payload savings from the headers needed to decode them. The toy uses positive, absolute first values for clarity. The production formats often use source-relative first offsets instead.

\subsection{One graph, two labelings}
Take two groups of three vertices. Each vertex points to the other two vertices in its group. The graph has twelve directed edges. In the interleaved labeling, the groups are $\{1,3,5\}$ and $\{2,4,6\}$. Renumbering old vertices in the order $(1,3,5,2,4,6)$ makes the groups consecutive.

\begin{center}\small
\begin{tabular}{crr}\toprule
Row number & Interleaved labeling & Grouped labeling\\\midrule
1 & $\{3,5\}$ & $\{2,3\}$\\
2 & $\{4,6\}$ & $\{1,3\}$\\
3 & $\{1,5\}$ & $\{1,2\}$\\
4 & $\{2,6\}$ & $\{5,6\}$\\
5 & $\{1,3\}$ & $\{4,6\}$\\
6 & $\{2,4\}$ & $\{4,5\}$\\\bottomrule
\end{tabular}\end{center}
A row number in one column need not identify the same original vertex as that row number in the other. Both sources and targets have been relabeled. The edges themselves have not changed.

\subsection{Write differences instead of identifiers}
The sorted list $(4,6)$ becomes $(4,2)$: write the first value, then the gap. Elias gamma coding writes a positive integer $x$ in $2\lfloor\log_2x\rfloor+1$ bits. It writes $\lfloor\log_2x\rfloor$ zeros followed by the binary representation of $x$. For example, $1$, $2$, and $4$ have codewords \texttt{1}, \texttt{010}, and \texttt{00100}.

\begin{center}\small
\begin{tabular}{crr}\toprule
Row & Interleaved gap payload & Grouped gap payload\\\midrule
1 & $(3,2): 6$ bits & $(2,1): 4$ bits\\
2 & $(4,2): 8$ bits & $(1,2): 4$ bits\\
3 & $(1,4): 6$ bits & $(1,1): 2$ bits\\
4 & $(2,4): 8$ bits & $(5,1): 6$ bits\\
5 & $(1,2): 4$ bits & $(4,2): 8$ bits\\
6 & $(2,2): 6$ bits & $(4,1): 6$ bits\\\midrule
Total & 38 bits & 30 bits\\\bottomrule
\end{tabular}\end{center}
Grouping reduces this payload, but not every row becomes cheaper. Row five becomes more expensive. A favorable total is not a row-by-row guarantee. These totals also exclude row lengths, action headers, and any stored permutation.

\subsection{Copy a shared neighbor}
In the grouped order, row two is $\{1,3\}$ and row one is $\{2,3\}$. With a one-row window, row two can copy the value $3$ from row one. A bitmap \texttt{01} says to skip the first reference entry and copy the second. The residual list is then $\{1\}$.

For this toy, a gamma-coded reference distance costs one bit, the bitmap costs two, and the residual costs one. That is four payload bits, the same as direct coding. The full representation still needs a way to identify its action and delimit the list. Copying is therefore not automatically a saving, even when two lists overlap.

The decoder reverses the process: read the reference distance, recover the reference list, copy marked entries, decode residuals, and merge the two sorted sets. Reference depth controls how many earlier lists this may require.

\subsection{Copy blocks and intervals}
A long copy bitmap may have runs. The bitmap \texttt{111100001111} can be described by run lengths $(4,4,4)$ plus its starting state. This is a different representation, not necessarily a shorter one: gamma-coding these three lengths already costs fifteen bits, more than the twelve raw bitmap bits. The encoder must charge the representation actually used.

An interval describes consecutive targets by a start and length. For targets $100,\ldots,108$, absolute-first gamma gaps cost $13+8=21$ payload bits. With minimum interval length four, encoding the start and the adjusted length $9-4+1=6$ costs $13+5=18$ bits. The interval count and mode still have to be paid. A short interval may lose after those fields are included.

\subsection{From row cost to an order choice}
An ordering can shorten gaps but move a useful reference outside the window. It can also expose more copies while leaving less regular residuals. The encoder's total includes reference descriptions, residuals, and headers. No one of these quantities determines the total by itself.

The selection problem adds another distinction. A score may choose the best of the orders it sees while missing an order that was never generated. This motivates the earlier comparison of regret within five-candidate and eight-candidate sets, retained in Appendix~\ref{sec:legacy-results}. The current campaign uses a different portfolio.

\section{Earlier evaluation protocol and supporting measurements}
\label{app:protocol}
This appendix retains the settings and measurements of the earlier three-encoder studies. Their preprocessing, candidate sets, timing protocol and byte accounting differ from the current campaign in Section~\ref{sec:results}. Statements about development-only merging or prospective settings below apply to those earlier studies. These rows are not additional campaign replicates.

\subsection{Inputs and preprocessing}
The inputs come from SNAP and the Laboratory for Web Algorithmics~\cite{leskovec2014snap,law_datasets}. Table~\ref{tab:allgraphs} reports the graph sizes recorded by the selection drivers. ``Earlier'' includes development and the subsequent six-graph holdout. It does not mean every setting was fixed before those graphs were inspected.

\begin{table}[p]\centering\small
\caption{Graphs in the selection comparison, plus enwiki-2013 from the older size tables. The latter was not part of the selection test.}
\label{tab:allgraphs}
\begin{tabular}{lrrl}\toprule
Graph & Vertices & Directed edges & Stage\\\midrule
amazon-0601 & 395,234 & 3,301,092 & Earlier \\
amazon-2008 & 735,323 & 5,158,388 & Earlier \\
arxiv-hep-ph & 34,546 & 421,578 & Earlier \\
cnr-2000 & 325,557 & 3,216,152 & Earlier \\
dblp-2010 & 326,183 & 1,615,400 & Earlier \\
eat & 7,754 & 247,172 & Earlier \\
email-EuAll & 265,009 & 418,956 & Prospective \\
enron & 69,244 & 276,143 & Earlier \\
enwiki-2013 & 4,206,785 & 101,355,853 & Historical only \\
eswiki-2013 & 972,839 & 23,041,488 & Prospective \\
eu-2005 & 862,664 & 19,235,140 & Earlier \\
in-2004 & 1,382,908 & 16,917,053 & Earlier \\
soc-Epinions1 & 75,879 & 508,837 & Prospective \\
soc-Slashdot0902 & 82,168 & 870,161 & Prospective \\
UK-2007-100K & 100,000 & 3,050,615 & Earlier \\
uk-2014-tpd & 1,766,010 & 18,244,650 & Earlier \\
web-BerkStan & 685,230 & 7,600,595 & Prospective \\
web-NotreDame & 325,729 & 1,469,679 & Prospective \\
web-Stanford & 281,903 & 2,312,497 & Prospective \\
web-google & 434,818 & 3,419,124 & Earlier \\
wiki-Vote & 7,115 & 103,689 & Prospective \\
 
\bottomrule\end{tabular}
\end{table}

The six development graphs are Amazon-0601, HEP-PH, EAT, Web-Google, CNR-2000, and in-2004. Amazon, EAT, and Web-Google use extracted cores: the subgraphs induced by their largest strongly connected components, with vertex identifiers compacted in input order. HEP-PH uses the full loaded graph. The two crawls preserve the identifiers in their URL-ordered input files.

The next six inputs are Amazon-2008, DBLP-2010, Enron, EU-2005, UK-2007-100K (the 100,000-vertex subset of uk-2007-05), and UK-2014-tpd. The prospective SNAP graphs are dense-ranked in source-identifier order, with duplicate arcs and self-loops removed. The prospective eswiki-2013 input preserves the supplied identifiers. Here input order refers to the identifiers after this stated preprocessing.

\subsection{Configurations and candidate identities}
The common held-out and prospective settings use BG window 64 with two-reference support enabled and residual splitting disabled; CS window 256 without splitting; and CG window eight with intervals, minimum interval length four, and splitting enabled. CG uses one group for these sequential files. BVGraph uses window seven and depth three.

Development settings differ. BG enables splitting on Amazon, HEP-PH, and EAT. CS uses window 64 on Amazon and 256 otherwise. CG windows are 8, 8, 256, 256, 128, and 8 on Web-Google, Amazon, HEP-PH, EAT, CNR, and in-2004, respectively. Its minimum interval length is five on CNR and four elsewhere. Development merging uses thresholds 100 on Web-Google and Amazon, 20 on HEP-PH, and none on the other three. Fresh and held-out runs have no merge.

The direct guard uses the no-reference BVGraph-style cost. The exact guard mirrors BVGraph at window seven and depth three, including its offset stream. The mirror was checked against the Java implementation. This does not make it an exact cost model for the other encoders.

The RGB comparison uses the Petri--Mackenzie implementation at revision \texttt{ac9fde47}, with the recorded serial compatibility layer, minimum list length one, and twenty iterations. The minimum list length matters: filtering out short rows would change the graph presented to the orderer.

The ordering experiments use library revision \texttt{a2379512} with local changes that are not publicly archived; the revision alone is insufficient for reproduction. The prospective run uses seed zero. The sensitivity study uses seeds zero through four on EAT, HEP-PH, CNR-2000, Web-Google, DBLP-2010, and Enron. The original five-candidate tie order is input, direct-guard LLP, direct-guard community+LLP, exact-guard LLP, exact-guard community+LLP. The eight-candidate list appends RGB, unguarded LLP, and unguarded community+LLP.

\subsection{Prospective outcomes by graph}
\begin{table}[t]\centering\small
\caption{The five-candidate selector on the prospective graphs. ``Direct'' and ``BV'' name its guard. Each worst regret is over the five target encoders for that graph.}
\label{tab:freshgraphs}
\begin{tabular}{llrr}\toprule
Graph & Selected order & Worst vs five (\%) & Worst vs eight (\%)\\\midrule
email-EuAll & LLP/BV & 0.74 & 0.74 \\
eswiki-2013 & LLP/BV & 0.00 & 5.02 \\
soc-Epinions1 & LLP/BV & 0.21 & 0.21 \\
soc-Slashdot0902 & LLP/direct & 0.00 & 2.01 \\
web-BerkStan & Community/direct & 0.18 & 0.18 \\
web-NotreDame & LLP/direct & 0.00 & 0.00 \\
web-Stanford & LLP/BV & 0.27 & 0.27 \\
wiki-Vote & LLP/direct & 0.00 & 5.52 \\
 
\bottomrule\end{tabular}
\end{table}
Table~\ref{tab:freshgraphs} gives the prospective choices by graph. The within-five mean regret is $0.077\%$. This is distinct from the $0.868\%$ mean against all eight candidates. The prospective script and scorer hashes were checked against their frozen values before scoring. Candidate thresholds and the replacement rule were not changed after the results.

The directly guarded community construction wins two fresh graphs against directly guarded LLP. Under the exact BVGraph guard it wins one; without either guard it wins three. Four different candidates win under BG-ctx across the fresh set. These counts support variation in the best order, but do not rescue the stronger non-one-sidedness prediction under every guard.

\subsection{Winners and workflow outcomes}
Table~\ref{tab:winners} shows that the best order can change with the encoder. Table~\ref{tab:expanded} reports the resulting losses of the expanded selector, and Table~\ref{tab:workflow-outcomes} gives the size and memory outcomes of the workflow comparison. Here BV-w7 means BVGraph with reference window seven and maximum reference depth three.
\begin{table}[t]\centering\small
\caption{Smallest measured order among all eight candidates, by target encoder. L/C mean plain/community LLP; D/B/U mean direct guard, exact BVGraph guard, or unguarded. I denotes input. A plus sign joins exact ties at retained precision.}
\label{tab:winners}
\begin{tabular}{lccccc}\toprule
Graph & BG-ctx & CS-ctx & CG-ctx & BG-Fib. & BV-w7\\\midrule
email-EuAll & C/D & C/D & C/D & L/B & C/B \\
eswiki-2013 & L/B & L/B & L/B & RGB & RGB \\
soc-Epinions1 & L/B & L/B & L/D & L/D & L/D \\
soc-Slashdot0902 & RGB & RGB & L/D & RGB & RGB \\
web-BerkStan & C/D & C/D & C/B & C/D & C/B \\
web-NotreDame & L/D & L/D & L/D & L/D & L/D \\
web-Stanford & L/B & L/B & L/B & L/B & L/D \\
wiki-Vote & RGB & RGB & RGB & RGB & RGB \\
 
\bottomrule\end{tabular}\end{table}

\begin{table}[t]\centering\small
\caption{Regret of the eight-order selector against its own candidates, in percent. Abbreviations are as in Table~\ref{tab:winners}. The choice is shared across encoders, but its loss is not.}
\label{tab:expanded}
\begin{tabular}{llrrrrr}\toprule
Graph & Choice & BG-ctx & CS-ctx & CG-ctx & BG-Fib. & BV-w7\\\midrule
email-EuAll & L/B & 0.36 & 0.74 & 0.74 & 0.00 & 0.50 \\
eswiki-2013 & RGB & 1.47 & 0.56 & 0.74 & 0.00 & 0.00 \\
soc-Epinions1 & L/B & 0.00 & 0.00 & 0.21 & 0.03 & 0.02 \\
soc-Slashdot0902 & RGB & 0.00 & 0.00 & 1.03 & 0.00 & 0.00 \\
web-BerkStan & C/D & 0.00 & 0.00 & 0.05 & 0.00 & 0.18 \\
web-NotreDame & L/D & 0.00 & 0.00 & 0.00 & 0.00 & 0.00 \\
web-Stanford & L/B & 0.00 & 0.00 & 0.00 & 0.00 & 0.27 \\
wiki-Vote & RGB & 0.00 & 0.00 & 0.00 & 0.00 & 0.00 \\
 
\bottomrule\end{tabular}\end{table}

\begin{table}[t]\centering\small
\caption{File sizes and peak resident memory for the workflow experiment. Size is BG-ctx bpe. Memory covers the complete process, not just the decoder. All-three and all-eight sizes are omitted here; the three-order choice matches all-three in each case.}
\label{tab:workflow-outcomes}
\begin{tabular}{lrrrrrrrr}\toprule
& \multicolumn{3}{c}{Size (bpe)} & \multicolumn{5}{c}{Peak memory (GB)}\\
Graph & Fixed & Select 3 & Select 8 & Fixed & Select 3 & All 3 & Select 8 & All 8\\\midrule
cnr-2000 & 1.6921 & 1.6783 & 1.6783 & 1.72 & 1.49 & 1.58 & 1.76 & 1.69 \\
eat & 8.4259 & 8.4259 & 8.4699 & 1.10 & 1.13 & 1.19 & 1.15 & 1.20 \\
uk-2014-tpd & 8.6652 & 8.6564 & 8.6564 & 4.67 & 4.83 & 4.66 & 5.06 & 5.39 \\
web-google & 3.0589 & 3.0589 & 3.0589 & 1.85 & 1.82 & 1.95 & 1.87 & 2.08 \\
 
\bottomrule\end{tabular}\end{table}

\subsection{What the component measurements do and do not explain}
For the six development graphs, Table~\ref{tab:components} separates modeled residual bits from reference-description bits. It also reports the share of arcs copied by that model and the mean source--target log distance
\[
 \frac1m\sum_{(u,v)\in E}\log_2(1+|\pi(u)-\pi(v)|).
\]
This locality statistic counts all edges. It is not the consecutive-target gap objective, and it does not exclude copied edges.

\begin{table}[t]\centering\small
\caption{Community+LLP minus LLP under the direct guard, seed zero. Residual and reference changes are modeled bits per edge under $C_{64}$; reference includes bitmap bits. Copy shares and log distances list LLP/community. BG is the actual BG-ctx size change. Component normalization uses the actual edge count.}
\label{tab:components}
\begin{tabular}{lrrrrr}\toprule
Graph & Residual & Reference & Copied (\%) & Log distance & BG\\\midrule
amazon-0601 & -0.637 & +0.051 & 44.7/46.3 & 6.27/5.47 & -0.417 \\
arxiv-hep-ph & -0.387 & +0.098 & 38.5/41.6 & 8.15/7.46 & -0.294 \\
web-google & -0.142 & +0.013 & 69.6/70.6 & 4.90/4.69 & -0.141 \\
eat & -0.156 & +0.122 & 13.0/15.9 & 9.37/8.99 & -0.079 \\
cnr-2000 & +0.013 & +0.012 & 74.5/75.4 & 7.28/7.36 & +0.014 \\
in-2004 & +0.007 & -0.002 & 57.4/57.2 & 6.72/6.68 & +0.007 \\
 
\bottomrule\end{tabular}
\end{table}
The modeled residual change has the same sign as the target-size change here. Reference costs often move against it. The log-distance change gives the wrong sign on in-2004. These observations locate a trade-off in these orders; they do not provide a statistic that predicts the winner before both orders are built. Copy fractions describe the surrogate's actions, not the copies made by every target encoder.

\FloatBarrier
\subsection{Matched implementation benchmark}
\label{app:matched}
All five inputs use their original order. EAT and Web-Google are the extracted cores in Table~\ref{tab:allgraphs}; HEP-PH, CNR-2000, and in-2004 use the retained full inputs. Each implementation receives the same adjacency lists. Ordering and selection are outside this benchmark.

\paragraph{Source and settings.}
The native base is public commit \texttt{d55ebb39da97}. The experiments use a frozen snapshot with documented ablation switches and query instrumentation, not that commit alone. This complete experimental snapshot is not publicly archived (Section~\ref{sec:discussion}). Julia is version 1.12.4. We enable \texttt{optimize\_decode=true} and CG's \texttt{optimize\_costs=true}. The shipped files remain byte-identical to the preceding size runs. The BG/CS specialized decoder requires residual splitting, disabled here, so those files use the general path. WebGraph is version 3.6.12 on Java 21.0.1. Zuckerli is public commit \texttt{874ac40705d1}, compiled with GCC 9.4.0 and \texttt{-O3}. The native, Java, and C++ measurements compare implementations, not isolated coding algorithms.

BG uses reference window 64 and two-reference search; CS uses window 256. Both use context coding without residual splitting. BG-Fibonacci changes the integer code. CG uses one group, blocks of 128 vertices, and window eight. Indexed BG/CS use chunks of 2,048 vertices. BVGraph-7 uses window seven, reference depth three, minimum interval length four, and zeta-3, a self-delimiting integer code with parameter three~\cite{boldi2004webgraph}. BV-HC changes the window to 16 and depth to $2^{31}-1$.

Zuckerli uses window 32 in archive and access modes. The default controls use one reference-selection round and hybrid-integer parameters $(k,i,j)=(4,2,1)$. The tuned configurations use the published settings: two rounds and $(4,1,0)$. Here $k$ sets the direct-token threshold, and $i,j$ count high and low bits included in the token. A compatibility patch enables the published flags, including the missing low-bit flag. All eight tuned files on the four smaller inputs match the upstream command-line encoder byte-for-byte. On each graph the headline baseline is the smallest raw file among both modes and both settings.

The CNR archive reproduction check failed: $1.848965$ bpe does not round to the published $1.84$. The cause is unidentified, and the publication supplies neither an input checksum nor a complete command line. We disclose the discrepancy and use the measured file size. The access result rounds to the published $2.19$. On in-2004, archive and access give $1.309304$ and $1.654176$ bpe, matching the published $1.31$ and $1.65$. This agreement does not resolve the CNR discrepancy.

\paragraph{Timing and validation.}
All new timings use the same AMD EPYC 9655 allocation, pinned to logical CPU 6 with one application thread. Five independent processes run in randomized order for each configuration and operation. Each has three timed repetitions after one native/C++ warmup or three Java warmups. Tables report the median of process medians and their full range, not confidence intervals. Java uses \texttt{UseSerialGC}, \texttt{ActiveProcessorCount=1}, a 4\,GiB heap limit, and one WebGraph worker. Julia and Java request garbage collection before repetitions, outside the timer. Warmup limits compilation cost without proving steady state.

Each process follows 30 quiet seconds. During the run, the activity check sums container-visible external processes individually using more than 5\% of one logical CPU; attempts with a combined total above 50\% of one logical CPU are rejected. Preflight checks total external activity. We retain the first accepted attempt, not the fastest. This cannot rule out contention from other users of the physical host. Historical laptop times are not pooled with these results.

Every encoded file is hash-checked and decoded against every input list outside timing. Full decodes and all query answers are checked. Decoding materializes independent neighbor lists. Peak resident memory includes runtime, inputs, validation, and returned lists; it is not encoder workspace alone. The 3,690 accepted processes comprise 600 encoding/decoding ablation runs, 2,800 four-graph query runs, and 290 in-2004 runs. Interrupted and failed runs are excluded from timing aggregates.

\paragraph{Size accounting.}
Table~\ref{tab:matched-all} separates sequential and indexed sizes. BVGraph sequential bytes include its graph and properties, but exclude offsets; its indexed bytes include offsets. Encoding time includes writing them. Native and Zuckerli sizes count complete files. In-memory indexes and decoded caches are excluded from file size. On in-2004, BVGraph-7 is $2.17$ bpe sequential and $2.83$ indexed; BV-HC is $1.77$ and $2.39$, respectively. Zuckerli access files can also be fully decoded, so their stream size appears in both relevant columns. With access modes matched on in-2004, indexed CS is $23.70\%$ smaller than tuned Zuckerli access.

\begin{table}[p]\centering\footnotesize\setlength{\tabcolsep}{3pt}
\caption{Matched files, time ranges, and encoding-process memory. Times are median [minimum, maximum] across five process medians. A dash is an unmeasured or inapplicable entry. Indexed native sizes include the complete files used for queries. Their encoding times and process memory were measured only on in-2004; indexed native full decoding was not scheduled. ZK denotes Zuckerli.}\label{tab:matched-all}
\begin{tabular}{lrrrrr}\toprule
Configuration & Seq. bpe & Index bpe & Encode (s) & Decode (ms) & Peak MiB\\\midrule
\multicolumn{6}{l}{\textbf{EAT}}\\
BG-ctx & 9.113 & --- & 1.662 [1.626, 1.737] & 13.56 [13.16, 13.72] & 1037 \\
CS-ctx & 9.056 & --- & 2.801 [2.741, 2.844] & 13.70 [13.24, 14.19] & 1064 \\
CG-ctx & 9.140 & --- & 0.054 [0.054, 0.055] & 10.62 [10.12, 11.19] & 1179 \\
BG-Fib. & 11.045 & --- & 1.636 [1.618, 1.697] & 15.53 [15.14, 15.59] & 1043 \\
BVGraph-7 & 10.783 & 11.307 & 0.097 [0.093, 0.112] & 15.33 [13.10, 19.40] & 94 \\
BV-HC & 10.754 & 11.279 & 0.132 [0.123, 0.138] & 19.83 [12.55, 19.96] & 97 \\
ZK default archive & 9.703 & --- & 0.154 [0.153, 0.161] & 5.67 [5.59, 5.93] & 15 \\
ZK default access & 9.322 & 9.322 & 0.172 [0.162, 0.177] & 5.25 [5.18, 5.41] & 11 \\
ZK tuned archive & 9.413 & --- & 0.299 [0.296, 0.316] & 5.34 [5.27, 5.62] & 15 \\
ZK tuned access & 9.227 & 9.227 & 0.308 [0.306, 0.309] & 4.44 [4.36, 4.67] & 11 \\
BG indexed & --- & 9.135 & --- & --- & --- \\
CS indexed & --- & 9.081 & --- & --- & --- \\
\multicolumn{6}{l}{\textbf{HEP-PH}}\\
BG-ctx & 8.959 & --- & 1.362 [1.349, 1.381] & 28.70 [27.94, 29.52] & 1036 \\
CS-ctx & 8.918 & --- & 1.470 [1.463, 1.475] & 28.12 [27.78, 28.73] & 1036 \\
CG-ctx & 9.068 & --- & 0.103 [0.101, 0.105] & 20.11 [19.92, 21.04] & 1197 \\
BG-Fib. & 10.174 & --- & 1.342 [1.329, 1.424] & 30.44 [29.72, 33.75] & 1035 \\
BVGraph-7 & 10.280 & 11.308 & 0.090 [0.088, 0.090] & 19.05 [14.10, 25.17] & 110 \\
BV-HC & 10.150 & 11.175 & 0.169 [0.164, 0.175] & 18.67 [14.31, 19.08] & 112 \\
ZK default archive & 9.379 & --- & 0.222 [0.220, 0.231] & 11.57 [11.51, 12.26] & 21 \\
ZK default access & 9.130 & 9.130 & 0.292 [0.288, 0.308] & 10.33 [10.26, 11.12] & 13 \\
ZK tuned archive & 9.205 & --- & 0.423 [0.421, 0.424] & 11.17 [10.96, 11.88] & 20 \\
ZK tuned access & 9.084 & 9.084 & 0.534 [0.531, 0.562] & 9.48 [9.38, 10.09] & 15 \\
BG indexed & --- & 9.136 & --- & --- & --- \\
CS indexed & --- & 9.102 & --- & --- & --- \\
 
\bottomrule\end{tabular}\end{table}

\begin{table}[p]\ContinuedFloat\centering\footnotesize\setlength{\tabcolsep}{3pt}
\caption{Matched files, times, and memory (continued). Accounting is as above.}
\begin{tabular}{lrrrrr}\toprule
Configuration & Seq. bpe & Index bpe & Encode (s) & Decode (ms) & Peak MiB\\\midrule
\multicolumn{6}{l}{\textbf{CNR-2000}}\\
BG-ctx & 1.732 & --- & 20.940 [19.725, 22.827] & 126.29 [124.60, 139.53] & 1256 \\
CS-ctx & 1.695 & --- & 53.819 [52.238, 53.894] & 127.60 [124.08, 128.44] & 1207 \\
CG-ctx & 1.995 & --- & 0.700 [0.675, 0.740] & 75.44 [73.52, 80.15] & 1273 \\
BG-Fib. & 2.044 & --- & 21.121 [19.457, 21.400] & 147.17 [137.91, 163.93] & 1233 \\
BVGraph-7 & 2.900 & 3.709 & 0.205 [0.187, 0.213] & 33.55 [33.10, 36.02] & 251 \\
BV-HC & 2.451 & 3.218 & 0.585 [0.561, 0.587] & 33.46 [31.98, 33.64] & 248 \\
ZK default archive & 1.892 & --- & 0.876 [0.875, 0.880] & 38.68 [38.31, 41.01] & 76 \\
ZK default access & 2.291 & 2.291 & 1.074 [1.063, 1.081] & 36.61 [35.55, 37.88] & 74 \\
ZK tuned archive & 1.849 & --- & 1.730 [1.700, 1.743] & 39.81 [37.70, 41.29] & 77 \\
ZK tuned access & 2.185 & 2.185 & 2.054 [2.029, 2.106] & 35.98 [35.56, 37.15] & 74 \\
BG indexed & --- & 1.747 & --- & --- & --- \\
CS indexed & --- & 1.713 & --- & --- & --- \\
\multicolumn{6}{l}{\textbf{Web-Google}}\\
BG-ctx & 4.802 & --- & 12.139 [12.046, 12.545] & 261.34 [236.63, 300.13] & 1336 \\
CS-ctx & 4.800 & --- & 13.680 [12.673, 14.055] & 238.49 [231.41, 249.18] & 1302 \\
CG-ctx & 4.920 & --- & 1.163 [1.152, 1.177] & 126.64 [124.74, 129.62] & 1409 \\
BG-Fib. & 5.786 & --- & 12.414 [11.827, 14.013] & 301.46 [279.97, 337.28] & 1372 \\
BVGraph-7 & 6.720 & 8.057 & 0.486 [0.445, 0.504] & 59.09 [57.99, 59.83] & 276 \\
BV-HC & 6.167 & 7.464 & 1.211 [1.199, 1.230] & 56.93 [56.22, 57.52] & 271 \\
ZK default archive & 4.957 & --- & 1.471 [1.450, 1.581] & 69.97 [69.78, 71.45] & 106 \\
ZK default access & 5.526 & 5.526 & 1.832 [1.757, 1.855] & 62.66 [62.43, 62.97] & 92 \\
ZK tuned archive & 4.925 & --- & 2.820 [2.813, 2.863] & 71.28 [68.07, 74.05] & 106 \\
ZK tuned access & 5.366 & 5.366 & 3.391 [3.357, 3.563] & 63.18 [62.51, 65.49] & 92 \\
BG indexed & --- & 5.008 & --- & --- & --- \\
CS indexed & --- & 5.010 & --- & --- & --- \\
 
\bottomrule\end{tabular}\end{table}

\begin{table}[p]\ContinuedFloat\centering\footnotesize\setlength{\tabcolsep}{3pt}
\caption{Matched files, times, and memory (continued). Accounting is as above.}
\begin{tabular}{lrrrrr}\toprule
Configuration & Seq. bpe & Index bpe & Encode (s) & Decode (ms) & Peak MiB\\\midrule
\multicolumn{6}{l}{\textbf{in-2004}}\\
BG-ctx & 1.259 & --- & 121.518 [116.067, 126.279] & 987.54 [941.56, 1008.31] & 2382 \\
CS-ctx & 1.239 & --- & 288.050 [285.817, 299.236] & 1012.74 [954.61, 1172.38] & 2315 \\
CG-ctx & 1.415 & --- & 4.517 [3.991, 4.905] & 453.11 [434.75, 505.65] & 2486 \\
BG-Fib. & 1.524 & --- & 117.509 [115.274, 125.872] & 1232.62 [1100.49, 1323.57] & 2537 \\
BVGraph-7 & 2.173 & 2.830 & 0.863 [0.792, 0.891] & 143.18 [141.68, 144.97] & 784 \\
BV-HC & 1.767 & 2.386 & 2.536 [1.905, 2.617] & 137.79 [132.05, 147.20] & 755 \\
ZK default archive & 1.326 & --- & 3.880 [3.850, 4.109] & 167.08 [164.85, 181.23] & 341 \\
ZK default access & 1.777 & 1.777 & 4.551 [4.527, 4.912] & 163.28 [157.80, 171.95] & 339 \\
ZK tuned archive & 1.309 & --- & 7.660 [7.592, 8.654] & 163.74 [163.29, 169.82] & 341 \\
ZK tuned access & 1.654 & 1.654 & 8.906 [8.775, 9.295] & 165.40 [157.69, 171.51] & 339 \\
BG indexed & --- & 1.282 & 137.610 [137.158, 143.416] & --- & 2470 \\
CS indexed & --- & 1.262 & 320.882 [314.721, 331.009] & --- & 2558 \\
 
\bottomrule\end{tabular}\end{table}
\begin{table}[t]\centering\small
\caption{CS ablation costs on the four smaller inputs. Size changes and encoding-time ratios use shipped CS as the denominator. Negative size changes favor the variant. Timing ratios use the five-process medians. Fast search reduces search effort.}
\label{tab:ablation-times}
\begin{tabular}{lrrrrrr}\toprule
& \multicolumn{2}{c}{Copy + residual} & \multicolumn{2}{c}{Window 32} & \multicolumn{2}{c}{Fast search}\\
Graph & Size (\%) & Time ratio & Size (\%) & Time ratio & Size (\%) & Time ratio\\\midrule
EAT & -0.86 & 0.40 & +1.38 & 0.16 & -0.49 & 0.19 \\
HEP-PH & -2.13 & 0.50 & +1.86 & 0.35 & -0.93 & 0.39 \\
CNR-2000 & +1.61 & 0.26 & +6.31 & 0.15 & +10.97 & 0.06 \\
Web-Google & -0.54 & 0.48 & +0.45 & 0.56 & +0.83 & 0.38 \\
 
\bottomrule\end{tabular}\end{table}

\subsection{Query traces and decoded-cache budgets}
\label{app:access-protocol}
The four-graph grid uses random and breadth-first traces of 1, 10, 100, 1,000, and 2,000 queries. BG and CS use no cache, 64\,KiB, 1\,MiB, 16\,MiB, or unlimited whole-chunk caching. A separate control uses the older prefix cache, which stores decoded row prefixes rather than whole chunks. These are distinct access algorithms. The same traces are used for BVGraph-7 and tuned Zuckerli access. On in-2004 the grid is limited to 1, 100, and 2,000 queries, with no cache or unlimited whole chunks.

Each repetition starts with a fresh reader and retains returned lists for validation. Session totals sum opening and the query loop in that repetition. Setup between the timers is excluded. A second fresh reader measures individual request times, so those measurements do not warm the session measurement. Request clocks add overhead; session totals do not include those per-request clocks. Runtime warmup is used and the filesystem cache is not flushed.

The cache budget covers retained decoded chunks only. Compressed input, temporary decoding memory, and returned lists are outside it. A chunk larger than the budget is decoded without being cached. Zuckerli decodes at opening to construct its in-memory node-offset index. Its opening cost is included, but this runtime index is not added to file size.

The exploratory latency condition compares BG/CS median request times with tuned Zuckerli, excluding opening, for the zero and 64\,KiB budgets. Two encoders, four graphs, two traces, five lengths, and two budgets give 160 comparisons. Only two pass the $100\times$ condition: both encoders on CNR, breadth-first length 100, at 64\,KiB. The median ratio is about $4{,}810\times$ and the maximum about $38{,}144\times$. The protocol expected the broad condition to fail somewhere; no claim of uniformly competitive access was assumed.

\begin{figure}[p]\centering
\includegraphics[width=.94\linewidth]{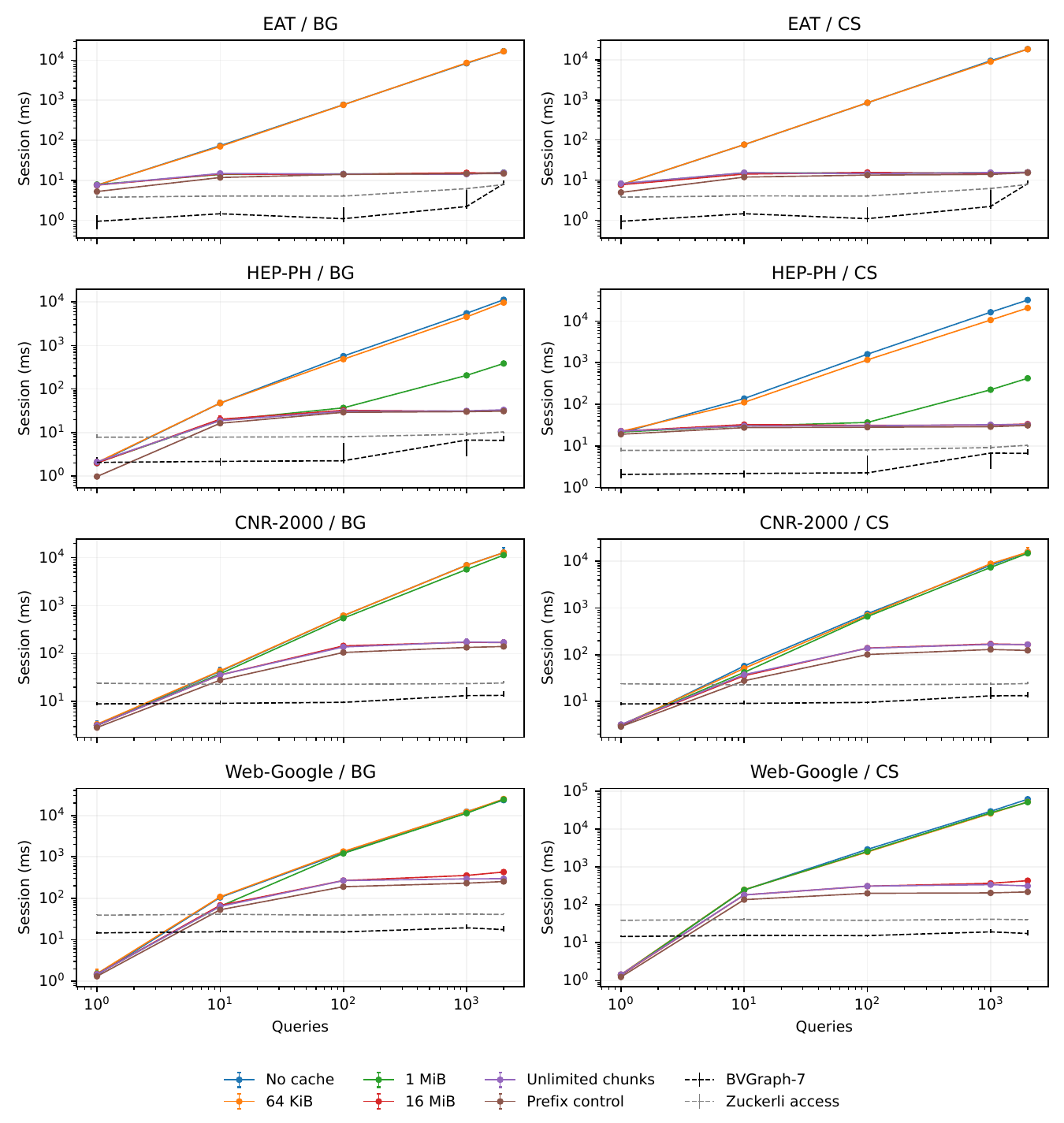}
\Description{Eight panels show random-query session cost for BG and CS on four graphs, all cache budgets, and both external readers. Small native caches cause repeated chunk decoding.}
\caption{Random-query sessions on four inputs. Both axes are logarithmic. Points and bars show medians and ranges across five processes. Totals include opening. Cache budgets cover retained decoded chunks only; the prefix control is a different access algorithm.}
\label{fig:cache-random}\end{figure}

\begin{figure}[p]\centering
\includegraphics[width=.94\linewidth]{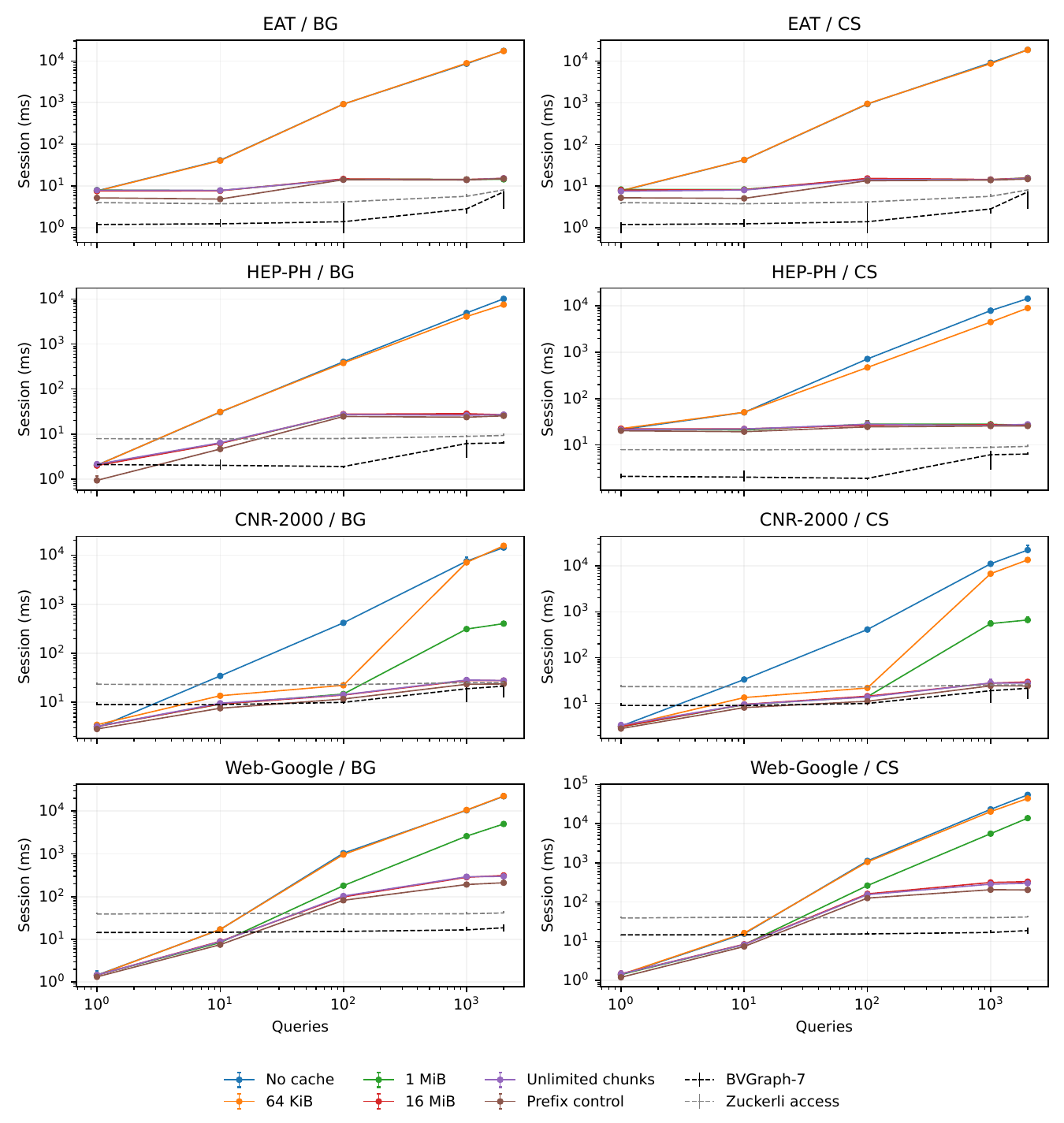}
\Description{Eight panels show breadth-first query sessions for BG and CS on four graphs, all cache budgets, and both external readers. Locality can reduce the cost when chunks are retained.}
\caption{Breadth-first query sessions, with the same accounting as Figure~\ref{fig:cache-random}. All cache choices and trace lengths are retained.}
\label{fig:cache-bfs}\end{figure}

Figures~\ref{fig:cache-random} and~\ref{fig:cache-bfs} retain every measured cache choice on the four smaller inputs. Table~\ref{tab:large-query} gives the larger graph's CS and external-reader session ranges. On its breadth-first trace, unlimited-chunk CS is faster; on its long random trace it is slower. Neither result is a general guarantee about locality or cache capacity.


\begin{table}[t]\centering\footnotesize\setlength{\tabcolsep}{3pt}
\caption{Opening plus query time on in-2004, in milliseconds: median [minimum, maximum] across five processes. BFS denotes breadth-first traversal. Native columns use indexed CS; external columns use BVGraph-7 and tuned Zuckerli access.}
\label{tab:large-query}
\begin{tabular}{llrrrr}\toprule
Trace & Queries & CS, no cache & CS, unlimited & BVGraph & Zuckerli\\\midrule
Random & 1 & 13.2 [13.0, 13.3] & 14.6 [14.0, 15.0] & 36.0 [33.9, 36.4] & 99.8 [96.9, 104.8] \\
Random & 100 & 898.0 [889.2, 937.7] & 604.5 [583.6, 623.6] & 36.0 [33.9, 38.1] & 100.1 [98.4, 112.4] \\
Random & 2,000 & 17172.8 [16811.9, 18560.2] & 1889.1 [1536.0, 2072.7] & 37.1 [36.7, 41.5] & 103.6 [98.7, 108.5] \\
BFS & 1 & 13.2 [13.1, 13.5] & 14.2 [13.9, 14.5] & 34.5 [34.1, 35.7] & 98.8 [97.6, 105.0] \\
BFS & 100 & 867.8 [850.6, 904.0] & 25.6 [25.3, 29.1] & 34.6 [34.5, 36.2] & 98.5 [97.3, 102.7] \\
BFS & 2,000 & 18522.4 [17981.9, 18796.7] & 25.7 [25.4, 26.2] & 38.0 [36.4, 41.3] & 105.2 [100.0, 109.8] \\
 
\bottomrule\end{tabular}\end{table}

\FloatBarrier
\subsection{Earlier workflow and order-timing measurements}
\label{app:older-timing}

Each complete workflow runs in its own process after warming the same workflow on EAT. The reported total excludes graph loading but includes construction, scoring when used, relabeling, writing, cleanup, and intervening garbage collection. Peak resident memory is measured for the whole process. It is not the decoder's workspace alone. The workflow driver does not include a roundtrip check in its timed path; separate size runs perform adjacency equality checks for the contributed encoders.

Table~\ref{tab:absolute-decode} reports the earlier decode study, which covers EAT, HEP-PH, Web-Google, CNR, and in-2004. It measures the input order, directly guarded LLP, directly guarded community+LLP, and RGB. Full decoding reports the minimum of three measurements. The first decoded graph is checked against every adjacency list; this is separate from the timing comparison.

Indexed BG and CS use chunks of 2,048 vertices. Query timing uses 20,000 uniformly drawn encoded vertex identifiers after a separate warmup. A newly opened reader starts the timed trace. This is not evidence that every query is cold: repeated queries and any internal state may affect later queries. The driver verifies the first 2,000 returned lists, not all 20,000. Opening time is recorded separately. For these earlier workflow files, the BG/CS indexed-size increments range from $0.0125$ to $0.3504$ bpe. These older measurements are not pooled with the matched benchmark above.

The older driver also opens a fresh reader after its separate warmup; it does not begin with that warmup reader's cached answers. Its longer trace spreads cache-filling work over more requests. It also discards each returned value rather than retaining independently owned lists as the matched benchmark does. These protocol differences prevent a direct comparison. The new measurements do not establish the cause of the earlier per-query discrepancy. Full decoding previously built a graph object and reported a minimum, while the matched benchmark builds neighbor lists and reports a median. Source versions, optimization switches, and some encoding orders also differ.

The earlier workflow and order-timing experiments were run on an AMD Ryzen 7 PRO 4750U laptop with 32 GB RAM, shared with other work. Table~\ref{tab:absolute-encode} reports the fixed-workflow encoding times. The reported workflow timings are single retained observations per arm, not confidence intervals. They do not estimate a numerical noise band. Comparing encode times under different orders cannot isolate contention. Different source snapshots and different query semantics in older campaigns are not silently pooled with this one.

\begin{table}[t]\centering\small
\caption{Absolute retained decoding and query times in input order. Full decoding builds the in-memory graph; each entry is the minimum of three timings. Query entries are one timed trace of 20,000 requests to indexed BG/CS files, as amortized microseconds per query. File opening is excluded. The fresh reader fills its cache during the trace; these are not isolated request latencies.}
\label{tab:absolute-decode}
\begin{tabular}{lrrrrrr}\toprule
& \multicolumn{4}{c}{Full decode (seconds)} & \multicolumn{2}{c}{Query ($\mu$s)}\\
Graph & BG-ctx & CS-ctx & CG-ctx & BG-Fib. & BG & CS\\\midrule
eat & 0.067 & 0.069 & 0.071 & 0.066 & 1.40 & 1.41 \\
arxiv-hep-ph & 0.159 & 0.156 & 0.173 & 0.161 & 2.76 & 2.69 \\
cnr-2000 & 1.153 & 1.175 & 1.432 & 1.181 & 17.77 & 16.65 \\
web-google & 1.829 & 1.806 & 2.399 & 2.221 & 26.24 & 22.51 \\
in-2004 & 8.608 & 8.154 & 8.575 & 8.550 & 85.54 & 75.23 \\
 
\bottomrule\end{tabular}\end{table}

\begin{table}[t]\centering\small
\caption{Absolute BG-ctx encoding times for the fixed, directly guarded community order in the workflow experiment. One retained timing per graph; loading and ordering excluded. Peak memory covers the whole fixed workflow, including graph construction. GB denotes $10^9$ bytes. These orders differ from the input orders in Table~\ref{tab:absolute-decode}.}
\label{tab:absolute-encode}
\begin{tabular}{lrrr}\toprule
Graph & Encode (s) & Encode ($\mu$s/edge) & Peak GB\\\midrule
cnr-2000 & 84.7 & 26.3 & 1.72 \\
eat & 11.3 & 45.7 & 1.10 \\
uk-2014-tpd & 885.5 & 48.5 & 4.67 \\
web-google & 110.9 & 32.4 & 1.85 \\
 
\bottomrule\end{tabular}\end{table}

\FloatBarrier
\subsection{Matched encoder ablations}
\label{app:matched-ablation}
The ablation uses the same four input graphs and native source as the matched benchmark. We first vary one option from the shipped settings, then add features cumulatively from a plain baseline. Requested flags and effective actions are recorded separately. A switch that leaves the file unchanged is reported as a no-op. Every file is checked against the full adjacency list; shipped controls also match the earlier file hashes.

BG uses window 64 and restricted two-reference search; CS uses window 256 and one reference. Both use context coding with no left/right split. Tested alternatives vary the integer backend, window, references, row types, search effort, and copying. Unsupported combinations are excluded explicitly. The BG exact-costing/context combination failed decoding and is excluded from the compression results.

Table~\ref{tab:ablation-changes} reports the feature changes relative to shipped settings, and Table~\ref{tab:venue-ablation} compares selected CS variants with Zuckerli. The context/Fibonacci pairs have identical row-choice hashes, including reference distances, copy masks, and residual identifiers. The difference therefore comes from writing those choices, not selecting different references. With Fibonacci coding, CS is larger than the strongest Zuckerli file on every input.

Reducing the window to 32 makes CS larger than its shipped version on every input, most strongly on CNR. Restricting CS to copy-plus-residual rows makes three files smaller, but hurts CNR. A cheaper search setting also gives smaller files on EAT and HEP-PH but hurts CNR substantially. Thus additional candidates do not guarantee a smaller context-coded file. The cost estimate is based on Fibonacci codes; the adaptive writer pays a different cost. This is an observed mismatch, not a tested explanation of every loss.

\begin{table}[t]\centering\small
\caption{Selected ablation size changes relative to each shipped encoder, in percent. Negative means smaller. ``Copy + residual'' restricts row types; fast search changes search effort. These are size results, not matched-budget speed comparisons.}
\label{tab:ablation-changes}
\begin{tabular}{llrrrrr}\toprule
Graph & Codec & Fibonacci & Window 32 & Copy + residual & Fast search & No copying\\\midrule
EAT & BG & +21.20 & +0.44 & -0.62 & -0.39 & +5.34\\
EAT & CS & +21.09 & +1.38 & -0.86 & -0.49 & +6.36\\
HEP-PH & BG & +13.55 & +0.63 & -1.45 & -0.48 & +11.14\\
HEP-PH & CS & +13.26 & +1.86 & -2.13 & -0.93 & +11.91\\
CNR-2000 & BG & +18.00 & +3.81 & +1.04 & +9.05 & +119.15\\
CNR-2000 & CS & +19.36 & +6.31 & +1.61 & +10.97 & +122.29\\
Web-Google & BG & +20.50 & +0.23 & -0.45 & +0.99 & +77.33\\
Web-Google & CS & +19.91 & +0.45 & -0.54 & +0.83 & +77.13\\
 
\bottomrule\end{tabular}\end{table}

\subsection{Separating the coding path}
\label{app:coding-split}
We replay fixed row choices with three independently switched components (Table~\ref{tab:path-split}). The first uses adaptive streams for residual tokens, reference distances, commands, and flags. The second uses binary copy bitmaps instead of the legacy copy descriptions. The third codes residual neighbors by their ranks among targets that were not copied, rather than by their original identifier gaps. A \emph{rank} here is the position in that remaining target set.

All eight combinations are measured. To allocate a saving shared by interacting components, we average each component's marginal byte saving over all six orders of enabling the three components. This is an accounting convention, often called a Shapley allocation; it is not a causal comparison with another encoder.

\begin{table}[t]\centering\small
\caption{CS coding-path allocation, in percent of the total saving between the factorial endpoints. Row choices are fixed. Each component weakly reduces size in every tested combination.}
\label{tab:path-split}
\begin{tabular}{lrrr}\toprule
Graph & Adaptive streams & Copy bitmaps & Leftover ranks\\\midrule
EAT & 90.13 & 9.85 & 0.03\\
HEP-PH & 91.29 & 8.24 & 0.47\\
CNR-2000 & 59.82 & 37.77 & 2.42\\
Web-Google & 81.28 & 16.47 & 2.24\\
 
\bottomrule\end{tabular}\end{table}

Adaptive streams account for most of the gain; copy coding matters more on the web graphs. The adaptive-stream switch groups several stream models. The bitmap switch changes both representation and coding. Neither is a pure test of the entropy coder alone. The replay container has a fixed framing difference from the native Fibonacci file; we account for it separately and do not assign it to a component. Native endpoints and all row decisions are checked against the original encoders.

\subsection{Structural measurements across all orders}
\label{app:structural}
The study covers the twenty graphs in Table~\ref{tab:allgraphs}, with the input order and seven candidate orders on each. Most missing candidate permutations were rebuilt with their recorded seeds. We accept a reconstruction only when both its selection score and BG-Fibonacci size match the historical record at its recorded precision. This verifies those quantities, not identity with an unavailable historical permutation. Six retained permutations were replayed directly.

All 140 candidates have structural measurements and BG, CS, and CG encodings. Copy counters leave the output bytes unchanged, and each file is fully decoded and checked. We include the input-order controls separately. These graphs have already been used; this is not a new prospective prediction test.

For a sorted row $v_1<\cdots<v_d$, the consecutive-gap measure averages $\log_2(1+v_{i+1}-v_i)$ over all consecutive pairs in all rows. Source--target distance instead averages $\log_2(1+|u-v|)$ over edges $(u,v)$. They measure different properties. The frozen consecutive-gap statistic uses $\log_2(1+\mathrm{gap})$, not RGB's exact logarithmic-gap objective. ID-block share counts edges whose endpoints lie in the same block of 64 or 1,024 consecutive identifiers. It is not a host statistic.

Adjacent-row similarity is the size of the intersection of two consecutive neighbor sets divided by the size of their union, also called Jaccard similarity. We report both the mean with empty--empty pairs assigned zero and the mean omitting those pairs. Copied-edge share counts actual edges recovered from references, rather than a surrogate copying estimate.

For each candidate, size gain is the input-order byte count minus the candidate byte count, divided by the input-order byte count. A feature change is its candidate value minus its input-order value. Table~\ref{tab:structural-correlations} correlates these changes with size gain using Spearman's rank correlation. Intervals resample whole graphs, keeping their candidates together, for 10,000 bootstrap draws. They describe uncertainty across this small graph collection; candidate pairs are not independent observations.

\begin{table}[t]\centering\small
\caption{All seven structural measures: pooled rank correlation with size gain and 95\% graph-bootstrap intervals. Each encoder has twenty graphs and 140 candidate comparisons. Overlapping intervals do not support ranking one measure above another under the prespecified rule.}
\label{tab:structural-correlations}
\begin{tabular}{lrrr}\toprule
Change in measure & BG & CS & CG\\\midrule
Consecutive-neighbor gaps & $-0.77$ [-0.92, -0.52] & $-0.78$ [-0.92, -0.53] & $-0.79$ [-0.94, -0.54]\\
Source--target distance & $-0.84$ [-0.93, -0.68] & $-0.85$ [-0.93, -0.70] & $-0.87$ [-0.93, -0.73]\\
Same ID block (64) & $0.84$ [0.68, 0.93] & $0.85$ [0.70, 0.93] & $0.86$ [0.72, 0.94]\\
Same ID block (1,024) & $0.83$ [0.66, 0.91] & $0.84$ [0.68, 0.92] & $0.85$ [0.70, 0.91]\\
Adjacent-row similarity & $0.68$ [0.38, 0.86] & $0.67$ [0.37, 0.85] & $0.70$ [0.42, 0.87]\\
Similarity, nonempty unions & $0.62$ [0.28, 0.84] & $0.61$ [0.26, 0.83] & $0.64$ [0.30, 0.85]\\
Actually copied edges & $0.66$ [0.27, 0.88] & $0.55$ [0.14, 0.82] & $0.74$ [0.44, 0.91]\\
 
\bottomrule\end{tabular}\end{table}

The expected gap and copying directions hold in the pooled data. The expectation that the simpler locality measures would be weaker is not supported: ID-block share and source--target distance have larger point-estimate magnitudes. All seven measures are reported.

\begin{table}[t]\centering\small
\caption{Within-graph rank correlation with size gain across all eight orders. Entries give the median across twenty graphs and, in parentheses, the number with the expected sign: negative for gaps and distance, positive otherwise. Zero correlations count toward neither sign.}
\label{tab:structural-within}
\begin{tabular}{lrrr}\toprule
Measure & BG & CS & CG\\\midrule
Consecutive-neighbor gaps & $-0.88$ (17/20) & $-0.88$ (18/20) & $-0.85$ (19/20)\\
Source--target distance & $-0.81$ (18/20) & $-0.80$ (18/20) & $-0.79$ (18/20)\\
Same ID block (64) & $0.61$ (18/20) & $0.69$ (18/20) & $0.64$ (17/20)\\
Same ID block (1,024) & $0.74$ (18/20) & $0.71$ (18/20) & $0.75$ (18/20)\\
Adjacent-row similarity & $0.49$ (17/20) & $0.46$ (16/20) & $0.24$ (14/20)\\
Similarity, nonempty unions & $0.52$ (16/20) & $0.50$ (15/20) & $0.32$ (12/20)\\
Actually copied edges & $0.61$ (15/20) & $0.46$ (13/20) & $0.31$ (13/20)\\
 
\bottomrule\end{tabular}\end{table}

The pooled copying result is especially weak as an order-choice explanation for CG. Including the input order gives a within-graph median of $0.31$; excluding it gives $0.14$, positive on thirteen graphs. The corresponding gap medians among constructed candidates are $-0.89$, $-0.89$, and $-0.82$ for BG, CS, and CG, with the expected sign on sixteen, seventeen, and eighteen graphs. The eight-order results are in Table~\ref{tab:structural-within}. Different graph properties and codec settings can contribute to the pooled associations. They do not prove that manipulating one statistic will improve a file.

URL and host lists were not recovered. True host-transition measurements therefore remain missing. The ID-block and row-similarity measures answer only the broader locality question.

\section{The score and the encoder implementation}
\label{app:implementation}
\subsection{A precise definition of the selection score}
\label{app:score}
For row position $s$, let $T$ be its sorted target positions. Define $D_s(T)$ as the no-reference BVGraph-style row cost. It includes the degree, a no-reference field, interval metadata, and residual coding. Runs of at least four targets are intervals. Residuals use zeta-3 coding; the first residual is source-relative and later values are gap-coded. Empty rows have their specified one-bit cost.

For a nonempty earlier list $U$ at distance $d$, with nonempty overlap, the copy alternative costs
\[
 D_s(T\setminus U)-1+\gamma_0(d)+|U|.
\]
Here $\gamma_0(x)=2\lfloor\log_2(x+1)\rfloor+1$. The subtraction follows the evaluator's convention for replacing the no-reference field, including its special empty-residual case. The bitmap costs $|U|$ bits, including skipped entries. The evaluator chooses the minimum of direct coding and these alternatives over the window.

Let $b_s$ be that minimum. The score adds a modeled offset stream, beginning with one bit and then gamma-coded row lengths. It rounds the graph and offset streams separately to bytes. With window zero, the evaluator reproduces the direct guard's byte count. This check validates that relationship; it does not validate the score against a different target encoder.

\subsection{Reference search and two-reference limits}
The BG full-search profile scans prior rows in the declared window and counts their overlap with the current row. It sorts candidates by overlap and then distance. Up to 1,024 surviving reference candidates receive full cost evaluation. The windows in the main generic BG configuration are below that cap. Direct representations remain available.

Two-reference search is narrower. It requires at least six targets and the compatible header mode. It considers up to five first references from the overlap ranking, requires at least three residual targets after that copy, and scans the window for a second reference that covers at least three of them. It then charges both pointers, both copy descriptions, and the remaining residual. It does not enumerate every pair or optimize a graph-wide reference structure.

An action can use stop-terminated gaps, intervals with a tested minimum length, or hybrid run-length coding. Copy descriptions may use a raw bitmap or run representations. ``Cheapest'' always means cheapest among the enabled candidates under the selected cost model. Adaptive stream size is not, in general, the sum of independently optimized static field lengths.

\subsection{Hybrid integers and contexts}
The native hybrid representation has parameters $K=5$, $M=0$, and $L=1$. Values below $2^K$ are direct tokens. For a larger integer $v$, let $b=\lfloor\log_2v\rfloor$. Its token combines $b-K$ with the low bit. The remaining $b-1$ middle bits are emitted with equal bit probabilities. Thus large values retain a substantial cost even if their token becomes predictable.

There are 150 token symbols. The model has four position buckets, $0,1,2,3+$, and 25 previous-magnitude buckets, including a reset sentinel. This gives 100 token-model addresses. The previous magnitude is the integer's bit length, clipped at 23; it is not the full previous value.

The range coder maintains a 32-bit interval width and a 64-bit accumulator for its lower endpoint and carry. It starts with width $2^{32}-1$ and lower endpoint zero. For cumulative frequency $c$, symbol frequency $f$, and total $t$, it sets $r=\lfloor\mathrm{width}/t\rfloor$, advances the lower endpoint by $rc$, and sets the new width to $rf$. While the width is below $2^{24}$, it emits a byte with deferred carry handling and shifts the interval by eight bits. Remaining integer bits split the interval into equal halves. Finalization performs five byte shifts. The decoder skips the initial carry byte and initializes its code register from the next four bytes.

At each residual-region boundary, position and previous magnitude reset. Frequency tables persist across rows. Every token starts with frequency one. Each observed token adds 32 to its frequency. When the total reaches $2^{23}$, frequencies are replaced by their halved values with the low bit set. Indexed chunk boundaries restart the relevant models. Reference, command, copy, and flag streams use their own models.

Persistent frequency tables also carry information across rows. Counting token-model addresses alone does not describe the complete adaptive state.

\subsection{Offsets, splitting, and access}
A signed source offset can be mapped to a positive integer by $z(a)=2a+1$ for $a\ge0$ and $z(a)=-2a$ otherwise. A production format may use a different zero-based convention; estimates and emitted fields must use the same one.

Residual splitting treats targets below and above the source separately. Each side becomes an increasing sequence of distances from the source. This may shorten the first value on each side, but requires the corresponding mode information. It is a coding option, not a guarantee.

A sampled index records where selected rows or chunks begin. Its space must be counted for random access. Restartable adaptive chunks may also change compression because their models start afresh. CG's historical group-based access representation is different from the single-group sequential configuration. Their sizes cannot be interpreted as merely the cost of adding one offset table.

\subsection{A retained CNR feature study}
\label{app:features}
This earlier BG experiment holds the CNR graph and its plain-LLP order fixed (symmetric updates, five passes, seed zero). Table~\ref{tab:cnr-features} adds options cumulatively. Each file was fully decoded and checked; one- and eight-worker outputs were byte-identical.

\begin{table}[t]\centering\small
\caption{Earlier cumulative BG settings on one fixed CNR order. This is not an ablation of the current context-coded encoders.}
\label{tab:cnr-features}
\begin{tabular}{lr}\toprule
Cumulative setting & bpe\\\midrule
Zeta-3, search window 7 & 3.388 \\
Adaptive copy representation & 2.872 \\
Search window 64 & 2.664 \\
Fibonacci integer code & 2.507 \\
Terminated gap sequences & 2.438 \\
Adaptive header and restricted actions & 2.511 \\
Left/right split and fixed-width references & 2.543 \\
Two references requested but inactive & 2.543 \\
 
\bottomrule\end{tabular}\end{table}

Adaptive copy representation, a wider window, and the integer-code changes reduce size in this sequence. Later steps need care: the adaptive header also restricts available actions, and the left/right split changes reference-distance coding. The requested two-reference option is inactive under that header. A separate active control reduces size from $2.534$ to $2.518$ bpe when two-reference search is enabled. These results show interactions on one input; they do not identify the cause of the matched CS advantage or isolate context coding.

\section{Earlier selection guarantees and their limits}
\label{app:analysis}
These elementary statements concern the earlier guard and candidate sets. The current campaign uses the nesting guarantee in Section~\ref{sec:analysis}; it does not include the input-order arm assumed below.

\subsection{Guard monotonicity and candidate selection}
If a sequence of accepted steps satisfies $Q(\pi_{t+1})\le Q(\pi_t)$, induction gives $Q(\pi_T)\le Q(\pi_0)$. This elementary guarantee concerns $Q$ alone. A target encoder $E$ inherits it only if the checked quantity is its actual counted size, or if a proved comparison supplies the missing link.

For a finite set containing the input order, selecting the smallest measured $B_E$ cannot produce a larger counted file than the input. This is an exact finite-set guarantee. It gives no bound relative to an order outside the set and does not avoid the evaluation cost.

\subsection{A conditional regret bound}
Let $\widehat B(\pi)=aC(\pi)+b$ with $a>0$, fitted for one graph and target encoder. Suppose $|B(\pi)-\widehat B(\pi)|\le\varepsilon$ for every candidate. If $\hat\pi$ minimizes $C$ and $\pi^*$ minimizes $B$, then
\[
 B(\hat\pi)\le\widehat B(\hat\pi)+\varepsilon
 \le\widehat B(\pi^*)+\varepsilon
 \le B(\pi^*)+2\varepsilon.
\]
The scale $a$ handles different units. A negative fitted slope would not preserve the minimizer and is excluded.

For positive costs, a multiplicative form assumes $B(\pi)=aC(\pi)e^{e(\pi)}$ and $|e(\pi)|\le\varepsilon$. The same argument gives $B(\hat\pi)/B(\pi^*)\le e^{2\varepsilon}$. Neither statement estimates its own error bound.

\subsection{Shared-order feasibility and missing candidates}
\label{app:portfolio}
For a fixed graph, let $A$ be a candidate set and $B_e(p)$ the measured size of order $p$ under encoder $e$. The smallest attainable worst regret for one shared order is
\[
 \rho(A)=\min_{p\in A}\max_e
 \left(\frac{B_e(p)}{\min_{q\in A}B_e(q)}-1\right).
\]
This is a finite-table calculation, not a lower bound over all vertex orders. On email-EuAll with the five candidates, $\rho(A)$ is about $0.742\%$. Allowing for rounding of the retained sizes still leaves it above $0.739\%$. Thus the declared $0.5\%$ target was infeasible for a common choice from that set. Choosing separately for each encoder is a different rule.

Let $p$ be the five-candidate selector's choice, and let $b_5$ and $b_8$ be the best measured sizes in the five- and eight-candidate sets for one encoder. The loss decomposes exactly as
\[
 \frac{B_e(p)-b_8}{b_8}
 =\underbrace{\frac{B_e(p)-b_5}{b_8}}_{\text{ranking}}
 +\underbrace{\frac{b_5-b_8}{b_8}}_{\text{missing candidates}}.
\]
Using the same denominator is essential. Across the forty measurements, the mean contributions are $0.077$ and $0.791$ percentage points, respectively. Missing candidates therefore account for about $91\%$ of total loss. This retrospective diagnosis explains the observed failures; it is not a prospective guarantee.

\section{Earlier size comparisons retained for reference}
\label{app:legacy}\label{app:ra}
BV-HC denotes BVGraph with window 16 and maximum reference depth $2^{31}-1$, an effectively unrestricted chain depth. The next tables preserve the earlier absolute-size measurements. They are supplementary configuration records, not the prospective selector test. Input and community orders are unguarded here. The six development graphs use first-appearance grouping. The older enwiki rows placed groups by decreasing size instead. The community columns also use the development merge settings. Their configurations differ from the generic, unmerged fresh-graph results and from the matched benchmark in Appendix~\ref{app:matched}; those sizes come from separate runs and are not alternative roundings of the same files.

Tables~\ref{tab:sota} and~\ref{tab:ra-sota} use Zuckerli default parameters: one reference-selection round and hybrid-integer parameters $(4,2,1)$, in archive and access modes respectively. Their margins compare with this historical default baseline, not the published-parameter baseline in Table~\ref{tab:final-space}, which supersedes them for the earlier fixed-order comparison. For example, the historical input-order in-2004 access margin is $29.0\%$ against defaults, whereas the matched margin is $23.70\%$ against published settings.

The bold entry is the smallest of three contributed encoders after measurement. It is not a fixed encoder selected in advance. BVGraph sidecar offsets are omitted in these historical tables, including the access table. That convention prevents a complete storage comparison for random access. CG's access layout differs from its sequential layout and can incur a large penalty.

\begin{table}[p]
\caption{Historical whole-graph sizes in bits per edge. BV-HC is WebGraph with window 16 and maximum reference depth $2^{31}-1$; Zuck.-def. is Zuckerli archive mode with default parameters (one round, $(4,2,1)$), not the published settings used in Table~\ref{tab:final-space}. BG, CS, and CG use the context-range backend. Bold selects the smallest of these three after measurement. The last column is its reduction relative to default-parameter Zuckerli, recomputed from the displayed sizes. Community denotes unguarded community+LLP with the development merge settings. $^\ddagger$The enwiki measurements use the older decreasing-group-size composition rule. BVGraph offsets are excluded.}
\label{tab:sota}
\centering
\small
\begin{tabular}{ll rr rrr r}
\toprule
Dataset & Ord. & \BV{}-HC & Zuck.-def. & \BG{} & \CS{} & \CG{} & vs default \\
\midrule
\multirow{2}{*}{amazon-0601}  & input & 12.853 & 10.254 & 10.155 & 10.224 & \textbf{9.914} & $+3.3\%$ \\
                              & community & 7.965 & 6.693 & 6.378 & 6.387 & \textbf{6.238} & $+6.8\%$ \\
\midrule
\multirow{2}{*}{arxiv-hep-ph} & input & 10.132 & 9.379 & 8.954 & 8.923 & \textbf{8.795} & $+6.2\%$ \\
                              & community & 7.663 & 7.353 & 6.677 & 6.649 & \textbf{6.458} & $+12.2\%$ \\
\midrule
\multirow{2}{*}{cnr-2000}     & input & 2.448 & 1.892 & 1.732 & \textbf{1.695} & 1.894 & $+10.4\%$ \\
                              & community & 2.489 & 1.902 & 1.773 & \textbf{1.740} & 1.957 & $+8.5\%$ \\
\midrule
\multirow{2}{*}{EAT}          & input & 10.725 & 9.703 & 9.102 & 9.041 & \textbf{8.978} & $+7.5\%$ \\
                              & community & 9.684 & 9.124 & 8.444 & 8.453 & \textbf{8.360} & $+8.4\%$ \\
\midrule
\multirow{2}{*}{enwiki-2013\rlap{$^\ddagger$}}  & input & 15.625 & 13.299 & 13.007 & \textbf{12.945} & 13.121 & $+2.7\%$ \\
                              & community & 12.412 & 10.934 & \textbf{10.561} & 10.668 & 10.615 & $+3.4\%$ \\
\midrule
\multirow{2}{*}{in-2004}      & input & 1.767 & 1.326 & 1.259 & \textbf{1.240} & 1.415 & $+6.5\%$ \\
                              & community & 1.795 & 1.313 & 1.256 & \textbf{1.240} & 1.416 & $+5.6\%$ \\
\midrule
\multirow{2}{*}{web-google}   & input & 6.165 & 4.957 & 4.802 & \textbf{4.800} & 4.905 & $+3.2\%$ \\
                              & community & 3.973 & 3.305 & 3.077 & \textbf{3.055} & 3.193 & $+7.6\%$ \\
\bottomrule
\end{tabular}
\end{table}

\begin{table}[p]
\caption{Historical random-access sizes in bits per edge. BV-w7 uses reference window seven and maximum chain depth three. ZK-def.-RA is Zuckerli random-access mode with default parameters (one round, $(4,2,1)$). BG and CS include their sampled index; CG includes its distinct group-based access layout; BVGraph offsets are excluded. Bold and the last column have the same meaning as in Table~\ref{tab:sota}. $^\ddagger$The enwiki measurements use the older decreasing-group-size composition rule.}
\label{tab:ra-sota}
\centering
\small
\begin{tabular}{ll rr rrr r}
\toprule
Dataset & Ord. & \BV{}-w7 & ZK-def.-RA & \BG{} & \CS{} & \CG{} & vs default \\
\midrule
\multirow{2}{*}{amazon-0601}  & input & 13.001 & 10.514 & \textbf{10.389} & 10.453 & 19.710 & $+1.2\%$ \\
                              & community & 8.105 & 6.873 & \textbf{6.670} & 6.671 & 8.131 & $+3.0\%$ \\
\midrule
\multirow{2}{*}{arxiv-hep-ph} & input & 10.262 & 9.130 & 9.149 & \textbf{9.121} & 14.535 & $+0.1\%$ \\
                              & community & 7.908 & 7.151 & 6.921 & \textbf{6.894} & 8.282 & $+3.6\%$ \\
\midrule
\multirow{2}{*}{cnr-2000}     & input & 2.897 & 2.291 & 1.747 & \textbf{1.713} & 4.295 & $+25.2\%$ \\
                              & community & 2.981 & 2.325 & 1.773 & \textbf{1.742} & 2.426 & $+25.1\%$ \\
\midrule
\multirow{2}{*}{EAT}          & input & 10.754 & 9.322 & 9.130 & \textbf{9.071} & 12.592 & $+2.7\%$ \\
                              & community & 9.722 & 8.726 & \textbf{8.489} & 8.505 & 10.405 & $+2.7\%$ \\
\midrule
\multirow{2}{*}{enwiki-2013\rlap{$^\ddagger$}}  & input & 16.195 & 13.722 & \textbf{13.001} & 13.017 & 25.799 & $+5.3\%$ \\
                              & community & 12.867 & 11.304 & \textbf{10.505} & 10.672 & 13.775 & $+7.1\%$ \\
\midrule
\multirow{2}{*}{in-2004}      & input & 2.172 & 1.777 & 1.282 & \textbf{1.262} & 3.145 & $+29.0\%$ \\
                              & community & 2.212 & 1.775 & 1.268 & \textbf{1.251} & 1.941 & $+29.5\%$ \\
\midrule
\multirow{2}{*}{web-google}   & input & 6.717 & 5.526 & \textbf{5.008} & 5.010 & 11.188 & $+9.4\%$ \\
                              & community & 4.558 & 3.794 & 3.293 & \textbf{3.273} & 4.031 & $+13.7\%$ \\
\bottomrule
\end{tabular}
\end{table}

\section{Earlier order construction and surrogate selection}
\label{app:legacy-ordering}
This appendix describes the older guard and $C_{64}$ study, not the versioned campaign portfolio. In particular, its input-order candidate and no-merge restrictions do not describe pipeline v3.

\subsection{The candidate constructions}
Layered Label Propagation (LLP) groups vertices through repeated local label updates at several resolutions~\cite{boldi2011llp}. We use the symmetrized graph for these updates, while preserving the directed graph for encoding. The tested configurations use five passes. The input order itself is also a candidate.

The community variant first partitions the graph, then runs LLP within the resulting groups. The implementation uses Louvain-style local moves, connected-component refinement, and aggregation~\cite{blondel2008louvain}. It is not the full Leiden algorithm~\cite{traag2019leiden}. We therefore call it \emph{community+LLP}. The historical library name contains ``Leiden''; it does not justify transferring Leiden's guarantees to this implementation.

The measured pipeline does not seed the partition with a global LLP pass. That optional three-stage construction was described in an earlier version, but it did not produce the reported measurements. An optional merge of small groups is used only in the development configurations that declare it. The held-out and prospective configurations do not use it.

We concatenate groups by their first appearance in the current order. Each group preserves its internal order. This \emph{stable grouping} avoids an arbitrary size-based rearrangement, but does not guarantee a smaller compressed file. Even with stability, distances between groups can grow.

Recursive Graph Bisection (RGB) recursively partitions a graph to improve a logarithmic gap objective~\cite{dhulipala2016rgb}. The empirical comparison uses the Petri--Mackenzie implementation with the recorded serial compatibility changes, not the unrelated routine with the same name in Adjacently. Its revision and settings are given in Appendix~\ref{app:protocol}.

\subsection{Rejecting costly grouping steps}
A \emph{guard} is an acceptance check: keep an ordering step only if its declared cost does not increase. The direct-cost guard evaluates a BVGraph-style representation with references disabled. It can reject a whole grouping step or propose a subset of groups, but acceptance always uses the total cost.

The community variant is treated as one sequence of steps, including initial grouping and subsequent LLP refinements. Guarding can leave a community noncontiguous, so the output need not be a concatenation of intact communities.

A second guard uses an exact cost evaluator for BVGraph with window seven and reference depth three. Its acceptance target includes the graph and offset streams. This guard is exact for that declared configuration. It is not exact for our encoders, and its offset-inclusive acceptance target differs from a graph-stream-only BVGraph size column.

Neither guard proves improvement under another encoder. We keep guarded outputs as candidates and evaluate the final choice separately.

\subsection{Selecting after construction}
The score $C_{64}$ estimates the cost of an already constructed order. For each row it compares direct coding with copying from any of the previous 64 rows. A copied row pays a reference distance, a bitmap marking copied entries, and the cost of the residual list. The evaluator includes modeled row offsets. It scans the whole graph; it does not sample rows or emit the target encoder's file.

A bitmap uses one bit per entry of the reference list. Real encoders may compress long copied runs much more cheaply. They may also use different integer codes and adaptive probability models. Thus $C_{64}$ is a surrogate, not an exact target cost. Its detailed definition is in Appendix~\ref{app:score}.

The prospective selector uses five candidates: the input order, plain LLP and community+LLP under the direct-cost guard, and both constructions under the exact BVGraph guard. It selects the smallest $C_{64}$ score. Ties follow that listed order. The expanded comparison adds RGB and the two unguarded constructions.

The implementation's smaller three-candidate workflow uses only the input and the two directly guarded constructions. We distinguish its timing results from the five-candidate prospective test. Neither workflow is a predictor that avoids constructing candidate orders.

\begin{algorithm}[t]
\caption{Select an order with a copying-and-gap score}
\label{alg:select}
\begin{algorithmic}[1]
\REQUIRE Graph $G$; ordered candidate constructors $A_1,\ldots,A_k$, including the input order
\STATE $q^*\leftarrow+\infty$
\FOR{$i=1,\ldots,k$}
\STATE $\pi\leftarrow A_i(G)$
\STATE $q\leftarrow C_{64}(G,\pi)$
\IF{$q<q^*$}
\STATE $(q^*,\pi^*)\leftarrow(q,\pi)$
\ENDIF
\ENDFOR
\RETURN $\pi^*$
\end{algorithmic}
\end{algorithm}
Algorithm~\ref{alg:select} makes the tie rule and evaluation work explicit. Every candidate is constructed and scored; only the chosen order is finally encoded. If constructing candidate $i$ takes time $a_i$, score evaluation adds $O(k(n+Wm+\sum_v d_v\log(1+d_v)))$ for window $W$ and degrees $d_v$. The sorting term accounts for sorting targets after relabeling. Reference comparisons scan both lists. Total time also includes $\sum_i a_i$ and the final encoding; no near-linear bound on the guarded constructors is assumed.

\section{Earlier matched-codec and ordering experiments}
\label{sec:legacy-experiments}
This appendix retains the pre-campaign experiments, their configurations, and their failed predictions. Its graph preprocessing, encoder settings and candidate sets differ from the current campaign; its claims apply only to these earlier protocols. We use two complementary comparisons. The matched codec benchmark holds the input and vertex order fixed. The ordering study varies orders under common encoder settings.

The matched benchmark uses five inputs in their original order: EAT, HEP-PH, CNR-2000, Web-Google, and in-2004. EAT and Web-Google use their largest strongly connected components, not the full source datasets: 7,754 and 434,818 vertices, respectively. The largest matched input has 16.9 million directed edges. We measure encoding, full decoding, and neighbor-query sessions. Each timing uses five independent processes, run in randomized order on one logical CPU of an AMD EPYC 9655 server. Each process reports the median of three repetitions after warmup. We report the median and range across processes. The container shares its physical host; activity screening does not guarantee exclusive hardware access.

We call the baseline native configurations \emph{shipped}; \emph{fixed CS} uses the same CS settings on every input, without selecting an encoder or variant after measurement. \emph{Tuned Zuckerli} means its published settings (two reference-selection rounds and hybrid-integer parameters $(4,1,0)$), not a new per-input search. Archive mode targets sequential storage; access mode supports neighbor queries. BV-HC is BVGraph with window 16 and effectively unrestricted reference depth. Appendix~\ref{app:matched} gives the full settings.

The query study varies trace length and decoded-cache capacity. It includes random queries and queries that follow a breadth-first traversal, which tend to revisit nearby vertices. A fresh reader serves each trace. Session time includes opening; request latency is measured separately. Appendix~\ref{app:matched} gives the source snapshots, settings, checks, and timing protocol.

For ordering, six development graphs determined settings and six more tested hypotheses. A final prospective test used eight graphs: wiki-Vote, email-EuAll, soc-Epinions1, soc-Slashdot0902, web-NotreDame, web-Stanford, web-BerkStan, and eswiki-2013. Predictions and selection rules were frozen before measuring them. Eswiki had appeared in unrelated work, but not in selector development.

Each prospective graph uses five encoders, giving forty graph--encoder measurements. Means weight them equally. They are not forty independent graphs. The fresh protocol uses one seed and no small-group merge. A separate five-seed check covers six earlier graphs. Appendix~\ref{app:protocol} lists all inputs and settings, including extracted cores that differ from the full published datasets.

The complete-workflow study includes candidate construction, scoring, relabeling, and file writing. It excludes loading and compilation warmup. Some runs shared the machine with other work, so only large timing differences are informative. These timings are separate from the matched codec benchmark.

\section{Results under the earlier protocols}
\label{sec:legacy-results}
\subsection{File size and sequential processing}
Fixed CS produces a smaller sequential file than every tested Zuckerli configuration on all five matched inputs. Table~\ref{tab:final-space} pairs each saving with the encoding cost against that same baseline file. The smallest Zuckerli file uses random-access mode on EAT and HEP-PH, and archive mode on the other inputs. These modes have different access capabilities. The archive-only column gives the comparison with sequential use held fixed.

\begin{table}[t]\centering\small
\caption{Sequential CS versus public Zuckerli at its published settings. ``Smallest mode'' is chosen by raw bytes among both modes and both tested parameter settings; the published settings win on each input. Savings are relative to Zuckerli. The time ratio uses the same smallest-file mode, not the fastest one.}
\label{tab:final-space}
\begin{tabular}{lrlrrr}\toprule
& CS size & Smallest & \multicolumn{2}{c}{CS saving (\%)} & Encode time\\
Graph & (bpe) & ZK mode & vs smallest & vs archive & CS / ZK\\\midrule
EAT & 9.056 & Access & 1.86 & 3.79 & 9.09 \\
HEP-PH & 8.918 & Access & 1.83 & 3.12 & 2.75 \\
CNR-2000 & 1.695 & Archive & 8.33 & 8.33 & 31.11 \\
Web-Google & 4.800 & Archive & 2.53 & 2.53 & 4.85 \\
in-2004 & 1.239 & Archive & 5.34 & 5.34 & 37.60 \\
 
\bottomrule\end{tabular}\end{table}

\begin{figure}[t]
\centering\includegraphics[width=\linewidth]{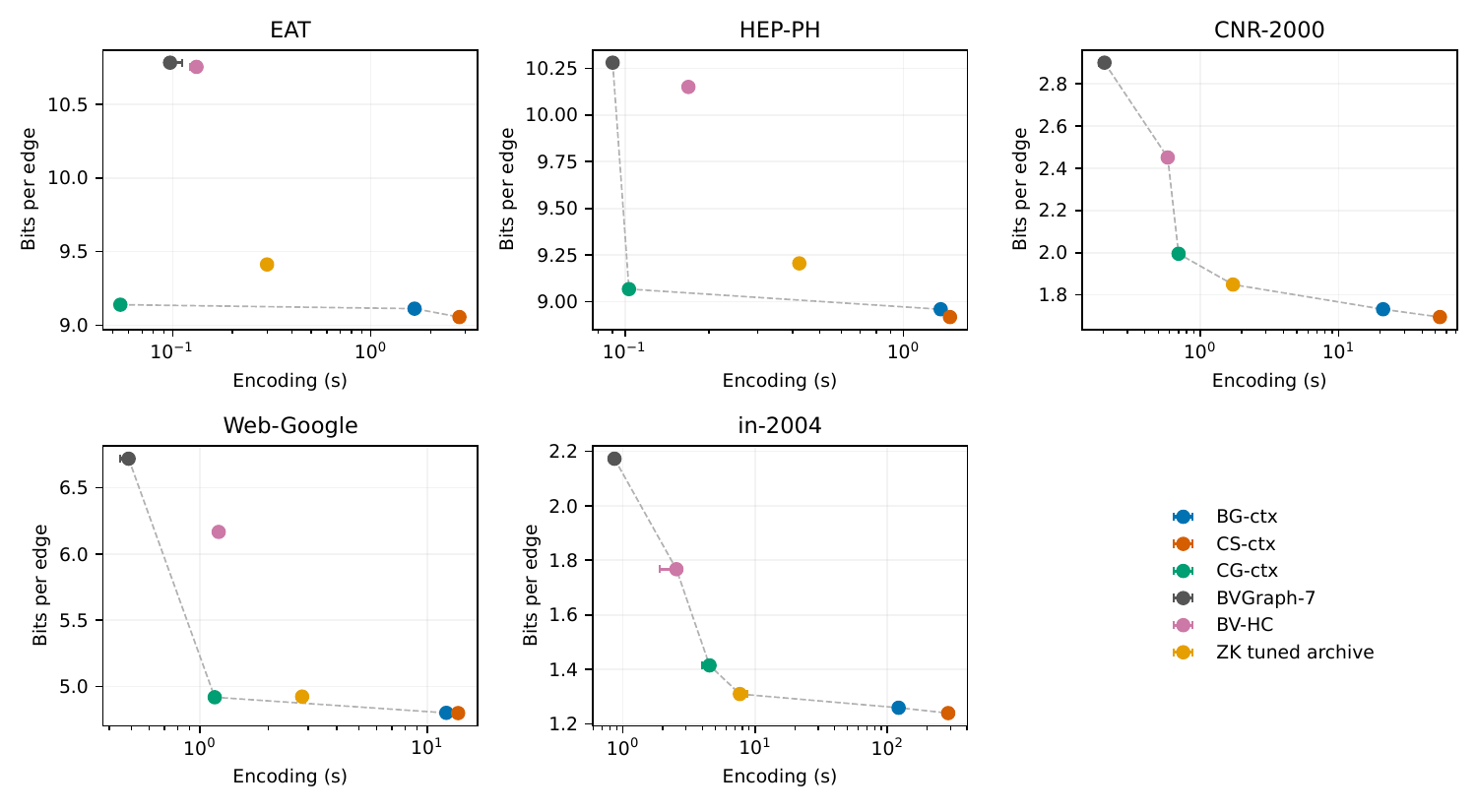}
\Description{Five panels show sequential bits per edge against encoding time. CS favors file size, CG favors native encoding speed, and BVGraph and Zuckerli provide faster external operating points.}
\caption{Sequential size and encoding time in input order. Time axes are logarithmic; bars show the range across five process medians. Dashed lines join measured Pareto points: no plotted choice is both smaller and faster. This is not a test of statistical dominance. BVGraph sizes exclude offsets; Zuckerli uses tuned archive mode.}
\label{fig:matched-tradeoff}
\end{figure}

The space saving comes at a substantial encoding cost: fixed CS takes about $16$--$334\times$ as long as BVGraph with window seven and depth three on the same inputs (Table~\ref{tab:matched-codecs}). BG and CS decode more slowly than tuned Zuckerli throughout. Against BVGraph, their EAT decode times overlap its process range; the other inputs are slower (Table~\ref{tab:matched-codecs}). CG offers a different trade-off: it encodes faster than tuned Zuckerli archive on all five inputs, but makes larger files on CNR and in-2004. Figure~\ref{fig:matched-tradeoff} places these choices on the same size--time axes.

\begin{table}[t]\centering\small
\caption{Matched times, medians across five independent processes. Encoding includes writing but excludes input graph construction. Full decoding reads the file and materializes all neighbor lists. BVGraph uses window seven and depth three; Zuckerli uses tuned archive mode on every row. Warmup and correctness checks are outside the timer. Appendix~\ref{app:matched} gives process ranges and memory.}
\label{tab:matched-codecs}
\begin{tabular}{lrrrrr}\toprule
Graph & BG-ctx & CS-ctx & CG-ctx & BVGraph & Zuckerli\\\midrule
\multicolumn{6}{l}{Encoding (seconds)}\\
EAT & 1.662 & 2.801 & 0.054 & 0.097 & 0.299 \\
HEP-PH & 1.362 & 1.470 & 0.103 & 0.090 & 0.423 \\
CNR-2000 & 20.940 & 53.819 & 0.700 & 0.205 & 1.730 \\
Web-Google & 12.139 & 13.680 & 1.163 & 0.486 & 2.820 \\
in-2004 & 121.518 & 288.050 & 4.517 & 0.863 & 7.660 \\
 
\midrule
\multicolumn{6}{l}{Full decoding (milliseconds)}\\
EAT & 13.562 & 13.701 & 10.619 & 15.334 & 5.336 \\
HEP-PH & 28.704 & 28.117 & 20.113 & 19.045 & 11.166 \\
CNR-2000 & 126.295 & 127.598 & 75.436 & 33.548 & 39.811 \\
Web-Google & 261.343 & 238.490 & 126.643 & 59.088 & 71.278 \\
in-2004 & 987.536 & 1012.742 & 453.112 & 143.179 & 163.744 \\
 
\bottomrule\end{tabular}\end{table}

Peak encoding-process memory for CS is about $1.0$--$2.3$\,GiB, versus $15$--$341$\,MiB for tuned Zuckerli archive (Table~\ref{tab:matched-all}). These peaks include runtime, input graphs, and validation, so they are not measurements of encoder workspace alone.

The larger in-2004 input preserves the main trade-off: CS saves space but takes much longer to encode. Its indexed file is also smaller than tuned Zuckerli's access file, by about $24\%$. On HEP-PH, however, indexed CS is larger than tuned Zuckerli access ($9.102$ versus $9.084$ bpe). Table~\ref{tab:matched-all} reports indexed BG/CS sizes for all five inputs. These size comparisons do not establish faster access.

The public Zuckerli code with published settings gives $1.849$ bpe on CNR, against the published $1.84$. The cause is unidentified. We report this reproduction discrepancy and use the measured baseline for comparisons. The CNR access result rounds to the published $2.19$; both in-2004 modes also match their published precision. Appendix~\ref{app:matched} gives the details.

\subsection{Random access depends on the workload}
\label{sec:access}
BG and CS decode chunks of up to 2,048 rows. A small cache can therefore cause repeated decoding. Table~\ref{tab:matched-query} shows the cost on random traces. An unlimited decoded cache helps, but it is a different memory choice, not free random access.

\begin{table}[t]\centering\small
\caption{Opening plus 2,000 random queries, in milliseconds. Each repetition starts with a fresh reader. The CS limited-cache column uses the stated budget; the next uses unlimited whole-chunk caching. BVGraph-7 and tuned Zuckerli access use their own readers. All entries are five-process medians.}
\label{tab:matched-query}
\begin{tabular}{llrrrr}\toprule
& Limited & \multicolumn{2}{c}{CS indexed} & & \\
Graph & cache & Limited & Unlimited & BVGraph & Zuckerli\\\midrule
EAT & 64 KiB & 18,257.3 & 15.6 & 8.2 & 7.8 \\
HEP-PH & 64 KiB & 20,770.0 & 32.6 & 6.6 & 10.3 \\
CNR-2000 & 64 KiB & 15,671.3 & 166.1 & 13.5 & 24.3 \\
Web-Google & 64 KiB & 52,276.3 & 316.6 & 17.6 & 40.7 \\
in-2004 & None & 17,172.8 & 1,889.1 & 37.1 & 103.6 \\
 
\bottomrule\end{tabular}\end{table}

The four-graph study also measures request latency without opening. Only two of 160 low-cache comparisons meet the exploratory condition of staying within $100\times$ Zuckerli's median latency. Both use CNR and a breadth-first trace of 100 queries. The median ratio over all comparisons is about $4{,}810\times$. The broad condition fails, as the protocol anticipated. These ratios are not whole-session ratios; per-request clock overhead also affects very fast responses.

\begin{figure}[t]\centering
\includegraphics[width=\linewidth]{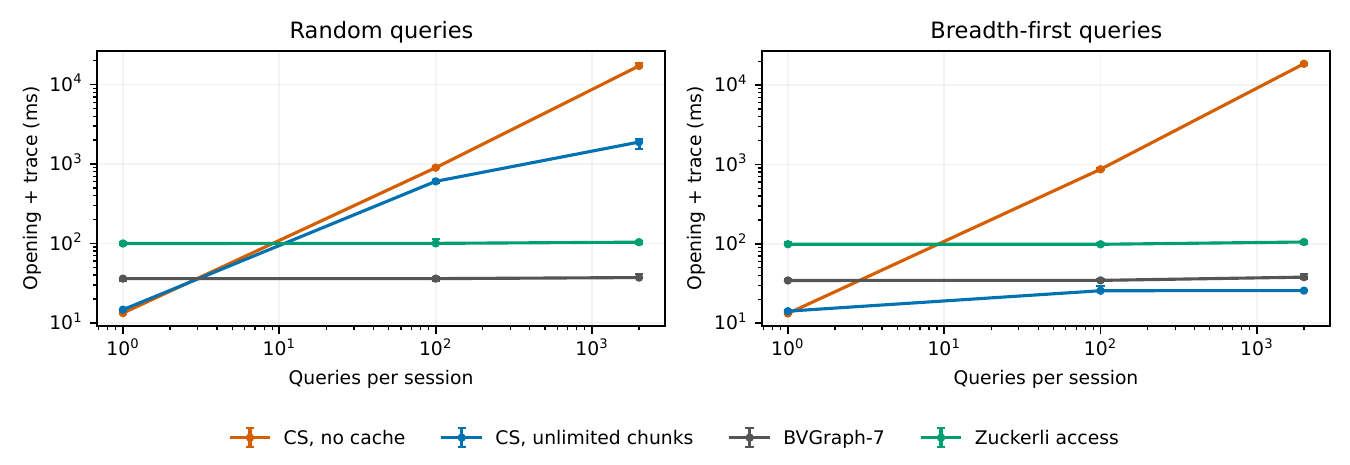}
\Description{Two panels compare fresh-reader query sessions on in-2004. Unlimited caching benefits CS on the breadth-first trace, while its random sessions remain slower for longer traces.}
\caption{Session cost on in-2004, including opening, for both measured trace types. Points are medians and bars are ranges across five processes. Both axes are logarithmic. Unlimited caching can favor CS for this breadth-first trace; it does not make random access uniformly competitive.}
\label{fig:large-access}\end{figure}

Locality can change the ranking. On in-2004, CS with unlimited caching completes the 2,000-query breadth-first trace faster than both external readers. It loses on the random trace of the same length (Figure~\ref{fig:large-access}). Opening matters too: Zuckerli builds its in-memory offset index at that stage, while BG and CS fill their decoded cache during queries. We count both costs in session totals.

The cache budget counts retained decoded chunks, not total process memory. Oversized chunks bypass the cache. A fresh reader is not a cold filesystem. The full cache grid and process ranges are in Appendix~\ref{app:access-protocol}; none of these measurements supports a general low-latency access claim.

\subsection{What produces the size saving}
\label{sec:ablation}
We change one feature at a time and also build up from a plain baseline. A feature counts as active only if the encoder actually uses it. Every file is decoded and checked. Table~\ref{tab:ablation-changes} reports the principal size contrasts; Appendix~\ref{app:matched-ablation} gives their settings and validation.

Context coding is decisive in this comparison. Switching CS to Fibonacci coding preserves every row choice but loses to Zuckerli on all four graphs. Matching Zuckerli's window leaves a much smaller advantage (Table~\ref{tab:venue-ablation}). Thus the shipped comparison combines a different coding path with a longer reference window.

\begin{table}[t]\centering\small
\caption{CS against the smallest of four Zuckerli files. Positive savings favor CS; the last column instead reports excess size. The HEP-PH window-32 difference is only 22 bytes, treated as a tie.}
\label{tab:venue-ablation}
\begin{tabular}{lrrr}\toprule
& \multicolumn{2}{c}{CS saving (\%)} & CS-Fibonacci\\
Graph & Shipped window & Window 32 & excess (\%)\\\midrule
EAT & 1.86 & 0.51 & 18.84\\
HEP-PH & 1.83 & 0.00 & 11.19\\
CNR-2000 & 8.33 & 2.55 & 9.42\\
Web-Google & 2.53 & 2.09 & 16.88\\
 
\bottomrule\end{tabular}\end{table}

A second experiment separates three parts of the coding path while keeping the row choices fixed. Adaptive streams account for most of the saving. Copy-bitmap coding also matters, especially on CNR. Changing the gaps between leftover neighbors contributes little. Appendix~\ref{app:coding-split} gives the full combination experiment and its accounting rule.

More search does not consistently help. Restricting CS to copying and direct gap coding makes three files smaller; CNR benefits from the full action set. A plausible reason is the cost mismatch: search estimates Fibonacci costs while the output uses adaptive context coding. The experiment identifies the mismatch, but does not test a replacement cost model. The matched timing runs show that this copy-plus-residual variant is $2.0$--$3.9\times$ faster to encode. The fast-search setting is both smaller and faster than shipped CS on EAT and HEP-PH. It is faster but larger on Web-Google and CNR, with about an $11\%$ size increase on CNR. Table~\ref{tab:ablation-times} reports these contrasts against shipped CS.

\subsection{The community advantage is conditional}

\begin{figure}[t]
\centering\includegraphics[width=.92\linewidth]{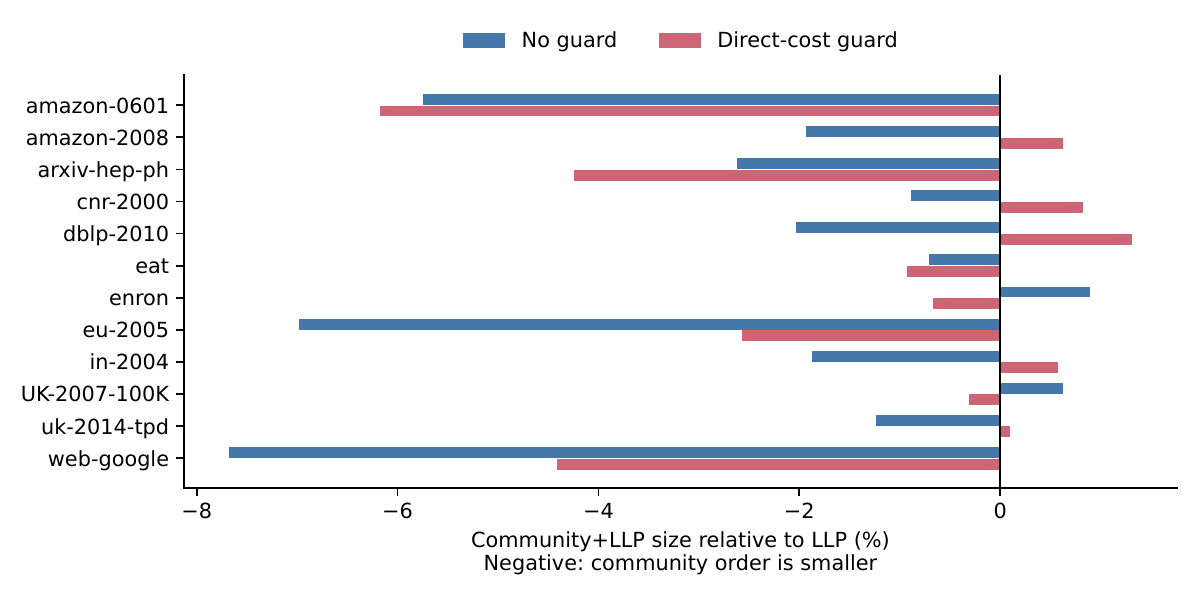}
\Description{Paired horizontal bars show the community-order size change relative to plain LLP, with and without the direct-cost guard, on twelve graphs. Several rankings reverse.}
\caption{The guard changes the ordering comparison. Negative values favor community+LLP. These are earlier graphs at seed zero, not the prospective test.}
\label{fig:guard}
\end{figure}

Without a guard, community+LLP beats plain LLP on ten of the twelve earlier graphs under BG-ctx. A direct-cost guard changes that comparison because it helps the two constructions by different amounts. Figure~\ref{fig:guard} shows both comparisons under the same target encoder.

On the prospective graphs, directly guarded community+LLP beats directly guarded LLP on only two of eight graphs. The largest losses are on web graphs. Within this earlier guarded candidate set, community+LLP is not a reliable default. This comparison does not test the merged spectral community orders in the later campaign.

The comparison is also sensitive to settings. Small-group merging was tuned on some development graphs but was absent from the fresh-graph protocol. Its contribution to the fresh losses in this earlier experiment is untested; the later campaign tests merged orders under a different protocol. The five-seed check also changes the interpretation of Enron: its sign varies, so it should be treated as a tie. Most of the other tested signs are stable, but their magnitude still varies.

The component measurements identify a trade-off, not a universal explanation (Table~\ref{tab:components}). Community grouping can shorten residual gaps while making reference descriptions or inter-group gaps more costly. Counting all graph gaps is insufficient because copied edges do not pay those residual costs. We do not attribute these reversals to one structural graph statistic.

\subsection{Structure and compression across orders}
\label{sec:structural}
We measure eight orders on each of twenty graphs. A \emph{consecutive-neighbor gap} is the difference between successive sorted targets in one row. The \emph{copied-edge share} is the fraction of edges actually recovered from references. Counting these edges leaves the encoded bytes unchanged.

Smaller logarithmic gaps and more copying accompany larger size gains in the pooled data. However, ID-block locality and source--target distance have larger correlation magnitudes than gaps and copied-edge share. This contradicts our expectation that the simpler measures would be weaker. Their uncertainty intervals overlap, so the prespecified comparison does not rank one measure above another. Appendix~\ref{app:structural} reports all seven measures.

The within-graph result is more informative for choosing an order. Gap reductions have the expected sign on at least seventeen graphs for every encoder. Copied-edge share is much weaker for CG within a graph: its median correlation with size gain is $0.31$ across all eight orders, or $0.14$ across the seven constructed candidates alone, despite a pooled correlation of $0.74$. The pooled association should therefore not be read as an equally strong within-graph relationship. These results describe the measured orders; they are neither causal evidence nor a validated prediction rule for new graphs.

True host transitions could not be measured because URL and host lists were unavailable. ID blocks are a locality proxy, not recovered host names.

\subsection{Selection helps, but two predictions fail}
Within its five candidates, the prospective selector's worst regret is about $0.74\%$. This exceeds the declared $0.5\%$ target. Against all eight candidates, it misses larger gains from RGB. Table~\ref{tab:predictions} reports both predictions as failures.

A post hoc check explains the two failures. On email-EuAll, even the best shared choice among the five candidates has worst regret of about $0.74\%$ across encoders. No ranking rule returning one of those orders could meet the $0.5\%$ target. This concerns the measured candidates, not all possible orders. Against the wider eight-order set, missing candidates account for about $91\%$ of the selector's aggregate loss. Appendix~\ref{app:portfolio} gives the calculation. These explanations do not turn the frozen failures into passes.

\begin{table}[t]
\centering\small
\caption{Prospective predictions. Thresholds were fixed before the eight-graph test. Regret is excess size over the stated reference set.}
\label{tab:predictions}
\begin{tabular}{p{6.6cm}rl}\toprule
Prediction & Observed & Outcome\\\midrule
Worst five-order regret over its own set $\le0.5\%$ & $0.74\%$ & Failed\\
Worst five-order regret over all eight $\le1\%$ & $5.52\%$ & Failed\\
Mean regret below every fixed candidate & Yes & Passed\\
Winner varies; each construction wins $\ge2$ graphs per guard & Guard-dependent & Partial\\
Adopt eight orders only if mean improves and worst $\le1\%$ & Worst $1.47\%$ & Not adopted\\\bottomrule
\end{tabular}
\end{table}

\begin{figure}[t]
\centering\includegraphics[width=.9\linewidth]{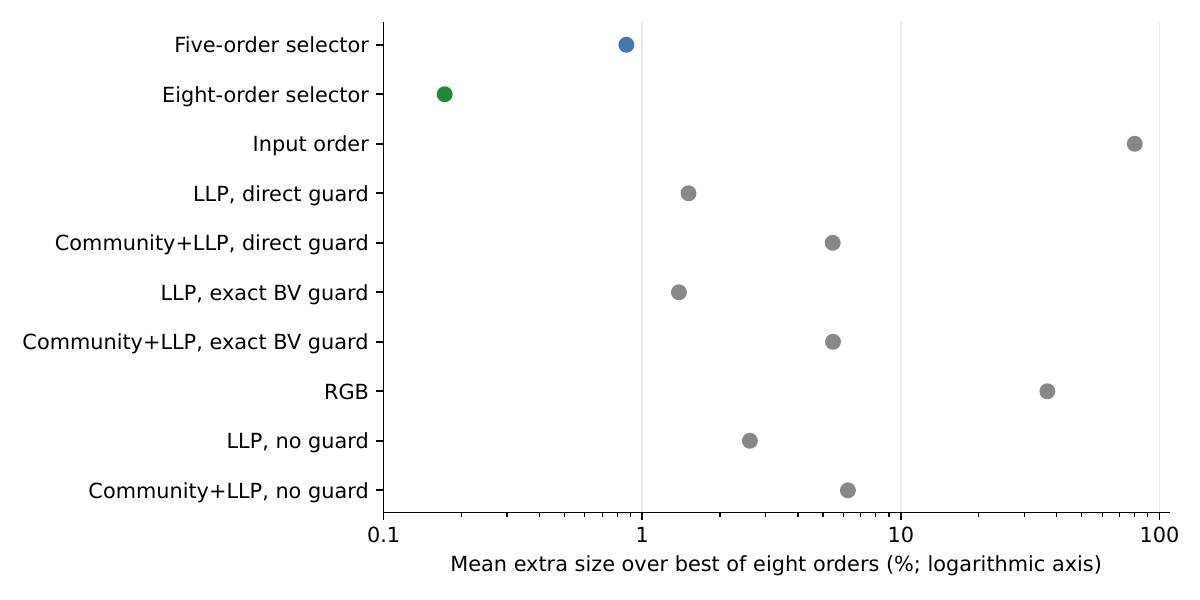}
\Description{Mean regret for both selectors and all eight fixed candidates on a logarithmic horizontal axis.}
\caption{Prospective mean regret for both selectors and all eight fixed candidates. The horizontal axis is logarithmic. The eight-order rule improves the mean but fails its prespecified worst-regret condition.}
\label{fig:regret}
\end{figure}

The selector nevertheless has lower mean regret than every fixed candidate on the fresh measurements. Figure~\ref{fig:regret} and Table~\ref{tab:regret} use the same eight-order reference set, so their comparisons have a common denominator.

\begin{table}[t]
\centering\small
\caption{Prospective regret against all eight candidates, in percent. The best fixed order is identified after measurement. Its selection is not an extra fitted predictor.}
\label{tab:regret}
\begin{tabular}{lrr}\toprule
Rule or fixed order & Mean & Worst\\\midrule
Five-order selector & 0.87 & 5.52 \\
Eight-order selector & 0.17 & 1.47 \\
Input order & 80.24 & 478.86 \\
LLP, direct guard & 1.51 & 5.52 \\
Community+LLP, direct guard & 5.45 & 23.21 \\
LLP, exact BV guard & 1.39 & 5.59 \\
Community+LLP, exact BV guard & 5.47 & 23.60 \\
RGB & 36.85 & 114.83 \\
LLP, no guard & 2.61 & 7.05 \\
Community+LLP, no guard & 6.24 & 24.96 \\
 
\bottomrule\end{tabular}
\end{table}

RGB is the smallest order in eleven of the forty graph--encoder measurements. It wins every encoder on wiki-Vote. The expanded selector chooses RGB on eswiki-2013. That choice is best for BG-Fibonacci and BVGraph, but loses under BG-ctx, CS-ctx, and CG-ctx. Table~\ref{tab:winners} shows the winners separately for all five encoders. The frozen rule therefore retains the five-candidate selector.

The prediction that both constructions would win needs a qualification. Under the direct guard, both constructions win at least twice. Under the exact BVGraph guard, community+LLP wins only once. Thus the prediction holds under the direct guard but fails under the exact guard; we report it as partial.

\subsection{Selection has a measurable cost}
The smaller three-order workflow matches the smallest file obtained by encoding all three candidates on each of the four timing graphs. It costs between $1.16$ and $1.61$ times the fixed community workflow (Table~\ref{tab:workflow}). Encoding all three costs more. These are measured outcomes, not a guarantee that the score always selects the smallest file.

\begin{table}[t]
\centering\small
\caption{Complete workflow costs under BG-ctx. Fixed is the directly guarded community order. Other columns are multiples of its time. Loading and compilation warmup are excluded. Concurrent machine use limits timing precision.}
\label{tab:workflow}
\begin{tabular}{lrrrrr}\toprule
Graph & Fixed (s) & Select 3 & Encode 3 & Select 8 & Encode 8\\\midrule
cnr-2000 & 150 & 1.61 & 2.47 & 3.59 & 7.13 \\
eat & 13 & 1.24 & 2.69 & 1.75 & 7.40 \\
uk-2014-tpd & 3012 & 1.16 & 2.24 & 3.10 & 6.20 \\
web-google & 174 & 1.44 & 2.36 & 3.96 & 8.07 \\
 
\bottomrule\end{tabular}
\end{table}

Selecting three orders also produces smaller files than the fixed choice on CNR-2000 and UK-2014-tpd. It ties the fixed choice on the other two graphs. Table~\ref{tab:workflow-outcomes} reports the sizes and peak process memory.

The eight-order workflow costs more and does not improve on the three-order choice in these cases. It must construct RGB and the exactly guarded candidates. On EAT it selects a slightly larger file. The comparison therefore includes both the cost and the error of selection.

These savings are relative to encoding every candidate. Selection remains slower than committing to one fixed order.

The earlier ordering study remains separate. Its absolute decode and query times, measured under a different protocol, are retained in Appendix~\ref{app:older-timing}. The older external size tables remain in Appendix~\ref{app:legacy}. Their best-of-three column is separate from the fixed-CS result above.

\section{Earlier KEEP/REORDER evidence and its provenance limits}
\label{sec:decision-results}

\paragraph{Evidence status.}
This is the retained earlier decision audit, with FULL/FAST referring to its historical encoder search settings, not pipeline versions v1--v3. The reproduction and workflow measurements below predate the present campaign rewrite; they were not rerun for it.
This section audits saved experimental records; it does not yet present a
fully provenance-validated benchmark. The frozen source hashes match the
archived design, but the historical evaluator did not retain complete
graph/order/encoder hashes and checked only the decoded arc count.
Its CSV loader call did not explicitly preserve publisher IDs. A filename
ending in \texttt{-nat} is therefore not sufficient to establish publisher
order. We retain the saved results below with this qualification, rather
than treating an arithmetic audit as experimental reproduction. A later
inspection of the remote machine recovered a second distinction: its four label encodes
used FAST, not the frozen FULL cost model. We report those targets separately.

\subsection{Evaluation units and controls}

The development report's 34 correct decisions out of 34 are not held-out
accuracy. The later main TSV contains 17 graph/order records, while its
saved textual summary covers only 13. One added record, in-2004, belongs
to a development graph family and is excluded from the following counts.
Four separate remote-run records concern Indochina-2004 and UK-2002, each in two
orders. The main records use FULL; the remote-run records use FAST. Together they
describe seven named graphs, but they are neither one encoder benchmark nor
independent trials: orders of one
graph are paired, UK subsets share a crawl, and later evaluation-set
amendments explicitly sought KEEP examples.

For the 16 non-development FULL records, the rule makes 15 correct decisions,
with mean regret 0.000118125~\bpe{} and worst regret 0.00189~\bpe{}.
Every true label is REORDER: always reordering has zero regret, whereas
the surrogate makes one error. KEEP recall is undefined on this cohort.

The four later FAST records have two KEEP and two REORDER labels, all
matched by the frozen rule. Its regret is zero on this cohort; always
reordering has mean regret 0.043445~\bpe{}. These provide evidence for
selective retention under FAST, not confirmation of KEEP under FULL.
The initial FULL remote attempts were stopped before recording results; no
completed FULL confirmation was recovered. We do not pool the two label
targets or infer that FAST and FULL must make the same ordering decision.

\subsection{KEEP cases and the decision boundary}

Table~\ref{tab:decision-audit} exposes both the near-tie error and the four
later remote-run records. Under FAST, in the saved native cases, replacing the order increases
file size by 0.10951~\bpe{} on Indochina and 0.06427~\bpe{} on UK-2002.
The frozen rule retains both. After randomization it selects the candidate
for both graphs. These are promising paired observations, conditional on
the provenance limitations above; they do not establish a general rule
that crawl orders should always be retained.

\begin{table}[t]
\centering
\small
\caption{Historical boundary cases and later remote results, generated from
the saved TSVs. Gains and regret are in \bpe{}; positive gain favors
reordering. ``Saved native'' repeats the record's label, not a verified
publisher-ID claim. FULL and FAST are different label targets, not a pooled
benchmark. The cohort denominators are discussed in the text.}
\label{tab:decision-audit}
\begin{tabular}{lllrrrl}
\toprule
Graph & Delivery & Target & $\widehat\Delta$ & $\Delta_E$ & Regret & Decision \\
\midrule
UK-2007 100K & pipeline probe & FULL & -0.00571 & +0.00189 & 0.00189 & KEEP (error) \\
indochina-2004 & saved native & FAST & -0.03155 & -0.10951 & 0.00000 & KEEP \\
indochina-2004 & random & FAST & +19.43942 & +20.30834 & 0.00000 & REORDER \\
uk-2002 & saved native & FAST & -0.03490 & -0.06427 & 0.00000 & KEEP \\
uk-2002 & random & FAST & +24.80474 & +25.59935 & 0.00000 & REORDER \\
\bottomrule
\end{tabular}

\end{table}

\begin{figure}[t]
\centering
\includegraphics[width=\linewidth]{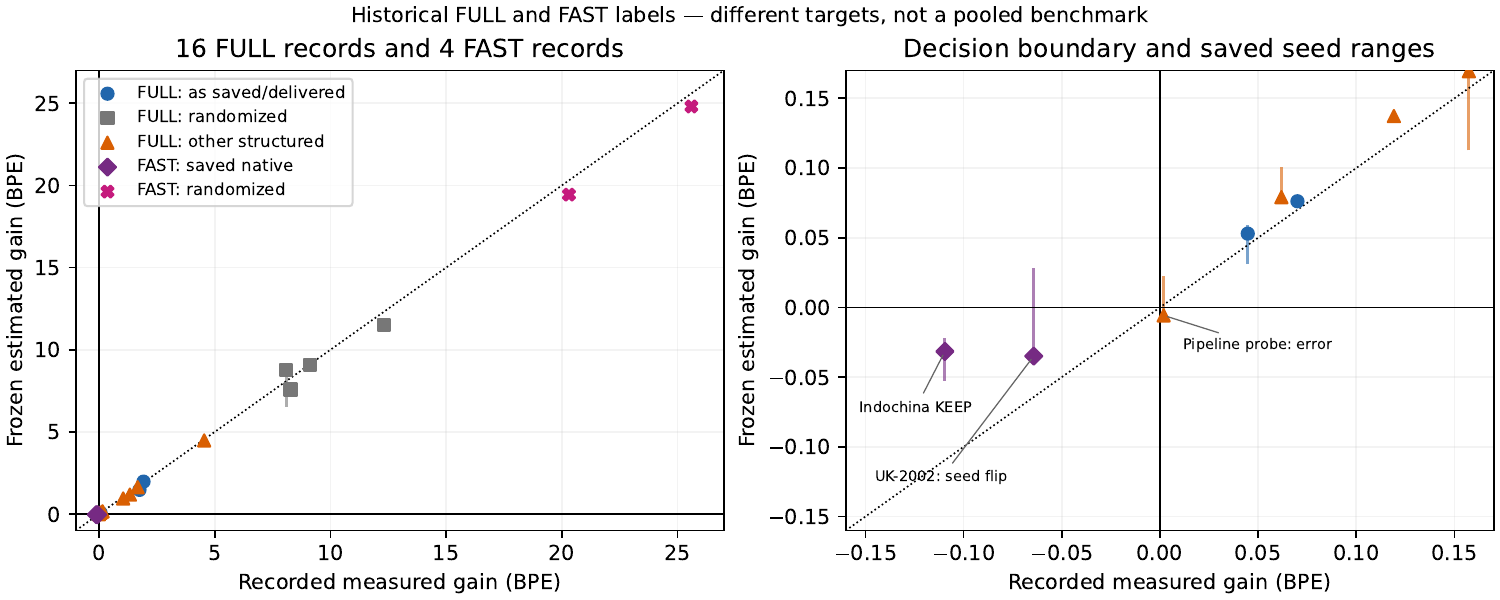}
\Description{Two scatter plots compare recorded and estimated gain. The full
range is dominated by randomized orders; a boundary zoom shows two saved KEEP
cases and the near-zero pipeline error. UK-2002's sampling range crosses zero.}
\caption{Recorded measured versus frozen estimated gain, distinguishing
FULL and FAST label encoders. The right panel
shows the decision boundary. Vertical segments are saved ranges over
sampling seeds where available, not confidence intervals; the UK-2007 1M
records do not include the extra-seed study. Large randomized gains can
dominate an overall correlation while obscuring near-zero errors.
Historical graph/order provenance and full roundtrip checks remain pending.}
\label{fig:decision-calibration}
\end{figure}

The pipeline-ordered UK-2007 100K probe has estimated gain $-0.00571$ but
measured gain $+0.00189$~\bpe{}. Retaining its order is a small but real
decision error. Applying an orderer twice is not guaranteed to be
idempotent; this probe must not be labeled a KEEP case by construction.
Its saved sampling range also crosses zero.

For UK-2002 in the saved native order, the estimated gain ranges from
$-0.03490$ to $+0.02838$~\bpe{} across the saved seeds. Thus a correct KEEP
decision at the frozen seed is not stable under resampling. Indochina's
corresponding range, $[-0.05269,-0.02188]$, stays negative. The archive
stores extrema rather than individual seed observations, so it does not
support a per-seed error rate. We leave the zero threshold unchanged.

\subsection{A retained reproducibility check}

We re-evaluated UK-2007 100K using the source snapshot recorded for that check and the frozen rule and
explicit first-encounter loading. The graph has 100{,}000 vertices and
3{,}050{,}615 arcs. The delivered and candidate files occupy 555{,}972 and
538{,}914 bytes, respectively: 1.457993 and 1.413260~\bpe{}.
The estimated gain is 0.052957~\bpe{} and the measured gain is
0.044733~\bpe{}, giving the same correct REORDER decision as the saved row.
Both sizes and gains match that record at its printed precision.

For each order, encodes with one and eight search workers are byte-identical.
All four files pass full successor-list equality, and the retained permutation
is checked against every original successor list. The validation record pins
the input, ordered adjacencies, permutation, source and dependency hashes.
This establishes a fresh reproduction for one first-encounter-order case;
it does not certify publisher-order provenance or validate the later KEEP
cases. The run contains only one timing observation per setting, with cold
candidate construction and some residual encoder compilation. We therefore
do not use it as a timing headline.

\subsection{Matched end-to-end decision time}

We measured the four complete workflows on this UK-100K input, using eight
search workers for every encode. Candidate construction and scoring remain
serial. Each arm receives a full warmup, followed by three paired blocks
with shuffled arm order. Measurements begin with the graph already loaded,
include candidate construction, scoring and file writes, and exclude
verification and hashing. GC is enabled. The machine is a Ryzen 7 PRO 4750U
running Julia 1.12.4, with eight application threads, one GC mark thread and
affinity to eight physical cores; no exclusive-machine or durable-write
guarantee is claimed.

\begin{figure}[t]
\centering
\includegraphics[width=\linewidth]{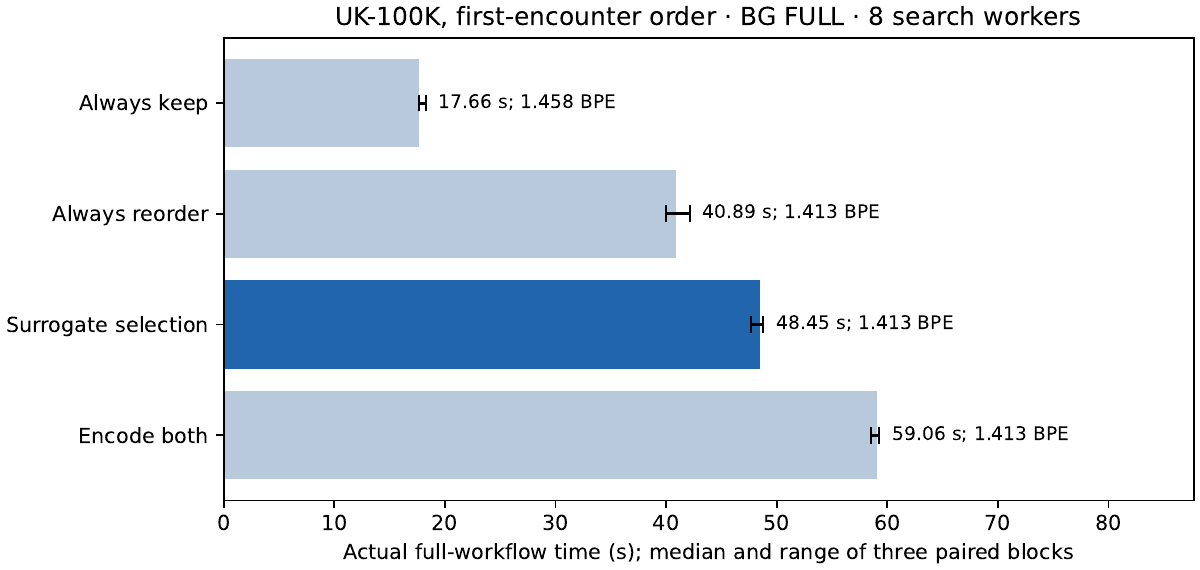}
\Description{Four workflow timings on UK-100K. Always keep is fastest but
produces a larger file. Always reorder, surrogate selection and encoding both
produce the same smaller file, with increasing time in that order.}
\caption{Actual full-workflow times on one first-encounter-order graph, with
BG FULL and eight search workers. Bars show medians and whiskers the observed
range across three paired blocks, not confidence intervals. These are not
sums of separately timed stages.}
\label{fig:decision-workflow}
\end{figure}

Surrogate selection takes a median 48.446 s, compared with 59.058 s for
encoding both; the median of the three paired speedup ratios is 1.219.
Always reordering takes 40.886 s and produces the same smaller file on this
REORDER case. Always keeping takes 17.659 s but costs an additional
0.044733~\bpe{}. All 20 warmup/measured files pass full adjacency checks
and pinned-byte checks. The 12 measured workflows report zero compilation.
This is evidence of reduced two-encode selection cost on one graph, not
a general speedup over always reordering or a substitute for KEEP validation.

\subsection{What remains before a headline result}

FULL confirmation and complete adjacency equality must be established for
the KEEP cases, with source and permutation identities recorded. Their
source properties contain 98 additional vertices on Indochina and 36{,}369
on UK-2002 compared with the saved evaluations. A publisher-order rerun must
preserve these isolated vertices as well as their IDs; equal arc counts alone
do not establish the intended graph contract.
The matched workflow comparison must be extended beyond UK-100K, including
validated KEEP cases and further worker budgets. The historical
evaluator's encoding durations include its decode check and are not
fresh encode-only measurements. Neither its ratio of scoring to two
encodes nor the current codec-only acceleration is an end-to-end
order-selection speedup. These limitations constrain the claim to the
observed decision behavior until revalidation is complete.

\section{Current campaign provenance and revision history}
\label{app:campaign-protocol}
\subsection{Campaign inputs and retained records}
\begin{table}[p]\centering\small
\caption{Processed input counts for the current campaign. Evaluation denotes a version-frozen role, not independence from earlier pipeline feedback. Enron-nat is excluded from the strict count-matched tally. Historical inputs are previously studied graphs added to this campaign. Citation tuning inputs have a separate comparison.}
\label{tab:campaign-inputs}
\begin{tabular}{lrrl}\toprule
Graph & Vertices & Directed arcs & Role\\\midrule
Human1 & 493 & 15,546 & evaluation \\
Drosophila2 & 1,770 & 17,810 & evaluation \\
openflights & 3,397 & 38,460 & evaluation \\
grqc & 4,158 & 26,844 & evaluation \\
Mouse3 & 1,076 & 181,622 & evaluation \\
facebook & 4,039 & 176,468 & evaluation \\
oregon & 11,174 & 46,818 & evaluation \\
hepph & 11,204 & 235,238 & evaluation \\
wordassociation-2011 & 10,617 & 72,172 & evaluation \\
condmat & 21,363 & 182,572 & evaluation \\
astroph & 17,903 & 393,944 & evaluation \\
enron & 33,696 & 361,622 & evaluation \\
enron-nat & 67,521 & 273,649 & evaluation \\
slashdot & 77,360 & 828,161 & evaluation \\
amazon & 334,863 & 1,851,744 & evaluation \\
google & 855,802 & 5,066,842 & evaluation \\
uk-2014-tpd & 1,766,010 & 16,861,750 & evaluation \\
patents & 3,764,117 & 16,511,740 & development \\
enwiki & 4,203,294 & 101,311,589 & development \\
uk-2002 & 18,459,128 & 291,989,310 & evaluation \\
eat & 7,754 & 247,172 & historical \\
cnr-2000 & 325,557 & 3,216,152 & historical \\
amazon0601 & 403,364 & 3,387,224 & historical \\
in-2004 & 1,382,908 & 16,917,053 & historical \\
 
\bottomrule\end{tabular}\end{table}

The retained record for an arm includes the processed input path, order/direction, codec parameters, vertex and arc counts, byte count, file hash and full-adjacency-check result. Version manifests fix candidate generation, seeds, direction decisions, the timeout and selection rule. The audit checks these records against each other; it does not substitute a graph-count check for a full comparison with an unavailable baseline input.

The rewrite freezes small metadata files and their hashes separately from bulk graph and codec artifacts. Current table sizes count actual native bytes, without the earlier synthetic one-byte allowance. G-direction files retain their original bytes; T-direction winners count the verified MGS transpose extension, adding two bytes when an extension block already exists and three otherwise. Their payloads are unchanged. Selection records, original hashes and flagged hashes are retained together.

\paragraph{Direction and history.}
The campaign initially encoded explicitly transposed CSR inputs with the transpose option off. The completed follow-up added the native transpose extension to the current transposed winners and verified their decoded adjacency against G. Direction selection is part of our encoder; the WebGraph and Zuckerli comparisons use G only. The cumulative-history appendix retains the unflagged oriented file for the superseded enwiki round-2 CB result, whose costly recheck was omitted; it is not a current winner. The post-freeze CS extension on in-2004 likewise checks the explicitly transposed input and remains outside the portfolio. Dedicated payload timing records predate the extension and exclude orientation restoration. The extension checks establish correct output, not a repeated timing benchmark. Historical appendices retain their original accounting.

The enron-nat recheck identifies the local input as the loop-free largest weakly connected component of the LAW source: 67,521 vertices and 273,649 arcs. The publication reports 67,251 vertices at that same arc count. The tested preprocessing alternatives do not reproduce the published vertex count; the reason for the discrepancy remains unknown, and the graph is excluded from the published-baseline campaign tally. It is also absent from the completed common-input access pass. Comedy, comedy-full, fork-10, metroidvania-5 and ias were unavailable. Full selected-arm histories and baseline identities are in Appendix~\ref{app:round-history} and Appendix~\ref{app:baseline-provenance}.

\subsection{Corrections carried forward from earlier manuscript versions}
\label{app:revision-history}
This corrective revision changes the title from \emph{Community-Aware Vertex Ordering for Reference-Based Graph Compression: A Cross-Encoder Empirical Study}. It supersedes the following claims and descriptions in v1/v2.
\begin{itemize}
\item The global LLP seeding stage described previously did not run in the pipeline that produced the reported measurements. The implemented partitioner uses Louvain-style moves and connected-component refinement, not the full Leiden algorithm. We withdraw the attribution of the measured construction to Leiden and the claim of encoder-independent ordering gains.
\item We withdraw v2's ``28/28'' best-of-three superiority claim. Its Zuckerli comparison used default parameters rather than the published settings. The retained five-input comparison uses one fixed CS configuration and tests both Zuckerli parameter settings and both modes, with measured speed costs alongside.
\item The old CNR ``Orig.'' row mixed vertex orderings: the CG result used first-encounter CSV identifiers followed by cluster relabeling, while the other results used URL order. It was not a matched input-order comparison; v3 replaces it with explicitly identified configurations.
\item The earlier feature-ablation ladder did not reproduce. We replace its attribution of gains with the matched ablations and the separately qualified CNR study in Appendix~\ref{app:features}. We also withdraw the claimed $-0.30$ bpe gain attributed to low-degree reference search: BVGraph's default limit of three bounds reference-chain depth, not vertex degree.
\item We withdraw uniformly low-overhead random access and a general $O(k)$ query-time claim based on sample stride $k$. Index layout, reference dependencies, decoded list sizes, and cache state all affect access cost.
\end{itemize}

\section{Coder details and extended teaching examples}
\label{app:candidate-details}
\subsection{Completing the candidate-row arithmetic}
The worked row in Section~\ref{sec:candidate-toy} uses two proposal generators and fixed shared illustrative probabilities. It is a fully specified toy coding comparison, separate from the production adaptive format. Place the reference $\{3,5,8\}$ at vertex $6$, the known row $\{9,10\}$ at vertex $2$, and the current row $\{3,5,9,14\}$ at vertex $10$. Other decoded rows can be empty. The reciprocal and co-citation generators propose $2$ and $9$; self-target $10$ is excluded by the toy's loop-free convention. The candidate schedule and probabilities are fixed for both parties.

The toy fixes reference use and assumes loop-free rows; it has no production self-loop, has-reference or same-reference controls. A toy common prefix consists of gamma-coded degree $4$, gamma-coded reference distance $10-6=4$, and the three-bit copy bitmap \texttt{110}. Its length is $5+5+3=13$ bits. After copying, the direct-residual path writes gamma$(9)$ followed by gamma$(14-9)$, costing $7+5=12$ bits, for a total of 25 bits. Gamma$(9)$ is \texttt{0001001}, gamma$(5)$ is \texttt{00101}, and gamma$(14)$ is \texttt{0001110}.

The candidate path charges $-\log_2(1-1/4)$ for rejecting $2$, $-\log_2(3/4)$ for accepting $9$, and seven bits for remainder $14$. Its ideal total is
\[
13-2\log_2(3/4)+7\approx20.830\ \text{bits}.
\]
The difference is approximately $4.170$ bits before any additional mode/model description or finite-precision termination. The probabilities and toy syntax are fixed here; a deployed adaptive format must charge the information and finalization costs it actually needs. Fractional ideal lengths are not literal standalone byte counts.

Prediction can also lose. Giving the absent target $2$ probability $0.99$ costs $-\log_2(0.01)\approx6.644$ bits for its rejection. With the other costs unchanged, the post-copy path costs about $14.059$ rather than 12 bits. The example therefore includes both successful prediction and the cost of an overconfident false proposal.

The popularity example in the main text is conditional on a known distance bucket. Four of its 64 identifiers share a bucket assigned probability $1/2$, and their rank is uniform. The illustrated target's probability is $(1/2)(1/4)=1/8$, hence three bits. The other sixty targets share the remaining probability. The model is normalized; a saving for the favored group comes with a larger cost for the other group. Production popgap estimates these probabilities online.

\subsection{\CB{} parameters and contexts}
The default candidate reference window is 64. The reference score is $4|(N(v)\cap N(r))\setminus\{v\}|-|N(r)|-4$, and a reference is used only for a positive best score. Generation considers at most 256 recent incoming neighbors of the current vertex and at most 32 recent incoming neighbors of each newly found target. Rows longer than 512 are skipped for expansion. The list-entry work threshold is $2048+1024d$ for current degree $d$, checked before each row expansion; a last expansion can cross that threshold by at most one eligible row. Reciprocal proposals and list-expansion work are distinct counters.

Eight feature-conditioned inputs and a global adaptive base-rate input combine reciprocal status, binned total and per-generator proposal counts, round number, observed target indegree, target outdegree when decoded, source--target distance, remaining degree and a proposal/popularity contrast. Fixed-point stretch/squash tables, clipped weights and deterministic updates keep encoder and decoder synchronized. Probabilities are clipped away from zero and one. The source snapshot and serialized parameter block define the exact arithmetic.

Popgap keeps gaps below 32 in the exact-token path. Larger gaps use a distance bucket, one of 22 popularity buckets determined by decoded incoming counts, and a rank among all identifiers in that distance/popularity bucket, including already tested identifiers. Bucket frequencies depend on hits and exposure, with a unit prior. Fenwick trees support rank and selection. The \CB{} pipeline uses the same optional reference and generators with either remainder backend.

\subsection{Pseudocode arithmetic and source mapping}
\label{app:candidate-pseudocode}
Algorithms~\ref{alg:cand-decode}--\ref{alg:cand-aux} describe the campaign's default generators and gap/popgap backends. They omit allocation, corruption checks and the experimental cell backend. Counts are cleared between proposal rounds; within a round, equal scores keep first-proposal order. The previous reference identifier persists even across rows with no reference. Degree, self, reference and copy models are separate from the candidate mixer.

For each selected context $j$, let $a_j$ be its probability numerator on scale $2^{22}$ and $n_j$ its capped observation count. Its initial values are $209715$ and zero. Let $s_j$ be the fixed stretch-table value of the clipped 12-bit probability $a_j\mathbin{\gg}10$. With initial weights $w_j=19661$, the mixed probability numerator is
\[
p=\operatorname{clip}_{1,4095}\!\left(\operatorname{squash}\left(
\operatorname{clip}_{-2047,2047}\left(\left(\sum_{j=1}^{9}w_js_j\right)\mathbin{\gg}16\right)\right)\right).
\]
The presence bit $y$ is coded at probability $p/4096$. The selected weight set and all nine active contexts then update by
\begin{align*}
e&=6(4096y-p),\\
w_j&\leftarrow\operatorname{clip}_{-2^{24},2^{24}}
\left(w_j+((s_je+32768)\mathbin{\gg}16)\right),\\
a_j&\leftarrow\operatorname{clip}_{1,2^{22}-1}
\left(a_j+\operatorname{trunc}\!\left(\frac{2(2^{22}y-a_j)}{2n_j+3}\right)\right),\qquad
n_j\leftarrow\min(n_j+1,1023).
\end{align*}
Here right shift is arithmetic and truncation is toward zero, matching the implementation. Fourteen weight sets distinguish reciprocal status and seven total-count buckets. The ninth input is one adaptive global base rate, rather than another graph-feature partition.

Popgap orders rank identifiers in the above-base interval first, then the below-source interval for a first remainder. For at most $2^{16}$ identifiers, the rank code uses equal integer frequencies. For a larger set, the implementation recursively codes a high-16-bit block index and then a rank within that block. A shorter last block therefore has the same block probability as a full block: the resulting rank distribution is hierarchical, not exactly uniform over all identifiers. The pedagogical 64-target example stays in the equal-frequency case. Hits and exposure update only on the large-gap popgap path; identifiers move between popularity buckets after the entire row is recovered.

\begin{table}[t]\centering\small
\caption{Source map for the inspected implementation revision of \texttt{src/compression/candidate\_coder.jl}. SHA-256 hashes are supplied in \texttt{generated/coder\_access\_provenance.json}; the recorded hashes match the cited public source revision.}
\begin{tabular}{ll}\toprule
Procedure & Source lines\\\midrule
Parameters and serialization & 16--101\\
Rank coding and Fenwick operations & 207--251\\
Degree, self and reference/copy & 300--365\\
Proposal generation and ordering & 366--445\\
Contexts, prediction and mixer updates & 446--485\\
Explicit target and gap/popgap remainder & 486--580\\
Row completion and incoming-state updates & 584--598\\\bottomrule
\end{tabular}\end{table}

\subsection{BG and CS}
\label{app:bgcs}
BG uses variable-length action headers and a restricted two-reference option. CS uses fixed command prefixes and one reference per row. Their indexed readers decode only the requested chunk prefix by default and resolve cross-chunk references recursively; the access studies here explicitly request whole chunks. Both compare gaps, intervals, repeated-gap patterns and reference descriptions. In the earlier studies, search uses Fibonacci estimates even when output uses adaptive context coding; the cheapest estimated local description can differ from the one minimizing final stream bytes. The search restrictions, integer coding and restartable indexed layouts are detailed in Appendix~\ref{app:implementation}; their measurements remain in Appendix~\ref{sec:legacy-experiments}.

\section{Strategies, baseline records and development rounds}
\label{app:strategy-details}
BP uses the Petri--Mackenzie implementation of recursive graph bisection, with 20 iterations per level, leaf size 16, seed one and 16 threads. The spectral order starts from the community CSR, recomputes groups with merge threshold 64, median placement and seed one, and symmetrizes inter-group arc counts. The angles of two leading nontrivial eigenvectors of the normalized quotient adjacency order whole clusters within each component. Components with fewer than four groups keep their order. Internal LLP order is preserved.

The CG direction score models copy blocks, minimum-length-four intervals and adaptive integer streams from at most 100,000 sampled rows with window 256. A predicted relative margin of at least two percent selects one direction; a smaller magnitude retains both. This is an empirical decision rule. The \CB{} coder evaluates both directions because its preference can differ.

\subsection{Selected baseline configurations}
\label{app:baseline-provenance}
The records below select the minimum validated byte count within each external codec over the supplied Community and BP orders in G. Transposed baseline experiments are excluded. Window/depth values describe WebGraph; archive-default and access-default describe the Zuckerli command lines. Sequential WebGraph sizes exclude \texttt{.properties} and offset sidecars; the separate access table counts offsets. Native and Zuckerli sizes count the codec file. The historical matched study separately evaluates Zuckerli's published parameter settings. The completed timing/access pass uses community order only and therefore has different files from some of these minima.

The first uk-2002 WebGraph run exhausted disk space while writing its temporary arc list. Its partial counts are superseded by the successful streaming rerun's \texttt{wg\_stream.tsv} records. The report values agree with the latter. Baseline logs/byte records and successful Zuckerli decoder exits are distinct from fresh full-adjacency validation, which this manuscript rewrite does not rerun.

\begin{table}[p]\centering\scriptsize
\caption{Selected campaign baseline records, including the excluded enron-nat record for traceability. ``Default'' refers to the recorded executable's command-line defaults.}
\begin{tabular}{lllll}\toprule
Graph & WG order & WG window/depth & ZK order & ZK mode\\\midrule
Human1 & community & w16-unlimited & community & access-default \\
Drosophila2 & bp & w16-unlimited & bp & access-default \\
openflights & bp & w16-unlimited & bp & access-default \\
grqc & community & w16-unlimited & community & access-default \\
Mouse3 & bp & w16-unlimited & bp & access-default \\
facebook & community & w16-unlimited & community & access-default \\
oregon & bp & w16-unlimited & bp & access-default \\
hepph & community & w16-unlimited & community & access-default \\
wordassociation-2011 & bp & w16-unlimited & bp & access-default \\
condmat & community & w16-unlimited & community & access-default \\
astroph & community & w16-unlimited & community & access-default \\
enron & bp & w16-unlimited & bp & access-default \\
enron-nat & bp & w16-unlimited & bp & access-default \\
slashdot & bp & w16-unlimited & bp & access-default \\
amazon & bp & w16-unlimited & community & archive-default \\
google & community & w16-unlimited & community & archive-default \\
uk-2014-tpd & bp & w16-unlimited & bp & archive-default \\
patents & bp & w16-unlimited & bp & archive-default \\
enwiki & bp & w16-unlimited & bp & archive-default \\
uk-2002 & community & w16-unlimited & community & archive-default \\
eat & bp & w16-unlimited & bp & access-default \\
cnr-2000 & community & w16-unlimited & community & archive-default \\
amazon0601 & community & w16-unlimited & community & archive-default \\
in-2004 & community & w16-unlimited & community & archive-default \\
 
\bottomrule\end{tabular}\end{table}

\subsection{Published HyperFast configurations}
\label{app:hf-configurations}
The main campaign compares against the smallest published central estimate in HyperFast's Table 6, over four embedders and two bucket-count settings. Confidence-interval endpoints are not used as configuration estimates. The table below preserves HMCS with $B_r=B_d=16$ and identifies the best configuration, including ties. Historical/citation results marked R are separately reconstructed CLOVE estimates and are not minima over this published menu.

\begin{table}[p]\centering\small
\caption{Published HyperFast bpe: the previous HMCS-16 comparator and the best Table 6 configuration. Configuration suffixes give $B_r=B_d$. Values are the publication's rounded central estimates.}
\label{tab:hf-configurations}
\begin{tabular}{lrrl}\toprule
Graph & HMCS-16 & Best & Best embedder--buckets\\\midrule
Human1 & 3.640 & 3.610 & BFKL-8 \\
Drosophila2 & 5.070 & 5.000 & CLOVE-16 \\
openflights & 3.880 & 3.830 & CLOVE-16 \\
grqc & 3.850 & 3.720 & CLOVE-16 \\
Mouse3 & 2.600 & 2.420 & BFKL-16 \\
facebook & 2.800 & 2.800 & HMCS-16 \\
oregon & 4.290 & 4.200 & CLOVE-16 \\
hepph & 3.240 & 3.210 & CLOVE-16 \\
wordassociation-2011 & 7.540 & 7.480 & CLOVE-16 \\
condmat & 5.610 & 5.470 & CLOVE-16 \\
astroph & 5.480 & 5.480 & HMCS-16, CLOVE-16 \\
enron & 5.670 & 5.610 & CLOVE-16 \\
slashdot & 8.320 & 8.320 & HMCS-16 \\
amazon & 5.180 & 5.110 & CLOVE-16 \\
google & 4.090 & 4.090 & HMCS-16 \\
uk-2014-tpd & 8.750 & 8.750 & HMCS-16 \\
patents & 10.570 & 10.570 & HMCS-16 \\
enwiki & 10.140 & 10.140 & HMCS-16 \\
uk-2002 & 2.990 & 2.990 & HMCS-16 \\
 
\bottomrule\end{tabular}\end{table}

\subsection{Complete cumulative history}
\label{app:round-history}
Round 1 was frozen on 20 November 2026; rounds 2 and 3 on 21 November. Their manifests preserve earlier outcomes and record changes before each round's new encodes. Google informed the second-round order choice. Patents was reclassified for development on 21 November after its round-2 loss; the loss remains in the table. Enwiki was a development input from the first round. These experimental rounds differ from manuscript release numbers.

\begin{table}[p]\centering\small
\caption{All count-eligible campaign and previously studied inputs in the cumulative history. Definitions and bundled-change qualification are as in Table~\ref{tab:method-ablation}. All final selections are complete; the superseded enwiki round-2 file has the oriented-file exception described in Appendix~\ref{app:campaign-protocol}.}
\begin{tabular}{lrrrl}\toprule
Graph & Round 1 & Round 2 & Round 3 & Selected arm\\\midrule
Human1 & 1.801 & 1.709 & 1.709 & Community/G/CB-gap \\
Drosophila2 & 3.387 & 3.331 & 3.324 & Community/G/CB-pop \\
openflights & 2.303 & 2.172 & 2.133 & Spectral/G/CB-pop \\
grqc & 2.625 & 2.162 & 2.079 & Spectral/G/CB-pop \\
Mouse3 & 1.305 & 1.146 & 1.144 & BP/G/CB-pop \\
facebook & 1.461 & 1.328 & 1.326 & Spectral/G/CB-pop \\
oregon & 3.024 & 2.860 & 2.721 & Spectral/G/CB-pop \\
hepph & 1.419 & 1.266 & 1.212 & Spectral/G/CB-pop \\
wordassociation-2011 & 8.767 & 7.418 & 7.350 & Community/G/CB-pop \\
condmat & 3.007 & 2.424 & 2.364 & Spectral/G/CB-pop \\
astroph & 2.280 & 1.946 & 1.919 & Spectral/G/CB-pop \\
enron & 3.136 & 2.724 & 2.705 & Spectral/G/CB-pop \\
slashdot & 6.024 & 5.953 & 5.927 & Spectral/G/CB-pop \\
amazon & 3.665 & 3.221 & 2.867 & Spectral/G/CB-pop \\
google & 4.588 & 3.345 & 3.118 & Spectral/G/CB-pop \\
uk-2014-tpd & 8.678 & 7.810 & 7.643 & Spectral/G/CB-pop \\
patents & 12.224 & 11.153 & 10.228 & Spectral/G/CB-pop \\
enwiki & 9.596 & 8.197 & 7.770 & Spectral/G/CB-pop \\
uk-2002 & 1.604 & 1.198 & 1.175 & Spectral/T/CB-pop \\
eat & 8.020 & 7.205 & 7.180 & Community/G/CB-pop \\
cnr-2000 & 2.137 & 1.609 & 1.541 & Spectral/G/CB-pop \\
amazon0601 & 6.097 & 4.809 & 4.602 & Spectral/G/CB-pop \\
in-2004 & 1.667 & 1.227 & 1.194 & Spectral/G/CB-pop \\
 
\bottomrule\end{tabular}\end{table}

\subsection{The pooled-precision gate}
\label{app:failed-gate}
A development gate estimated candidate cost from one pooled precision per graph. It predicted gains on only two of eight tuning inputs, whereas the \CB{} coder improved all eight. That prediction failed and the method therefore runs candidate coding on every input.

The completed replay holds the order, direction, reference rule and remainder backend fixed. Encode A enables the default generators and remainder seeding; B disables both. All sixteen instrumented streams are byte-identical to the original implementation and reproduce the recorded A/B sizes to their four printed decimals. An arc-set check confirms that B's explicit targets partition into A's successful proposals and A's surviving remainder targets.

For exact stream attribution, a mark records $8(b+c)-\log_2 R$, where $b$ is emitted bytes, $c$ is the range coder's cache count and $R$ its current range. Differences are assigned to degree, self, reference, copy, candidate and remainder streams; termination is charged separately. This includes finite-precision effects, unlike ideal log loss. The stream identity closes to the recorded numerical precision (reported residual below $10^{-6}$ bit per file). That equality checks bookkeeping; byte reproduction and the arc-set check provide the independent controls.

\begin{table}[t]\centering\small
\caption{Replay in the previously selected direction (G original, T transposed), in bits per arc. $S_G$ is the frozen gate prediction on G; four measured directions are T. $S_{\rm rem}$ and $\ell_{\rm code}$ are measured in the displayed direction. Net is actual B minus A; the small remaining overhead term is omitted from the rounded columns. The new T-direction gate indices are post-hoc and are not substituted for $S_G$.}
\label{tab:gate-replay}
\begin{tabular}{llrrrr}\toprule
Graph & Direction & $S_G$ & $S_{\rm rem}$ & $\ell_{\rm code}$ & Net\\\midrule
wiki-Vote & T & -0.059 & 4.438 & 3.541 & 0.898 \\
eat & G & 0.612 & 7.366 & 5.912 & 1.454 \\
soc-Epinions1 & G & 0.558 & 5.987 & 4.057 & 1.930 \\
email-EuAll & T & -0.104 & 3.331 & 1.571 & 1.760 \\
web-NotreDame & G & -0.134 & 1.313 & 0.531 & 0.783 \\
web-Stanford & T & -0.121 & 1.057 & 0.572 & 0.486 \\
cnr-2000 & G & -0.143 & 0.553 & 0.311 & 0.242 \\
web-BerkStan & T & -0.132 & 0.753 & 0.426 & 0.327 \\
 
\bottomrule\end{tabular}\end{table}

Against its own baseline, the gate's frozen G predictions have descriptive Spearman correlation $0.81$ with the recorded-best-direction gains over these eight graphs. For this frozen-G versus recorded-best-direction pairing, the errors are positive, from 0.385 to 1.864 bits per arc; the post-hoc same-direction pairing instead gives about 0.38--2.33 bits per arc. This is a sign/level failure, rather than a reversed ranking. Comparing to percentage improvements over CG changes both the baseline and denominator. For example, B is about 1.10 and 1.39 bpe larger than CG on EAT and Epinions.

Within each direction, remainder saving splits into the B-price of recovered targets minus the change in price of surviving remainder targets. On the web graphs the gate's aggregate candidate-cost estimate differs from measured cost by at most about 0.09 bpe: extra tests and lower conditional costs largely cancel. Underestimated target coverage and explicit-target cost explain the remaining saving-side error. The 64-bin empirical code, grouping events by the adaptive model's predicted probability, is within 0.018 bpe of coded candidate cost in the recorded-best directions. This measures calibration on an adaptive partition, not the overhead relative to an independently available static model. Full attribution tables and both directions are retained with the replay.

\subsection{Larger CG reference windows}
\label{app:large-windows}
The follow-up fixes the retained selected order/direction for each graph and tests three larger windows with plain and reciprocal modes. All eighteen files decode to the input adjacency. The table reports the smallest tested file per graph. These minima remain above the strongest recorded baseline; the reconstructed HyperFast caveat applies to EAT. The extended sweep has no uk-2002 record.
\begin{center}\small
\begin{tabular}{lllr}\toprule
Graph & Window & Mode & bpe\\\midrule
wordassociation-2011 & 8192 & Reciprocal & 8.396 \\
eat & 16384 & Reciprocal & 7.537 \\
in-2004 & 4096 & Reciprocal & 1.476 \\
 
\bottomrule\end{tabular}\end{center}

\subsection{Source and file identities}
The manuscript source bundle includes \texttt{generated/method\_provenance.json}: full selected native file hashes, byte counts and arm identities, the byte-record provenance of each baseline minimum, and the library digests from the frozen manifests. The source-tree digest prefixes for rounds 1--3 are listed below; the JSON carries their complete values. These identify the recorded source trees and artifacts, rather than asserting a fresh reproduction of them.

\begin{center}\small
\begin{tabular}{lll}\toprule
Round & Freeze date & Library source SHA-256 prefix\\\midrule
1 & 2026-11-20 & \texttt{12a856003e2c} \\
2 & 2026-11-21 & \texttt{a1b6f78daaaa} \\
3 & 2026-11-21 & \texttt{569c8bddf449} \\
 
\bottomrule\end{tabular}\end{center}

\subsection{Completed common-input decode measurements}
\label{app:completed-timing}
\begin{table}[p]\centering\small
\caption{All completed common-community-order, original-direction full-decode measurements in seconds. ZK max/access denotes Zuckerli's archive/access mode, WG denotes the indicated window/depth configuration, and CG is plain window 256. Zuckerli uses median-of-three process wall time; WG uses best-of-three in-JVM scans; CG uses median-of-three in-process decoding. These are configuration-specific times, not times for the size-selected portfolio.}
\begin{tabular}{lrrrrr}\toprule
Graph & ZK max & ZK access & WG w7/m3 & WG w16 & CG\\\midrule
Human1 & 0.0020 & 0.0020 & 0.0011 & 0.0016 & 0.0008 \\
Drosophila2 & 0.0022 & 0.0025 & 0.0016 & 0.0017 & 0.0022 \\
openflights & 0.0025 & 0.0026 & 0.0023 & 0.0026 & 0.0019 \\
grqc & 0.0022 & 0.0024 & 0.0017 & 0.0020 & 0.0013 \\
Mouse3 & 0.0037 & 0.0036 & 0.0055 & 0.0055 & 0.0055 \\
facebook & 0.0043 & 0.0043 & 0.0075 & 0.0056 & 0.0061 \\
oregon & 0.0031 & 0.0035 & 0.0021 & 0.0047 & 0.0027 \\
hepph & 0.0051 & 0.0051 & 0.0083 & 0.0055 & 0.0063 \\
wordassociation-2011 & 0.0044 & 0.0047 & 0.0023 & 0.0037 & 0.0063 \\
condmat & 0.0056 & 0.0056 & 0.0051 & 0.0053 & 0.0084 \\
astroph & 0.0075 & 0.0071 & 0.0141 & 0.0163 & 0.0153 \\
enron & 0.0093 & 0.0090 & 0.0119 & 0.0078 & 0.0212 \\
slashdot & 0.0175 & 0.0146 & 0.0226 & 0.0355 & 0.0445 \\
amazon & 0.0396 & 0.0329 & 0.0618 & 0.0498 & 0.1055 \\
google & 0.0716 & 0.0615 & 0.1147 & 0.1153 & 0.2091 \\
eat & 0.0065 & 0.0062 & 0.0076 & 0.0072 & 0.0122 \\
cnr-2000 & 0.0261 & 0.0228 & 0.0394 & 0.0356 & 0.0796 \\
amazon0601 & 0.0620 & 0.0503 & 0.0770 & 0.0771 & 0.1748 \\
in-2004 & 0.0932 & 0.0810 & 0.1730 & 0.1563 & 0.4257 \\
patents & 0.3538 & 0.2792 & 0.4539 & 0.4642 & 1.2564 \\
enwiki & 1.6987 & 1.2654 & 1.9628 & 1.9201 & 6.7648 \\
uk-2014-tpd & 0.2543 & 0.1833 & 0.2859 & 0.2821 & 0.8866 \\
uk-2002 & 1.4788 & 1.3022 & 2.2267 & 2.1315 & 9.0595 \\
 
\bottomrule\end{tabular}\end{table}

\subsection{Correctness of the uncached CS-RA reader}
\label{app:cold-reader-check}
The comparator pass's seek and full-adjacency checks used cached readers and compared sorted rows. A follow-up checks the reader used for cold timing: whole-chunk decoding with \texttt{cache\_budget\_bytes=0}. It reconstructs the same community-ordered G inputs with the same CS-RA settings. The regenerated files have the same byte counts as the comparator files; the check does not retain hashes establishing byte-for-byte identity.

Every vertex is checked on Human1, openflights and EAT. On the other six graphs, the script draws 20,000 uniformly random vertices with seed seven, adds each chunk's first and last vertex, and removes duplicates. Every checked row equals G's row as stored, before sorting (Table~\ref{tab:cold-reader-check}). This validates the uncached path on these nine graphs; it does not extend the sample to all 22 completed inputs or resolve the enwiki failure. The script, original progress log, counts and file lengths are retained with the evidence; \texttt{generated/cold\_reader\_provenance.json} supplies their identities.

\begin{table}[t]\centering\small
\caption{Follow-up verification of uncached CS-RA. Matched counts exact row equality before sorting. The larger-graph sets combine random draws with chunk boundaries.}
\label{tab:cold-reader-check}
\begin{tabular}{lrrrl}\toprule
Graph & Vertices & Checked & Matched & Coverage\\\midrule
Human1 & 493 & 493 & 493 & All vertices \\
openflights & 3397 & 3397 & 3397 & All vertices \\
eat & 7754 & 7754 & 7754 & All vertices \\
cnr-2000 & 325557 & 19743 & 19743 & Sample + boundaries \\
google & 855802 & 20583 & 20583 & Sample + boundaries \\
amazon0601 & 403364 & 19922 & 19922 & Sample + boundaries \\
in-2004 & 1382908 & 21177 & 21177 & Sample + boundaries \\
patents & 3764117 & 23589 & 23589 & Sample + boundaries \\
uk-2014-tpd & 1766010 & 21599 & 21599 & Sample + boundaries \\
 
\bottomrule\end{tabular}\end{table}

\end{document}